\documentclass[prl,twocolumn,showpacs,superscriptaddress,notitlepage]{revtex4-2}
\usepackage{makeidx}
\usepackage[utf8]{inputenc}
\usepackage{graphicx}
\usepackage{amsmath}
\usepackage{mathrsfs}
\usepackage{graphicx}
\usepackage{braket}
\usepackage[subrefformat=parens,labelformat=parens,caption=false]{subfig}
\usepackage{amsfonts}
\usepackage{dcolumn}
\usepackage{bm}
\usepackage{bbm}
\usepackage{color}
\usepackage[dvipsnames]{xcolor}
\usepackage{esint}
\usepackage{lineno}
\usepackage{verbatim}
\usepackage{cancel}
\usepackage[breaklinks,
            colorlinks,
            urlcolor=blue,
            linkcolor=blue,
            anchorcolor=blue,
            citecolor=blue]{hyperref}        
\usepackage{soul}
\usepackage{amssymb}

\newtheorem{theorem}{Theorem}

\newcommand\Osq{\mathbin{\text{\scalebox{.84}{$\square$}}}}

\begin{document}

\title{Dicke superradiance requires interactions beyond nearest-neighbors} 

\author{Wai-Keong Mok}
\affiliation{Centre for Quantum Technologies, National University of Singapore, 3 Science Drive 2, Singapore 117543}
\affiliation{California Institute of Technology, Pasadena, CA 91125, USA}
\author{Ana Asenjo-Garcia}
\affiliation{Department of Physics, Columbia University, New York, New York 10027, USA}
\author{Tze Chien Sum}
\affiliation{Division of Physics and Applied Physics, School of Physical and Mathematical Sciences, Nanyang Technological University,
Singapore 637371}
\author{Leong-Chuan Kwek}
\affiliation{Centre for Quantum Technologies, National University of Singapore, 3 Science Drive 2, Singapore 117543}
\affiliation{MajuLab, CNRS-UNS-NUS-NTU International Joint Research Unit, Singapore UMI 3654, Singapore}
\affiliation{National Institute of Education, Nanyang Technological University, Singapore 637616, Singapore}
\affiliation{Quantum Science and Engineering Centre (QSec), Nanyang Technological University, Singapore}

\begin{abstract}
Photon-mediated interactions within an excited ensemble of emitters can result in Dicke superradiance, where the emission rate is greatly enhanced, manifesting as a high-intensity burst at short times. The superradiant burst is most commonly observed in systems with long-range interactions between the emitters, although the minimal interaction range remains unknown. Here, we put forward a new theoretical method to bound the maximum emission rate by upper-bounding the spectral radius of an auxiliary Hamiltonian. We harness this tool to prove that for an arbitrary ordered array with only nearest-neighbor interactions in all dimensions, a superradiant burst is not physically observable. We show that Dicke superradiance requires minimally the inclusion of next-nearest-neighbor interactions. For exponentially-decaying interactions, the critical coupling is found to be asymptotically independent of the number of emitters in all dimensions, thereby defining the threshold interaction range where the collective enhancement balances out the decoherence effects. Our findings provide key physical insights to the understanding of collective decay in many-body quantum systems, and the designing of superradiant emission in physical systems for applications such as energy harvesting and quantum sensing.
\end{abstract}

\maketitle

\textit{Introduction.---} Collective spontaneous emission of $N$ initially-inverted atoms with identical all-to-all interactions mediated by the electromagnetic vacuum results in a burst of light with intensity scaling as $N^2$~\cite{Dicke1954coherence, Gross1982superradiance,rehler1971superradiance}. This phenomenon is commonly referred to as ``Dicke superradiance" or ``superradiant burst". Over the past decades, this many-body phenomenon has attracted a lot of interest in both theoretical~\cite{stroud1972superradiant,friedberg1972limited,coffey1978effect,banfi1975superfluorescence,Clemens2003collective,lin2012superradiance,Bhatti2015superbunching,Longo2016tailoring,Nefedkin2017superradiance,Pleinert2017hyperradiance,Shammah2017superradiance,Houde2017fast,Masson2020many,Fuchs2021superradiance,Lemberger2021radiation,Robicheaux2021theoretical,Rubies2021superradiance,malz2022large,Masson2022universality,Orioli2022emergent,Sierra2021dicke} and experimental studies~\cite{skribanowitz1973observation,raimond1982collective} using a multitude of physical platforms such as trapped ions~\cite{DeVoe1996observation}, molecular aggregates~\cite{spano1989superradiance,spano1990temperature,spano1991cooperative,meier1997temperature}, solid-state emitters~\cite{Scheibner2007superradiance,Yukalov2010dynamics,Bradac2017room,Haider2021superradiant,Mello2022extended}, cold atoms and molecules~\cite{Wang2007superradiance,Ferioli2021laser,Liedl2022collective,trebbia2022tailoring}, and superconducting qubits~\cite{Lambert2016superradiance,Wang2020controllable,Orell2021collective}, with wide-ranging applications including the generation of multi-photon states with  improved metrological properties~\cite{gonzalez2015deterministic,paulisch2019quantum,groiseau2021proposal,perarnau2020multimode,Lemberger2021radiation}, energy harvesting~\cite{monshouwer1997superradiance,scholes2002designing,celardo2014cooperative}, ultrabright LEDs~\cite{raino2020superradiant} and quantum sensing~\cite{yang2021realization,higgins2014superabsorption}.

The atoms in Dicke's original model were assumed to be confined within a spatial extent smaller than the emission wavelength $\lambda$. Consequently, the atoms become indistinguishable with respect to the absorption or emission of photons, such that their quantum state $\ket{j=N/2, m}$ (with $-N/2 \leq m \leq N/2$) is permutation-invariant. This permutation symmetry greatly reduces the complexity of the problem, as it constrains the dynamics to $N+1$ states, instead of exploring the full Hilbert space (which scales as $2^N$). Recently, there has been substantial research progress with extended systems where atoms are distributed over a region larger than $\lambda$, thus breaking this symmetry. Of particular interest are ordered atomic arrays~\cite{Masson2020atomic,ruks2022greens}, in which the superradiant properties can be greatly affected by the geometry and dimensionality of the lattice~\cite{Masson2022universality,Sierra2021dicke,Robicheaux2021theoretical,Rubies2021superradiance,Masson2020many,silvia2022many}. The interactions between the emitters are typically modelled by long-range dipole-dipole interactions mediated via the electromagnetic vacuum~\cite{carmichael2000quantum,mattiotti2020thermal}.

A long-standing fundamental question is the \textit{minimal} interaction range required for the occurrence of a superradiant burst. Intuitively, superradiance can be thought of as a  competition between (transient) phase synchronization, which leads to the buildup of atomic correlations, and decoherence~\cite{bellomo2017quantum}. Both effects stem from the same dissipative interactions~\cite{Clemens2003collective,Masson2022universality}. Since synchronization of nonlinear classical phase oscillators has been demonstrated with nearest-neighbor (NN) coupling~\cite{hassan2003synchronization}, one may expect the atomic phases to synchronize for sufficiently strong NN interactions resulting in a superradiant burst~\cite{bellomo2017quantum}. Moreover, for a fixed interaction range, higher dimensionality was reported to result in stronger superradiance due to long-range order~\cite{Robicheaux2021theoretical, Sierra2021dicke}. On the flip side, it could also be argued that for short-range interactions, the buildup of correlations is not strong enough to overcome decoherence, thereby preventing superradiance. 

In this Letter, we prove that superradiant burst is impossible in an arbitrary $D$-dimensional array with only nearest-neighbor interactions, for arbitrary times and initial states. That is, we show that, in all cases, the emission rate is upper bounded by that of independent emitters, resulting in no enhancement from collective dynamics. Including next-nearest-neighbor interactions, we show that a superradiant burst can be physically observed for certain values of the interaction strengths, thereby defining a minimal interaction range for superradiance. Another question is the \textit{threshold} interaction range, which we define to be such that the critical coupling required for a burst becomes independent of the number of emitters, for any $D$. We show that exponentially-decaying interactions lie on the threshold interaction range for which the synchronization of the dipoles arising from the emission balances the decoherence effects.
\begin{figure}
\centering
\subfloat{%
  \includegraphics[width=0.95\linewidth]{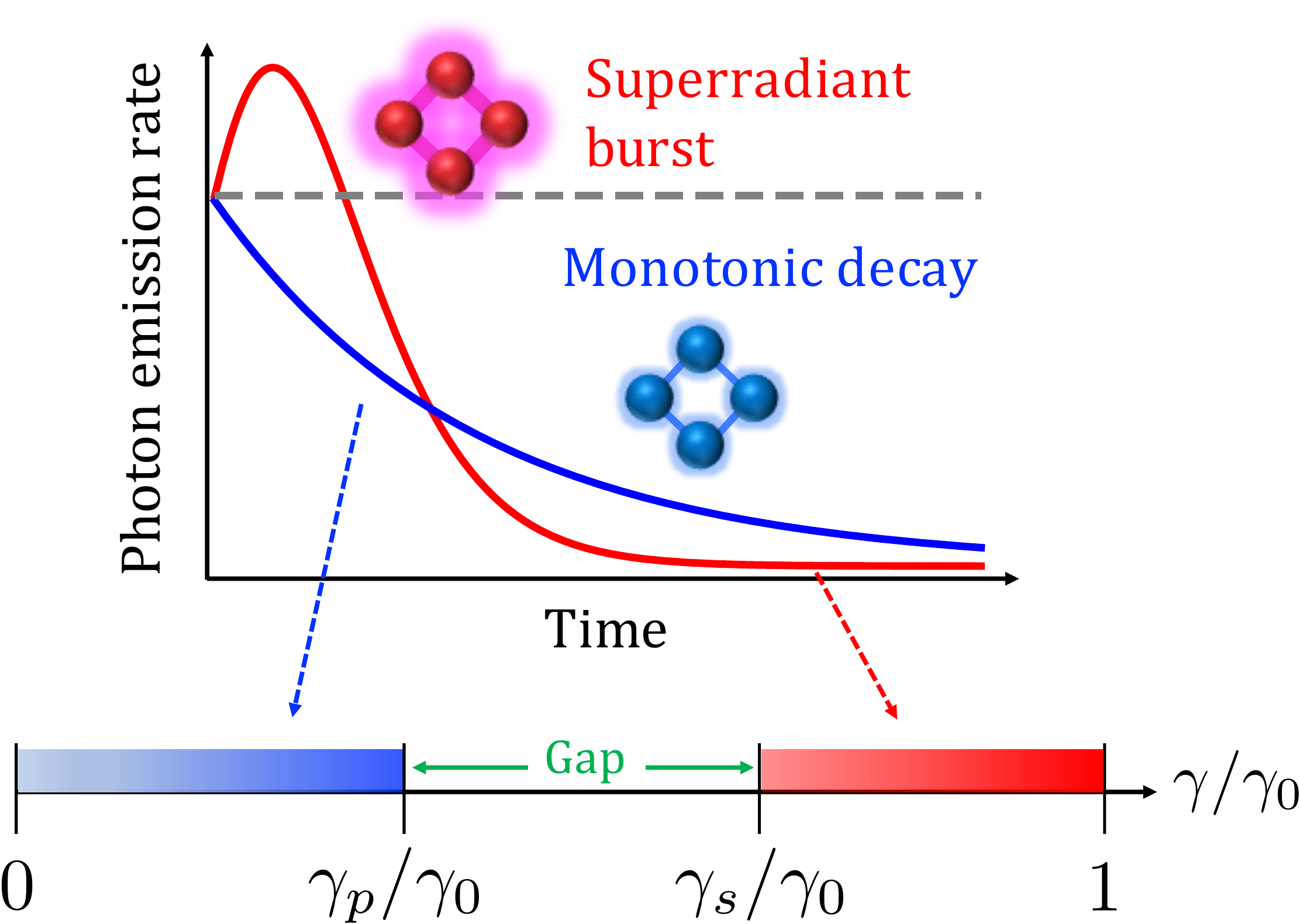}%
  \label{}%
}
	\caption{Dynamics of the photon emission rate $R(t)$ for emitter arrays with only  nearest-neighbor interactions of strength $\gamma$, normalized by the individual emitter decay rate $\gamma_0$. For $\gamma/\gamma_0 < \gamma_s$, $\dot{R}(0) < 0$ and the photon emission rate decays monotonically without a superradiant burst (blue). Superradiance occurs for $\gamma/\gamma_0 > \gamma_s$ (red). The physically-valid regime is defined by $0< \gamma/\gamma_0 \leq \gamma_p$. For nearest-neighbor interactions, $\gamma_p < \gamma_s$ (with a finite gap between $\gamma_p$ and $\gamma_s$) for any arbitrary emitter array in all dimensions, rendering Dicke superradiance physically impossible.}
	\label{fig:SR_NN_scheme}
\end{figure}

\textit{Model.---}The dynamics of an undriven ensemble of $N$ emitters can be described by the Lindblad master equation (setting $\hbar = 1$)
\begin{equation}
\begin{split}
    \dot{\rho} &= -i\sum_{i,j = 1}^{N}\left[  J_{ij} \sigma_i^+ \sigma_j^- , \rho \right] + \sum_{i,j = 1}^{N} \gamma_{ij} \mathcal{D}[\sigma_i^-, \sigma_j^-]\rho \equiv \mathcal{L}[\rho],
\end{split}
\label{eq:ME_original}
\end{equation}
with $J_{ij} = J_{ji}^*$ and $\gamma_{ij} = \gamma_{ji}^*$ to ensure Hermiticity. The raising and lowering operators for the $j^{\text{th}}$ emitter are denoted as $\sigma_j^+ \equiv \ket{e_i} \bra{g_i}$ and $\sigma_j^- \equiv \ket{g_i}\bra{e_i}$ which describe transitions between the ground state $\ket{g_i}$ and excited state $\ket{e_i}$. The first term contains the coherent Hamiltonian interactions between the emitters, while the second term captures processes such as collective and local dissipation of the emitters via the superoperator $\mathcal{D}[\sigma_i^-,\sigma_j^-]\rho = \sigma_i^- \rho \sigma_j^+ - \{\sigma_j^+ \sigma_i^-,\rho \}/2$. We assume $J_{ij}$ and $\gamma_{ij}$ to be time-independent, such that the superoperator $\mathcal{L}$ generates a dynamical semigroup describing the dynamics of a Markovian open quantum system. 

For a physically valid evolution (i.e., a completely positive and trace-preserving map), the matrix $\mathbf{\Gamma}$ containing the elements $\gamma_{ij}$ (which we will refer to as the \textit{decoherence matrix}) must be positive semi-definite~\cite{Lindblad1976on,Breuer2007theory,Lidar2019lecture}. The decoherence matrix can be diagonalized to yield $N$ decay rates $\Gamma_\nu \geq 0$, with $\nu \in \{1, \ldots, N\}$ and the corresponding collective jump operators $\hat{c}_\nu$. The total photon emission rate of the emitters, integrated over all emission directions, is defined for any state $\rho$ as 
\begin{equation}
    R_{\rho} \equiv \sum_{\nu=1}^N \Gamma_\nu \braket{\hat{c}_\nu^\dag \hat{c}_\nu} = \sum_{\nu=1}^N \Gamma_\nu \text{Tr}({\hat{c}_\nu^\dag \hat{c}_\nu}\rho).
\label{eq:rate_defn}
\end{equation}
For independent emitters with $\gamma_{ij} = \gamma_0 \delta_{ij}$, the total emission rate has a maximum of $N\gamma_0$ (saturated by the fully-excited state), and $R(t) \equiv R_{\rho(t)}$ decays exponentially. However, interactions between the emitters can cause $R(t)$ to increase beyond its initial value. This speedup in emission is commonly referred to as the superradiant burst, first discovered by Dicke~\cite{Dicke1954coherence} (see Fig.~\ref{fig:SR_NN_scheme}). Throughout this work, we refer to superradiant burst as the increase in the total emission rate beyond $N\gamma_0$, but the peak intensity need not scale as $N^2$. In general, characterizing the burst at arbitrary times can be difficult, hence one typically uses
\begin{equation}
    \dot{R}_\rho = i\sum_{\nu} \braket{[H,\hat{c}_\nu^\dag \hat{c}_\nu]} - \sum_{\mu,\nu} \Gamma_{\mu}\Gamma_{\nu}\braket{\hat{c}_{\mu}^\dag [\hat{c}_\mu,\hat{c}_\nu^\dag]\hat{c}_\nu]}
\label{eq:Rdot}
\end{equation}
evaluated at the fully-excited initial state $\rho(0)$, with $\dot{R}(0) \equiv \dot{R}_{\rho(0)} > 0$ a sufficient condition for a superradiant burst. While we consider the burst at $t = 0$, we will provide physical justification on why this is sufficient.

Here, we put forward a new (and complementary) criterion to preclude any possibility of a burst: by a simple change of basis, one can write Eq.~\eqref{eq:rate_defn} as the expectation value of an auxiliary spin Hamiltonian
\begin{equation}
    H_\Gamma =\sum_{j,k=1}^{N}\gamma_{kj} \sigma_j^+ \sigma_k^-,
    \label{hgamma}
\end{equation}
with $R_\rho = \text{tr}(H_\Gamma \rho)$. The maximum photon emission rate can thus be calculated by bounding the spectral radius of the auxiliary spin Hamiltonian. If the upper bound is equal or smaller than $N \gamma_0$, no burst can occur for all times and arbitrary initial states. While finding the largest eigenvalue of $H_\Gamma$ may be non-trivial,  this criterion allows one to definitively prove the absence of a burst for arbitrary times, thus going beyond the condition $\dot{R}(0) \equiv \dot{R}_{\rho(0)} > 0$. Furthermore, this approach opens up the possibility of finding theoretical limits for the emission rate arising from superradiant dynamics, as we show below and in the Supplementary Information~\cite{supp}.


\textit{No superradiance for nearest-neighbor coupling.---}
\label{sec:nogo_NN}Let us consider a hypercube array of $N$ emitters with arbitrary dimension $D$ $(N = n^D)$. For the case of NN interactions, $\gamma_{ii} \equiv \gamma_0 = 1$ and $\gamma_{ij} = \gamma$ if emitters $i$ and $j$ are nearest-neighbor ($\gamma_{ij}=0$ otherwise). The coupling $\gamma \in [0,1]$ is required for the matrix $\mathbf{\Gamma}$ to be positive semidefinite. Without loss of generality, we have assumed $\gamma_{ij}$ to be real and positive. We prove that for this model, superradiant burst cannot occur for any $t > 0$, for any arbitrary initial state and for any Hamiltonian coupling $J_{ij}$. To determine the physically valid regime, we impose the condition that $\mathbf{\Gamma}$ is positive semidefinite. Notice that the decoherence matrix can be expressed as $\mathbf{\Gamma} = \mathbf{I}_{N} + \gamma \mathbf{A}$, where $\mathbf{I}_{N}$ is the $N \times N$ identity matrix, and $\mathbf{A}$ is the adjacency matrix of a $n \times n$ grid graph. Using the fact that the grid graph is the Cartesian product of $D$ path graphs $P_n \Osq \cdots \Osq P_n$, it can be shown that the smallest eigenvalue of $\mathbf{\Gamma}$ is~\cite{supp}
\begin{equation}
    \Gamma_{\text{min}} = 1 - 2D \gamma \cos\left( \frac{\pi}{N^{1/D} + 1} \right), 
\label{eq:minGamma_d}
\end{equation}
which gives the physically valid regime as $\gamma \leq \gamma_p$,
\begin{equation}
    \gamma_p = \left[ 2D \cos\left( \frac{\pi}{N^{1/D} + 1} \right) \right]^{-1}.
\end{equation}
This rate reduces to $\gamma_p =  1/(2D)$ in the $N \to \infty$ limit, or when imposing periodic boundary conditions for a finite $N$. This can be regarded as coming from the coordination number for each emitter, which approaches $2D$ in the infinite-array limit. We now state our main result. 
\begin{theorem}
Let $\mathbf{\Gamma}$ be the decoherence matrix for a nearest-neighbor interaction model, with $\gamma_{ij} = \delta_{ij} + \gamma \delta_{\braket{ij}}$, where $\gamma \in [0,1]$, and $\delta_{\braket{ij}} = 1$ if the emitters indexed by $i$ and $j$ are nearest-neighbor on the $D-$dimensional regular lattice, and $0$ otherwise. For $\gamma \leq (2D)^{-1}$, the emission rate $R_\rho$ is maximized by the fully-excited state $\ket{e}^{\otimes N}$ with $R_\rho=N$.
\end{theorem}
We provide a sketch of the proof here, while the details can be found in the Supplementary Information~\cite{supp}. By expressing $H_\Gamma$ in the product-state basis and using the Gershgorin circle theorem~\cite{golub2013matrix}, we can upper bound $\max_t R(t) \leq N$ in the physically valid regime $\gamma < 1/(2D)$. This is saturated by $N$ independent emitters in the fully-excited state, with eigenvalue $N$. Hence, Theorem 1 implies that superradiant burst is impossible at all times. To gain a deeper physical understanding, we evaluate the superradiant regime $\gamma > \gamma_s$ for the fully-excited initial state, characterized by the transition at $\dot{R}(0) = 0$, for which~\cite{supp}
\begin{equation}
    \gamma_s = \left[2D(1-N^{-1/D})\right]^{-1/2}.
\label{eq:us_NNhypercube}
\end{equation}
For all $2 < N^{1/D} < \infty$, it can be shown that $\gamma_p < \gamma_s^2$ and therefore $\gamma_p < \gamma_s$. Hence, the superradiant regime does not overlap with the physically valid regime. Generalization to the hyper-rectangle configuration where the number of sites along each dimension can be different is straightforward, and the same conclusion is obtained~\cite{supp}. While our analysis of the NN model is valid for any initial state, we consider a fully-inverted initial state for the next two sections: the analysis of next-nearest neighbor and exponentially decaying interactions.
\begin{figure}
\centering
\subfloat{%
  \includegraphics[width=0.95\linewidth]{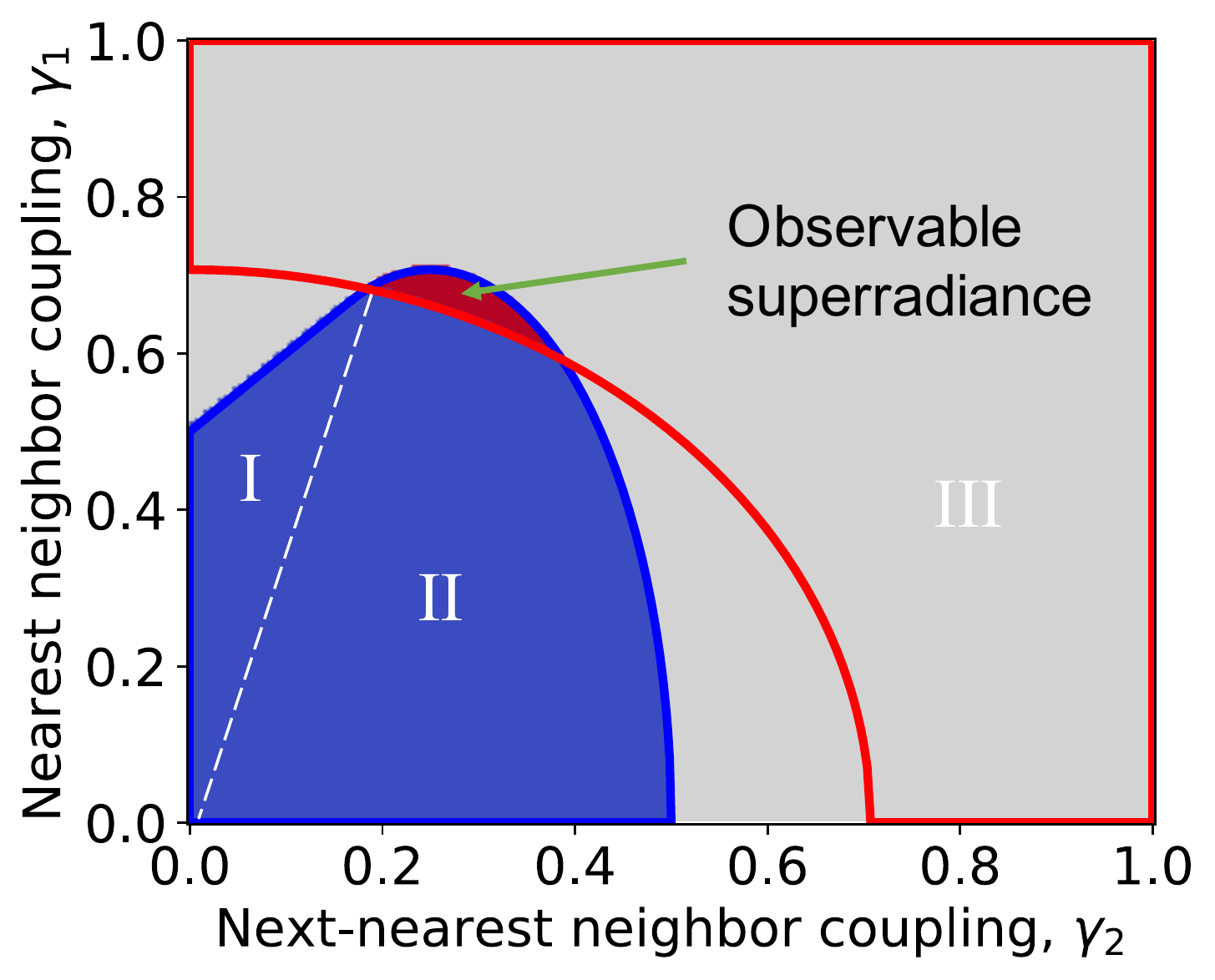}%
  \label{}%
}
	\caption{Region of superradiant burst in the $\gamma_2-\gamma_1$ plane. The physically valid (superradiant) regime is contained within the blue (red) boundary lines, with the conditions stated in the main text. Blue shaded region: Physically valid, but not superradiant. Regions $\text{I}$, $\text{II}$ and $\text{III}$ are defined in the main text. Red shaded region: Physically valid with superradiant burst. Grey shaded region: unphysical regime. The red shaded region requires a minimum of $\gamma_2 \approx 0.185$. All shaded regions here are obtained from numerical calculations for $N = 100$, which agree very well with the analytical results obtained in the infinite-array limit.}
	\label{fig:superradiance_region_NNNring}
\end{figure}
\begin{figure*}
\centering
\subfloat{%
  \includegraphics[width=0.95\linewidth]{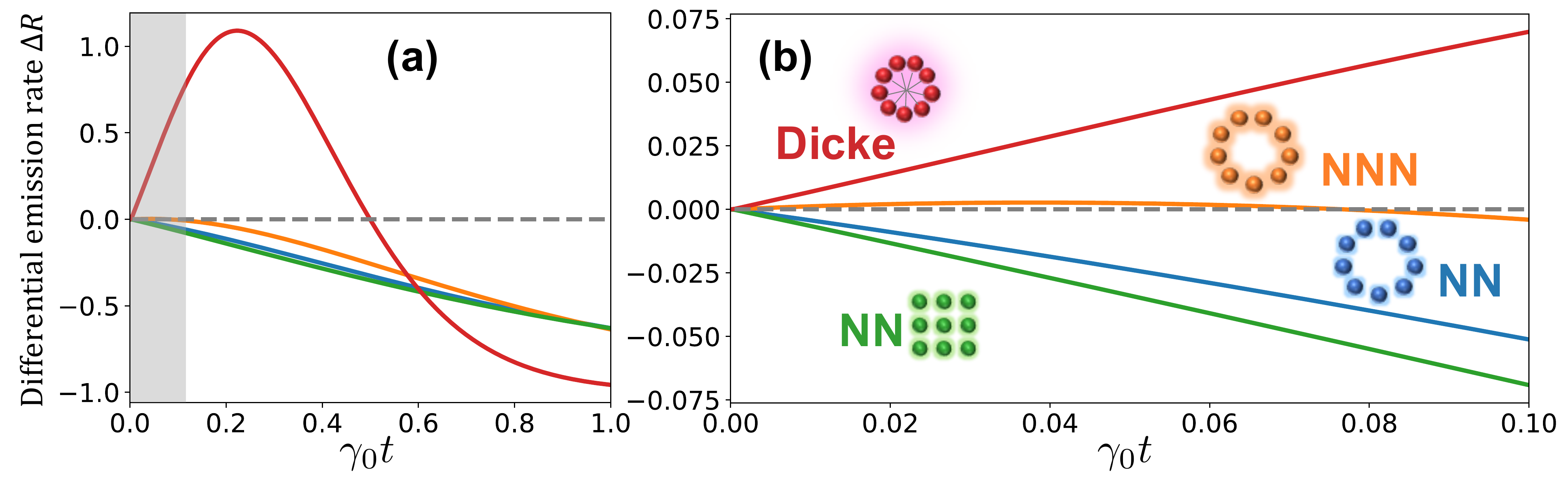}%
  \label{}%
}
	\caption{Differential emission rate $\Delta R = R(t)/R(0) - 1$ against time (in units of emitter lifetime), for $N=9$ emitters. $\Delta R > 0$ indicates superradiance. (a) Dynamical behavior of $\Delta R$ for the Dicke model (red), Next-nearest neighbor 1D ring (NNN, orange), Nearest-neighbor 1D ring (NN, blue) and Nearest-neighbor 2D square (NN, green) (see labels in (b)). The coupling parameters are chosen to maximize $g^{(2)}(0)$. (b) Short-time behavior obtained by zooming into the grey region of (a). Only the Dicke and the next-nearest neighbor models exhibit superradiance. The curve for the Dicke model is scaled down by a factor of 10 for visualization purposes.}
	\label{fig:emRates}
\end{figure*}

\textit{Next-nearest neighbor coupling.---}Including the NNN interactions, we now show that a superradiant burst is indeed possible. For simplicity, let us consider a 1D ring of $N$ emitters with periodic boundary conditions. In this configuration, $\mathbf{\Gamma}$ turns out to be a circulant matrix with the first column given by $(1,\gamma_1, \gamma_2, 0, \ldots, 0, \gamma_2, \gamma_1)^T$ with $0 \leq \{\gamma_1,\gamma_2\} \leq 1$. The subsequent columns are simply cyclic permutations of the first column. Diagonalizing $\mathbf{\Gamma}$ exactly yields the eigenvalues
\begin{equation}
    \Gamma_\nu = 1 + 2\gamma_1 \cos \left(\frac{2\pi \nu}{N}\right) + 2\gamma_2 \cos \left(\frac{4\pi \nu}{N}\right)
\end{equation}
for $\nu = 0, \ldots, N-1$. In the infinite-array limit $N \to \infty$, the eigenvalues form a continuous band in momentum space $\Gamma(k) = 1+2\gamma_1 \cos(k) + 2\gamma_2 \cos(2k)$, with the dimensionless wavevector $0 \leq k < 2\pi$. At the turning points where $\partial_k \Gamma = 0$, we have: $\Gamma(0) = 1 + 2(\gamma_1+\gamma_2)$ which is always positive, $\Gamma(\pi) = 1-2(\gamma_1 - \gamma_2)$ and $\Gamma(k_*) = 1-(\gamma_1^2 + 8\gamma_2^2)/4\gamma_2$ where $\cos k_* = -\gamma_1 / 4\gamma_2$. Demanding that $\Gamma(k) > 0$ thus produces the physically valid regimes: (I) $\gamma_1 - \gamma_2 \leq \frac{1}{2}, \gamma_1 > 4 \gamma_2$ and (II) $\gamma_1^2 + 8\gamma_2^2 \leq 4\gamma_2, \gamma_1 \leq 4 \gamma_2$, together with the bounds $\gamma_1, \gamma_2 \in [0,1]$ (blue regions in Fig.~\ref{fig:superradiance_region_NNNring}). The superradiant condition can be obtain from $\dot{R}(0)=0$ as (III) $\gamma_1^2 + \gamma_2^2 > 1/2$. 

There is an overlap region with the physically valid regime, as shown by the red shaded region in Fig.~\ref{fig:superradiance_region_NNNring}.
For certain values of $\gamma_1, \gamma_2$, superradiant burst can occur. Moreover, the fact that this overlap region requires $\gamma_2 > (4-\sqrt{2})/14 \approx 0.185$ is consistent with our previous conclusion of no superradiance using only NN coupling (i.e., $\gamma_2 = 0$). Superradiance is also forbidden by having only NNN coupling (i.e., $\gamma_1$ = 0). Results from numerical simulations of $N=9$ emitters are presented in Fig.~\ref{fig:emRates}, which show that the NNN model has a small superradiant burst compared to the Dicke model, and no superradiance for NN models. We remark that this superradiance arises from destructive interference leading to dark decay channels with suppressed decay rates $\Gamma_\nu \approx 0$ while the dominant decay channel has a rate that does not scale with $N$. This mechanism is generally true for all models with a sharp interaction cutoff beyond a certain range.

\textit{Threshold interaction range for a superradiant burst.---}
\label{sec:examples}
In many previous works~\cite{Masson2022universality,Robicheaux2021theoretical,Sierra2021dicke, Rubies2021superradiance,Masson2020many}, $\mathbf{\Gamma}$ is obtained from a realistic modelling of the atomic interactions mediated by electromagnetic vacuum using the appropriate Green's function. Our goal here, however, is to shed light on the essential physics of superradiance by considering analytically tractable models that still exhibit interesting behaviors.
\begin{figure}
\centering
\subfloat{%
  \includegraphics[width=0.95\linewidth]{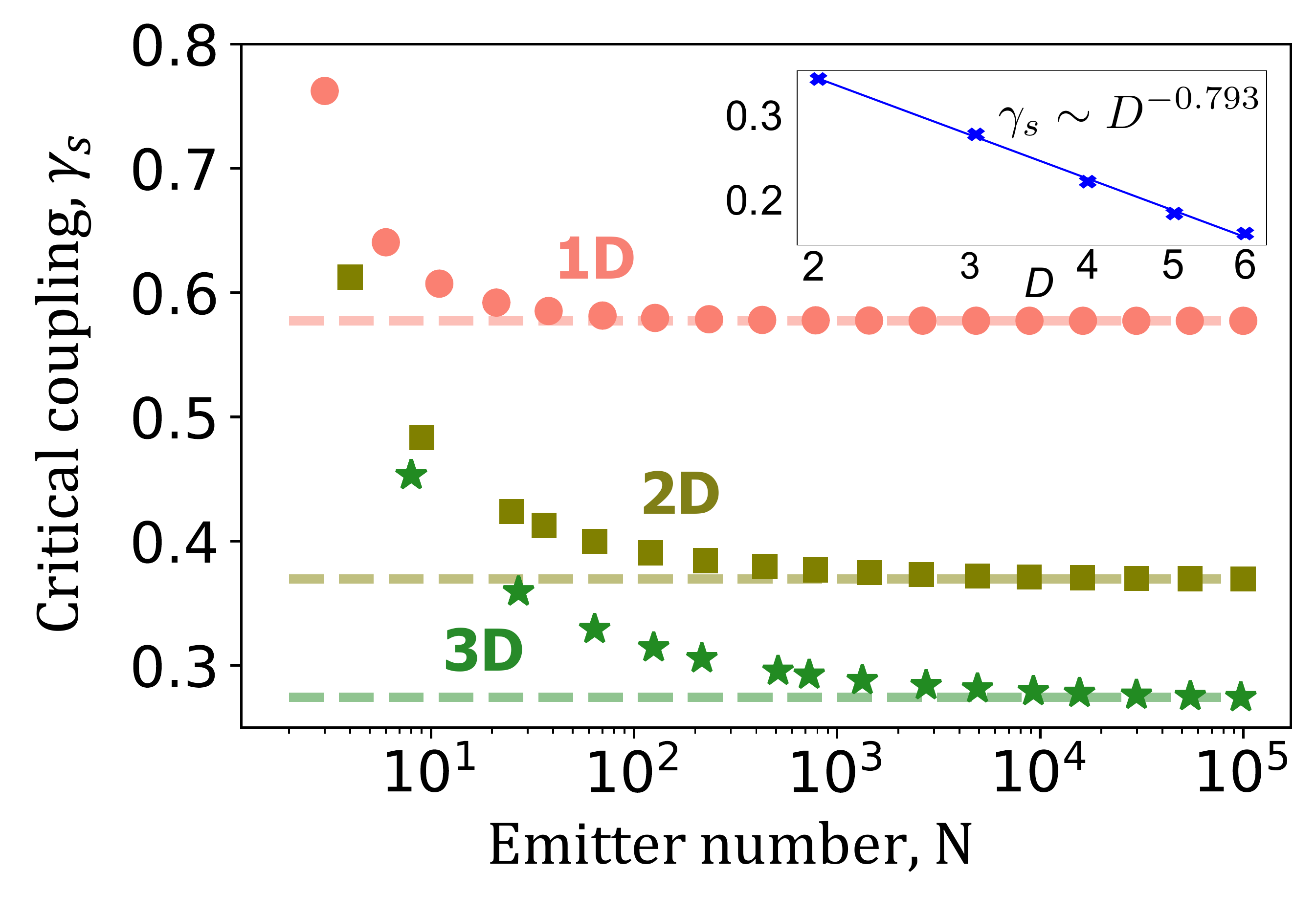}%
  \label{}%
}
	\caption{Critical coupling $\gamma_s$ for exponentially-decaying interactions in a 1D chain, a 2D square array and a 3D cubic array with $N$ emitters. Superradiance occurs for $\gamma > \gamma_s$. For all dimension $D$, $\gamma_s$ becomes independent of $N$ for large $N$. (Inset) Log-log plot of $\gamma_s$ against $D$ for $N \approx 10^6$. $\gamma_s$ decreases as $D$ increases with a power-law scaling $\gamma_s \sim D^{-0.793}$.}
	\label{fig:expCritU}
\end{figure}
Consider an interaction which decays exponentially with the separation $r_{ij}$ between the emitters: $\gamma_{ij} \propto e^{-\kappa r_{ij}}$, where $\kappa$ controls the decay of the interaction strength with emitter separation. We set the diagonal elements of $\mathbf{\Gamma}$ as $1$, and define $\gamma \equiv e^{-\kappa d}$ with $d$ the emitter NN separation such that $\gamma_{ij} = \gamma^{|\vec{x}_i-\vec{x}_j|}$, where $\vec{x}_i \in \mathbb{Z}^D$ is the position vector of the $i^{\text{th}}$ lattice site. Physically, this model describes exponentially-decaying interactions between the atoms. For a sufficiently large $N$ in $D$ dimensions such that $\gamma^N \ll 1$, $\dot{R}(0)$ is approximately given by the asymptotic form
\begin{equation}
    \dot{R}(0) \sim N \left( \frac{2D\gamma^2}{1-\gamma^2} - 1 + \frac{C}{(-\ln \gamma)^D} \right)
\end{equation}
for some constant $C$~\cite{supp}. Interestingly, this suggests that the critical coupling parameter $\gamma_s$ for superradiance is independent of $N$ as $N \to \infty$ for all dimension, agreeing with the numerical results shown in Fig.~\ref{fig:expCritU}. This is in stark contrast with previous results (primarily using long-range power-law interactions such as $\gamma_{ij} \propto 1/r_{ij}$), which predict that the critical emitter separation increases with $N$ in 2D and 3D arrays~\cite{Robicheaux2021theoretical,Sierra2021dicke}. Figure~\ref{fig:expCritU} also shows that for large $N$, $\gamma_s \sim D^{-0.793}$ exhibits a power-law scaling with the spatial dimension. This is intuitive as the average coupling per emitter increases with $D$ which in turn lowers the critical coupling required for superradiance~\cite{Sierra2021dicke}. The $N$-independence of $\gamma_s$ for our short-range exponential model can be physically interpreted as the threshold interaction range where the synchronization effects due to collective interactions scales similarly with $N$ as the local decoherence, such that adding more emitters do not affect the onset of the superradiant regime. For even shorter-range interactions such as the NN model, the local decoherence dominates which prevents superradiance. Longer-range models such as power-law interactions favor synchronization and thus enhance superradiance as $N$ increases. 

\textit{Scaling of the peak emission rate with number of emitters}--- Eq.~\eqref{hgamma} shows that the problem of calculating the emission rate is equivalent to finding the average energy of a state under the Hamiltonian $H_\Gamma$. This enables us to find upper bounds on the scaling of the peak emission rate with $N$, for arbitrary geometries and types of interactions. As we have shown before in Theorem 1, the maximum emission rate for arbitrary NN models is $N\gamma_0$. For 1D arrays with an exponentially-decaying interaction, the upper bound on the emission rate is found to scale as $O(N)$ for $\gamma < 1$~\cite{supp}. This bound increases to $O(N \log N)$ for 1D arrays with a power-law interaction of the form $1/r$~\cite{supp}. This latter scaling is consistent with the numerical results obtained in the literature which, in contrast to our bound, have only been obtained for relatively small systems and under certain approximations~\cite{Masson2022universality,Sierra2021dicke,Rubies2021superradiance}. While finding exact bounds may be exponentially hard, one could in principle upper-bound other models, as well as tighten the currently-obtained bounds.

{\textit{Discussion.}---}In this Letter, we addressed the fundamental problem of the minimal interaction range required for superradiance. Crucially, we proved that nearest-neighbor interactions cannot induce emitter correlations faster that the decoherence, resulting in the impossibility of superradiance. As shown, the minimal interaction range is therefore next-nearest neighbor, and longer-range interactions generally lead to stronger superradiance. We also found that the short-range exponential interaction marks the threshold interaction range in all dimensions where the emitter correlations and local decoherence scale similarly with the number of emitters such that the critical coupling required for superradiant burst becomes independent of the number of emitters, in stark contrast with previous conclusions using longer-range power-law interactions. We stress that, apart from the nearest-neighbor model, our classification of a superradiant burst is strictly speaking only valid at short times up to $\mathcal{O}((\gamma_0 t)^2)$ (if $\ddot{R}(0) < 0$ which is true for the models considered here~\cite{supp}), where the dynamics of the fully-excited emitters do not depend on the Hamiltonian. This can be physically justified for later times using second-order mean field theory~\cite{supp}.

The techniques used in this work have broader applications in determining the theoretical bounds for the emission rate of different models, thereby exposing the ultimate limitations of superradiance beyond the NN model. Beyond providing fundamental insights to the physics of superradiance, our results can also motivate the design of atomic lattices in engineered baths such as nanophotonic crystals with engineered interactions or superconducting resonator arrays for qubits. Moreover, hypercube geometries should be within reach of state-of-the-art quantum simulators, given the recent advances in generating arbitrary networks in cavity~\cite{periwal2021programmable} and circuit~\cite{Kollar2019hyperbolic} quantum electrodynamics platforms. 

\nocite{Douglas2015quantum,Munro2018population,Solano2017super,Lodahl2017chiral,Pichler2015quantum,mok2020long,mok2020microresonators,mahmoodian2020dynamics,robicheaux2021beyond,oriol2022characterizing}

\begin{acknowledgements} The authors are grateful to Ben Grossmann, McCoy Lim, Chris Chen, Jasen Zion, Kishor Bharti, Davit Aghamalyan, Lewis Ruks, Thi Ha Kyaw, Tobias Haug, Steven Touzard, Klaus M{\o}lmer and Stuart Masson for insightful discussions. This research is supported by the National Research Foundation (NRF), Singapore, under its Competitive Research Program (CRP) (NRF-CRP25-2020-0004). A.A.-G. gratefully acknowledges support from the Air Force Office of Scientific Research through their Young Investigator Prize (grant No.~21RT0751), the National Science Foundation through their CAREER Award (No. 2047380), the A. P. Sloan foundation, and the David and Lucile Packard foundation. 
\end{acknowledgements}

\bibliography{bib}

\begin{thebibliography}{77}%
\makeatletter
\providecommand \@ifxundefined [1]{%
 \@ifx{#1\undefined}
}%
\providecommand \@ifnum [1]{%
 \ifnum #1\expandafter \@firstoftwo
 \else \expandafter \@secondoftwo
 \fi
}%
\providecommand \@ifx [1]{%
 \ifx #1\expandafter \@firstoftwo
 \else \expandafter \@secondoftwo
 \fi
}%
\providecommand \natexlab [1]{#1}%
\providecommand \enquote  [1]{``#1''}%
\providecommand \bibnamefont  [1]{#1}%
\providecommand \bibfnamefont [1]{#1}%
\providecommand \citenamefont [1]{#1}%
\providecommand \href@noop [0]{\@secondoftwo}%
\providecommand \href [0]{\begingroup \@sanitize@url \@href}%
\providecommand \@href[1]{\@@startlink{#1}\@@href}%
\providecommand \@@href[1]{\endgroup#1\@@endlink}%
\providecommand \@sanitize@url [0]{\catcode `\\12\catcode `\$12\catcode
  `\&12\catcode `\#12\catcode `\^12\catcode `\_12\catcode `\%12\relax}%
\providecommand \@@startlink[1]{}%
\providecommand \@@endlink[0]{}%
\providecommand \url  [0]{\begingroup\@sanitize@url \@url }%
\providecommand \@url [1]{\endgroup\@href {#1}{\urlprefix }}%
\providecommand \urlprefix  [0]{URL }%
\providecommand \Eprint [0]{\href }%
\providecommand \doibase [0]{https://doi.org/}%
\providecommand \selectlanguage [0]{\@gobble}%
\providecommand \bibinfo  [0]{\@secondoftwo}%
\providecommand \bibfield  [0]{\@secondoftwo}%
\providecommand \translation [1]{[#1]}%
\providecommand \BibitemOpen [0]{}%
\providecommand \bibitemStop [0]{}%
\providecommand \bibitemNoStop [0]{.\EOS\space}%
\providecommand \EOS [0]{\spacefactor3000\relax}%
\providecommand \BibitemShut  [1]{\csname bibitem#1\endcsname}%
\let\auto@bib@innerbib\@empty
\bibitem [{\citenamefont {Dicke}(1954)}]{Dicke1954coherence}%
  \BibitemOpen
  \bibfield  {author} {\bibinfo {author} {\bibfnamefont {R.~H.}\ \bibnamefont
  {Dicke}},\ }\bibfield  {title} {\bibinfo {title} {Coherence in spontaneous
  radiation processes},\ }\href {https://doi.org/10.1103/PhysRev.93.99}
  {\bibfield  {journal} {\bibinfo  {journal} {Phys. Rev.}\ }\textbf {\bibinfo
  {volume} {93}},\ \bibinfo {pages} {99} (\bibinfo {year} {1954})}\BibitemShut
  {NoStop}%
\bibitem [{\citenamefont {Gross}\ and\ \citenamefont
  {Haroche}(1982)}]{Gross1982superradiance}%
  \BibitemOpen
  \bibfield  {author} {\bibinfo {author} {\bibfnamefont {M.}~\bibnamefont
  {Gross}}\ and\ \bibinfo {author} {\bibfnamefont {S.}~\bibnamefont
  {Haroche}},\ }\bibfield  {title} {\bibinfo {title} {Superradiance: An essay
  on the theory of collective spontaneous emission},\ }\href
  {https://doi.org/https://doi.org/10.1016/0370-1573(82)90102-8} {\bibfield
  {journal} {\bibinfo  {journal} {Phys. Rep.}\ }\textbf {\bibinfo {volume}
  {93}},\ \bibinfo {pages} {301} (\bibinfo {year} {1982})}\BibitemShut
  {NoStop}%
\bibitem [{\citenamefont {Rehler}\ and\ \citenamefont
  {Eberly}(1971)}]{rehler1971superradiance}%
  \BibitemOpen
  \bibfield  {author} {\bibinfo {author} {\bibfnamefont {N.~E.}\ \bibnamefont
  {Rehler}}\ and\ \bibinfo {author} {\bibfnamefont {J.~H.}\ \bibnamefont
  {Eberly}},\ }\bibfield  {title} {\bibinfo {title} {Superradiance},\ }\href
  {https://doi.org/10.1103/PhysRevA.3.1735} {\bibfield  {journal} {\bibinfo
  {journal} {Phys. Rev. A}\ }\textbf {\bibinfo {volume} {3}},\ \bibinfo {pages}
  {1735} (\bibinfo {year} {1971})}\BibitemShut {NoStop}%
\bibitem [{\citenamefont {Stroud}\ \emph {et~al.}(1972)\citenamefont {Stroud},
  \citenamefont {Eberly}, \citenamefont {Lama},\ and\ \citenamefont
  {Mandel}}]{stroud1972superradiant}%
  \BibitemOpen
  \bibfield  {author} {\bibinfo {author} {\bibfnamefont {C.~R.}\ \bibnamefont
  {Stroud}}, \bibinfo {author} {\bibfnamefont {J.~H.}\ \bibnamefont {Eberly}},
  \bibinfo {author} {\bibfnamefont {W.~L.}\ \bibnamefont {Lama}},\ and\
  \bibinfo {author} {\bibfnamefont {L.}~\bibnamefont {Mandel}},\ }\bibfield
  {title} {\bibinfo {title} {Superradiant effects in systems of two-level
  atoms},\ }\href {https://doi.org/10.1103/PhysRevA.5.1094} {\bibfield
  {journal} {\bibinfo  {journal} {Phys. Rev. A}\ }\textbf {\bibinfo {volume}
  {5}},\ \bibinfo {pages} {1094} (\bibinfo {year} {1972})}\BibitemShut
  {NoStop}%
\bibitem [{\citenamefont {Friedberg}\ \emph {et~al.}(1972)\citenamefont
  {Friedberg}, \citenamefont {Hartmann},\ and\ \citenamefont
  {Manassah}}]{friedberg1972limited}%
  \BibitemOpen
  \bibfield  {author} {\bibinfo {author} {\bibfnamefont {R.}~\bibnamefont
  {Friedberg}}, \bibinfo {author} {\bibfnamefont {S.}~\bibnamefont
  {Hartmann}},\ and\ \bibinfo {author} {\bibfnamefont {J.}~\bibnamefont
  {Manassah}},\ }\bibfield  {title} {\bibinfo {title} {Limited superradiant
  damping of small samples},\ }\href
  {https://doi.org/https://doi.org/10.1016/0375-9601(72)90533-6} {\bibfield
  {journal} {\bibinfo  {journal} {Phys. Lett. A}\ }\textbf {\bibinfo {volume}
  {40}},\ \bibinfo {pages} {365} (\bibinfo {year} {1972})}\BibitemShut
  {NoStop}%
\bibitem [{\citenamefont {Coffey}\ and\ \citenamefont
  {Friedberg}(1978)}]{coffey1978effect}%
  \BibitemOpen
  \bibfield  {author} {\bibinfo {author} {\bibfnamefont {B.}~\bibnamefont
  {Coffey}}\ and\ \bibinfo {author} {\bibfnamefont {R.}~\bibnamefont
  {Friedberg}},\ }\bibfield  {title} {\bibinfo {title} {Effect of short-range
  coulomb interaction on cooperative spontaneous emission},\ }\href
  {https://doi.org/10.1103/PhysRevA.17.1033} {\bibfield  {journal} {\bibinfo
  {journal} {Phys. Rev. A}\ }\textbf {\bibinfo {volume} {17}},\ \bibinfo
  {pages} {1033} (\bibinfo {year} {1978})}\BibitemShut {NoStop}%
\bibitem [{\citenamefont {Banfi}\ and\ \citenamefont
  {Bonifacio}(1975)}]{banfi1975superfluorescence}%
  \BibitemOpen
  \bibfield  {author} {\bibinfo {author} {\bibfnamefont {G.}~\bibnamefont
  {Banfi}}\ and\ \bibinfo {author} {\bibfnamefont {R.}~\bibnamefont
  {Bonifacio}},\ }\bibfield  {title} {\bibinfo {title} {Superfluorescence and
  cooperative frequency shift},\ }\href
  {https://doi.org/10.1103/PhysRevA.12.2068} {\bibfield  {journal} {\bibinfo
  {journal} {Phys. Rev. A}\ }\textbf {\bibinfo {volume} {12}},\ \bibinfo
  {pages} {2068} (\bibinfo {year} {1975})}\BibitemShut {NoStop}%
\bibitem [{\citenamefont {Clemens}\ \emph {et~al.}(2003)\citenamefont
  {Clemens}, \citenamefont {Horvath}, \citenamefont {Sanders},\ and\
  \citenamefont {Carmichael}}]{Clemens2003collective}%
  \BibitemOpen
  \bibfield  {author} {\bibinfo {author} {\bibfnamefont {J.~P.}\ \bibnamefont
  {Clemens}}, \bibinfo {author} {\bibfnamefont {L.}~\bibnamefont {Horvath}},
  \bibinfo {author} {\bibfnamefont {B.~C.}\ \bibnamefont {Sanders}},\ and\
  \bibinfo {author} {\bibfnamefont {H.~J.}\ \bibnamefont {Carmichael}},\
  }\bibfield  {title} {\bibinfo {title} {Collective spontaneous emission from a
  line of atoms},\ }\href {https://doi.org/10.1103/PhysRevA.68.023809}
  {\bibfield  {journal} {\bibinfo  {journal} {Phys. Rev. A}\ }\textbf {\bibinfo
  {volume} {68}},\ \bibinfo {pages} {023809} (\bibinfo {year}
  {2003})}\BibitemShut {NoStop}%
\bibitem [{\citenamefont {Lin}\ and\ \citenamefont
  {Yelin}(2012)}]{lin2012superradiance}%
  \BibitemOpen
  \bibfield  {author} {\bibinfo {author} {\bibfnamefont {G.-D.}\ \bibnamefont
  {Lin}}\ and\ \bibinfo {author} {\bibfnamefont {S.~F.}\ \bibnamefont
  {Yelin}},\ }\bibfield  {title} {\bibinfo {title} {Superradiance in spin-$j$
  particles: Effects of multiple levels},\ }\href
  {https://doi.org/10.1103/PhysRevA.85.033831} {\bibfield  {journal} {\bibinfo
  {journal} {Phys. Rev. A}\ }\textbf {\bibinfo {volume} {85}},\ \bibinfo
  {pages} {033831} (\bibinfo {year} {2012})}\BibitemShut {NoStop}%
\bibitem [{\citenamefont {Bhatti}\ \emph {et~al.}(2015)\citenamefont {Bhatti},
  \citenamefont {von Zanthier},\ and\ \citenamefont
  {Agarwal}}]{Bhatti2015superbunching}%
  \BibitemOpen
  \bibfield  {author} {\bibinfo {author} {\bibfnamefont {D.}~\bibnamefont
  {Bhatti}}, \bibinfo {author} {\bibfnamefont {J.}~\bibnamefont {von
  Zanthier}},\ and\ \bibinfo {author} {\bibfnamefont {G.~S.}\ \bibnamefont
  {Agarwal}},\ }\bibfield  {title} {\bibinfo {title} {Superbunching and
  nonclassicality as new hallmarks of superradiance},\ }\href
  {https://doi.org/10.1038/srep17335} {\bibfield  {journal} {\bibinfo
  {journal} {Sci. Rep.}\ }\textbf {\bibinfo {volume} {5}},\ \bibinfo {pages}
  {17335} (\bibinfo {year} {2015})}\BibitemShut {NoStop}%
\bibitem [{\citenamefont {Longo}\ \emph {et~al.}(2016)\citenamefont {Longo},
  \citenamefont {Keitel},\ and\ \citenamefont {Evers}}]{Longo2016tailoring}%
  \BibitemOpen
  \bibfield  {author} {\bibinfo {author} {\bibfnamefont {P.}~\bibnamefont
  {Longo}}, \bibinfo {author} {\bibfnamefont {C.~H.}\ \bibnamefont {Keitel}},\
  and\ \bibinfo {author} {\bibfnamefont {J.}~\bibnamefont {Evers}},\ }\bibfield
   {title} {\bibinfo {title} {Tailoring superradiance to design artificial
  quantum systems},\ }\href {https://doi.org/10.1038/srep23628} {\bibfield
  {journal} {\bibinfo  {journal} {Sci. Rep.}\ }\textbf {\bibinfo {volume}
  {6}},\ \bibinfo {pages} {23628} (\bibinfo {year} {2016})}\BibitemShut
  {NoStop}%
\bibitem [{\citenamefont {Nefedkin}\ \emph {et~al.}(2017)\citenamefont
  {Nefedkin}, \citenamefont {Andrianov}, \citenamefont {Zyablovsky},
  \citenamefont {Pukhov}, \citenamefont {Vinogradov},\ and\ \citenamefont
  {Lisyansky}}]{Nefedkin2017superradiance}%
  \BibitemOpen
  \bibfield  {author} {\bibinfo {author} {\bibfnamefont {N.~E.}\ \bibnamefont
  {Nefedkin}}, \bibinfo {author} {\bibfnamefont {E.~S.}\ \bibnamefont
  {Andrianov}}, \bibinfo {author} {\bibfnamefont {A.~A.}\ \bibnamefont
  {Zyablovsky}}, \bibinfo {author} {\bibfnamefont {A.~A.}\ \bibnamefont
  {Pukhov}}, \bibinfo {author} {\bibfnamefont {A.~P.}\ \bibnamefont
  {Vinogradov}},\ and\ \bibinfo {author} {\bibfnamefont {A.~A.}\ \bibnamefont
  {Lisyansky}},\ }\bibfield  {title} {\bibinfo {title} {Superradiance of
  non-dicke states},\ }\href {https://doi.org/10.1364/OE.25.002790} {\bibfield
  {journal} {\bibinfo  {journal} {Opt. Express}\ }\textbf {\bibinfo {volume}
  {25}},\ \bibinfo {pages} {2790} (\bibinfo {year} {2017})}\BibitemShut
  {NoStop}%
\bibitem [{\citenamefont {Pleinert}\ \emph {et~al.}(2017)\citenamefont
  {Pleinert}, \citenamefont {von Zanthier},\ and\ \citenamefont
  {Agarwal}}]{Pleinert2017hyperradiance}%
  \BibitemOpen
  \bibfield  {author} {\bibinfo {author} {\bibfnamefont {M.-O.}\ \bibnamefont
  {Pleinert}}, \bibinfo {author} {\bibfnamefont {J.}~\bibnamefont {von
  Zanthier}},\ and\ \bibinfo {author} {\bibfnamefont {G.~S.}\ \bibnamefont
  {Agarwal}},\ }\bibfield  {title} {\bibinfo {title} {Hyperradiance from
  collective behavior of coherently driven atoms},\ }\href
  {https://doi.org/10.1364/OPTICA.4.000779} {\bibfield  {journal} {\bibinfo
  {journal} {Optica}\ }\textbf {\bibinfo {volume} {4}},\ \bibinfo {pages} {779}
  (\bibinfo {year} {2017})}\BibitemShut {NoStop}%
\bibitem [{\citenamefont {Shammah}\ \emph {et~al.}(2017)\citenamefont
  {Shammah}, \citenamefont {Lambert}, \citenamefont {Nori},\ and\ \citenamefont
  {De~Liberato}}]{Shammah2017superradiance}%
  \BibitemOpen
  \bibfield  {author} {\bibinfo {author} {\bibfnamefont {N.}~\bibnamefont
  {Shammah}}, \bibinfo {author} {\bibfnamefont {N.}~\bibnamefont {Lambert}},
  \bibinfo {author} {\bibfnamefont {F.}~\bibnamefont {Nori}},\ and\ \bibinfo
  {author} {\bibfnamefont {S.}~\bibnamefont {De~Liberato}},\ }\bibfield
  {title} {\bibinfo {title} {Superradiance with local phase-breaking effects},\
  }\href {https://doi.org/10.1103/PhysRevA.96.023863} {\bibfield  {journal}
  {\bibinfo  {journal} {Phys. Rev. A}\ }\textbf {\bibinfo {volume} {96}},\
  \bibinfo {pages} {023863} (\bibinfo {year} {2017})}\BibitemShut {NoStop}%
\bibitem [{\citenamefont {Houde}\ \emph {et~al.}(2017)\citenamefont {Houde},
  \citenamefont {Mathews},\ and\ \citenamefont {Rajabi}}]{Houde2017fast}%
  \BibitemOpen
  \bibfield  {author} {\bibinfo {author} {\bibfnamefont {M.}~\bibnamefont
  {Houde}}, \bibinfo {author} {\bibfnamefont {A.}~\bibnamefont {Mathews}},\
  and\ \bibinfo {author} {\bibfnamefont {F.}~\bibnamefont {Rajabi}},\
  }\bibfield  {title} {\bibinfo {title} {{Explaining fast radio bursts through
  Dicke's superradiance}},\ }\href {https://doi.org/10.1093/mnras/stx3205}
  {\bibfield  {journal} {\bibinfo  {journal} {Mon. Not. R. Astron. Soc.}\
  }\textbf {\bibinfo {volume} {475}},\ \bibinfo {pages} {514} (\bibinfo {year}
  {2017})}\BibitemShut {NoStop}%
\bibitem [{\citenamefont {Masson}\ \emph {et~al.}(2020)\citenamefont {Masson},
  \citenamefont {Ferrier-Barbut}, \citenamefont {Orozco}, \citenamefont
  {Browaeys},\ and\ \citenamefont {Asenjo-Garcia}}]{Masson2020many}%
  \BibitemOpen
  \bibfield  {author} {\bibinfo {author} {\bibfnamefont {S.~J.}\ \bibnamefont
  {Masson}}, \bibinfo {author} {\bibfnamefont {I.}~\bibnamefont
  {Ferrier-Barbut}}, \bibinfo {author} {\bibfnamefont {L.~A.}\ \bibnamefont
  {Orozco}}, \bibinfo {author} {\bibfnamefont {A.}~\bibnamefont {Browaeys}},\
  and\ \bibinfo {author} {\bibfnamefont {A.}~\bibnamefont {Asenjo-Garcia}},\
  }\bibfield  {title} {\bibinfo {title} {Many-body signatures of collective
  decay in atomic chains},\ }\href
  {https://doi.org/10.1103/PhysRevLett.125.263601} {\bibfield  {journal}
  {\bibinfo  {journal} {Phys. Rev. Lett.}\ }\textbf {\bibinfo {volume} {125}},\
  \bibinfo {pages} {263601} (\bibinfo {year} {2020})}\BibitemShut {NoStop}%
\bibitem [{\citenamefont {Fuchs}\ \emph {et~al.}(2021)\citenamefont {Fuchs},
  \citenamefont {Vukics},\ and\ \citenamefont
  {Buhmann}}]{Fuchs2021superradiance}%
  \BibitemOpen
  \bibfield  {author} {\bibinfo {author} {\bibfnamefont {S.}~\bibnamefont
  {Fuchs}}, \bibinfo {author} {\bibfnamefont {A.}~\bibnamefont {Vukics}},\ and\
  \bibinfo {author} {\bibfnamefont {S.~Y.}\ \bibnamefont {Buhmann}},\
  }\bibfield  {title} {\bibinfo {title} {Superradiance from nonideal initial
  states: A quantum trajectory approach},\ }\href
  {https://doi.org/10.1103/PhysRevA.103.043712} {\bibfield  {journal} {\bibinfo
   {journal} {Phys. Rev. A}\ }\textbf {\bibinfo {volume} {103}},\ \bibinfo
  {pages} {043712} (\bibinfo {year} {2021})}\BibitemShut {NoStop}%
\bibitem [{\citenamefont {Lemberger}\ and\ \citenamefont
  {M\o{}lmer}(2021)}]{Lemberger2021radiation}%
  \BibitemOpen
  \bibfield  {author} {\bibinfo {author} {\bibfnamefont {B.}~\bibnamefont
  {Lemberger}}\ and\ \bibinfo {author} {\bibfnamefont {K.}~\bibnamefont
  {M\o{}lmer}},\ }\bibfield  {title} {\bibinfo {title} {Radiation eigenmodes of
  dicke superradiance},\ }\href {https://doi.org/10.1103/PhysRevA.103.033713}
  {\bibfield  {journal} {\bibinfo  {journal} {Phys. Rev. A}\ }\textbf {\bibinfo
  {volume} {103}},\ \bibinfo {pages} {033713} (\bibinfo {year}
  {2021})}\BibitemShut {NoStop}%
\bibitem [{\citenamefont {Robicheaux}(2021)}]{Robicheaux2021theoretical}%
  \BibitemOpen
  \bibfield  {author} {\bibinfo {author} {\bibfnamefont {F.}~\bibnamefont
  {Robicheaux}},\ }\bibfield  {title} {\bibinfo {title} {Theoretical study of
  early-time superradiance for atom clouds and arrays},\ }\href
  {https://doi.org/10.1103/PhysRevA.104.063706} {\bibfield  {journal} {\bibinfo
   {journal} {Phys. Rev. A}\ }\textbf {\bibinfo {volume} {104}},\ \bibinfo
  {pages} {063706} (\bibinfo {year} {2021})}\BibitemShut {NoStop}%
\bibitem [{\citenamefont {Rubies-Bigorda}\ and\ \citenamefont
  {Yelin}(2022)}]{Rubies2021superradiance}%
  \BibitemOpen
  \bibfield  {author} {\bibinfo {author} {\bibfnamefont {O.}~\bibnamefont
  {Rubies-Bigorda}}\ and\ \bibinfo {author} {\bibfnamefont {S.~F.}\
  \bibnamefont {Yelin}},\ }\bibfield  {title} {\bibinfo {title} {Superradiance
  and subradiance in inverted atomic arrays},\ }\href
  {https://doi.org/10.1103/PhysRevA.106.053717} {\bibfield  {journal} {\bibinfo
   {journal} {Phys. Rev. A}\ }\textbf {\bibinfo {volume} {106}},\ \bibinfo
  {pages} {053717} (\bibinfo {year} {2022})}\BibitemShut {NoStop}%
\bibitem [{\citenamefont {Malz}\ \emph {et~al.}(2022)\citenamefont {Malz},
  \citenamefont {Trivedi},\ and\ \citenamefont {Cirac}}]{malz2022large}%
  \BibitemOpen
  \bibfield  {author} {\bibinfo {author} {\bibfnamefont {D.}~\bibnamefont
  {Malz}}, \bibinfo {author} {\bibfnamefont {R.}~\bibnamefont {Trivedi}},\ and\
  \bibinfo {author} {\bibfnamefont {J.~I.}\ \bibnamefont {Cirac}},\ }\bibfield
  {title} {\bibinfo {title} {Large-$n$ limit of dicke superradiance},\ }\href
  {https://doi.org/10.1103/PhysRevA.106.013716} {\bibfield  {journal} {\bibinfo
   {journal} {Phys. Rev. A}\ }\textbf {\bibinfo {volume} {106}},\ \bibinfo
  {pages} {013716} (\bibinfo {year} {2022})}\BibitemShut {NoStop}%
\bibitem [{\citenamefont {Masson}\ and\ \citenamefont
  {Asenjo-Garcia}(2022)}]{Masson2022universality}%
  \BibitemOpen
  \bibfield  {author} {\bibinfo {author} {\bibfnamefont {S.~J.}\ \bibnamefont
  {Masson}}\ and\ \bibinfo {author} {\bibfnamefont {A.}~\bibnamefont
  {Asenjo-Garcia}},\ }\bibfield  {title} {\bibinfo {title} {Universality of
  dicke superradiance in arrays of quantum emitters},\ }\href
  {https://doi.org/10.1038/s41467-022-29805-4} {\bibfield  {journal} {\bibinfo
  {journal} {Nat. Commun.}\ }\textbf {\bibinfo {volume} {13}},\ \bibinfo
  {pages} {2285} (\bibinfo {year} {2022})}\BibitemShut {NoStop}%
\bibitem [{\citenamefont {Pi\~neiro Orioli}\ \emph {et~al.}(2022)\citenamefont
  {Pi\~neiro Orioli}, \citenamefont {Thompson},\ and\ \citenamefont
  {Rey}}]{Orioli2022emergent}%
  \BibitemOpen
  \bibfield  {author} {\bibinfo {author} {\bibfnamefont {A.}~\bibnamefont
  {Pi\~neiro Orioli}}, \bibinfo {author} {\bibfnamefont {J.~K.}\ \bibnamefont
  {Thompson}},\ and\ \bibinfo {author} {\bibfnamefont {A.~M.}\ \bibnamefont
  {Rey}},\ }\bibfield  {title} {\bibinfo {title} {Emergent dark states from
  superradiant dynamics in multilevel atoms in a cavity},\ }\href
  {https://doi.org/10.1103/PhysRevX.12.011054} {\bibfield  {journal} {\bibinfo
  {journal} {Phys. Rev. X}\ }\textbf {\bibinfo {volume} {12}},\ \bibinfo
  {pages} {011054} (\bibinfo {year} {2022})}\BibitemShut {NoStop}%
\bibitem [{\citenamefont {Sierra}\ \emph {et~al.}(2022)\citenamefont {Sierra},
  \citenamefont {Masson},\ and\ \citenamefont
  {Asenjo-Garcia}}]{Sierra2021dicke}%
  \BibitemOpen
  \bibfield  {author} {\bibinfo {author} {\bibfnamefont {E.}~\bibnamefont
  {Sierra}}, \bibinfo {author} {\bibfnamefont {S.~J.}\ \bibnamefont {Masson}},\
  and\ \bibinfo {author} {\bibfnamefont {A.}~\bibnamefont {Asenjo-Garcia}},\
  }\bibfield  {title} {\bibinfo {title} {Dicke superradiance in ordered
  lattices: Dimensionality matters},\ }\href
  {https://doi.org/10.1103/PhysRevResearch.4.023207} {\bibfield  {journal}
  {\bibinfo  {journal} {Phys. Rev. Research}\ }\textbf {\bibinfo {volume}
  {4}},\ \bibinfo {pages} {023207} (\bibinfo {year} {2022})}\BibitemShut
  {NoStop}%
\bibitem [{\citenamefont {Skribanowitz}\ \emph {et~al.}(1973)\citenamefont
  {Skribanowitz}, \citenamefont {Herman}, \citenamefont {MacGillivray},\ and\
  \citenamefont {Feld}}]{skribanowitz1973observation}%
  \BibitemOpen
  \bibfield  {author} {\bibinfo {author} {\bibfnamefont {N.}~\bibnamefont
  {Skribanowitz}}, \bibinfo {author} {\bibfnamefont {I.~P.}\ \bibnamefont
  {Herman}}, \bibinfo {author} {\bibfnamefont {J.~C.}\ \bibnamefont
  {MacGillivray}},\ and\ \bibinfo {author} {\bibfnamefont {M.~S.}\ \bibnamefont
  {Feld}},\ }\bibfield  {title} {\bibinfo {title} {Observation of dicke
  superradiance in optically pumped hf gas},\ }\href
  {https://doi.org/10.1103/PhysRevLett.30.309} {\bibfield  {journal} {\bibinfo
  {journal} {Phys. Rev. Lett.}\ }\textbf {\bibinfo {volume} {30}},\ \bibinfo
  {pages} {309} (\bibinfo {year} {1973})}\BibitemShut {NoStop}%
\bibitem [{\citenamefont {Raimond}\ \emph {et~al.}(1982)\citenamefont
  {Raimond}, \citenamefont {Goy}, \citenamefont {Gross}, \citenamefont
  {Fabre},\ and\ \citenamefont {Haroche}}]{raimond1982collective}%
  \BibitemOpen
  \bibfield  {author} {\bibinfo {author} {\bibfnamefont {J.~M.}\ \bibnamefont
  {Raimond}}, \bibinfo {author} {\bibfnamefont {P.}~\bibnamefont {Goy}},
  \bibinfo {author} {\bibfnamefont {M.}~\bibnamefont {Gross}}, \bibinfo
  {author} {\bibfnamefont {C.}~\bibnamefont {Fabre}},\ and\ \bibinfo {author}
  {\bibfnamefont {S.}~\bibnamefont {Haroche}},\ }\bibfield  {title} {\bibinfo
  {title} {Collective absorption of blackbody radiation by rydberg atoms in a
  cavity: An experiment on bose statistics and brownian motion},\ }\href
  {https://doi.org/10.1103/PhysRevLett.49.117} {\bibfield  {journal} {\bibinfo
  {journal} {Phys. Rev. Lett.}\ }\textbf {\bibinfo {volume} {49}},\ \bibinfo
  {pages} {117} (\bibinfo {year} {1982})}\BibitemShut {NoStop}%
\bibitem [{\citenamefont {DeVoe}\ and\ \citenamefont
  {Brewer}(1996)}]{DeVoe1996observation}%
  \BibitemOpen
  \bibfield  {author} {\bibinfo {author} {\bibfnamefont {R.~G.}\ \bibnamefont
  {DeVoe}}\ and\ \bibinfo {author} {\bibfnamefont {R.~G.}\ \bibnamefont
  {Brewer}},\ }\bibfield  {title} {\bibinfo {title} {Observation of
  superradiant and subradiant spontaneous emission of two trapped ions},\
  }\href {https://doi.org/10.1103/PhysRevLett.76.2049} {\bibfield  {journal}
  {\bibinfo  {journal} {Phys. Rev. Lett.}\ }\textbf {\bibinfo {volume} {76}},\
  \bibinfo {pages} {2049} (\bibinfo {year} {1996})}\BibitemShut {NoStop}%
\bibitem [{\citenamefont {Spano}\ and\ \citenamefont
  {Mukamel}(1989)}]{spano1989superradiance}%
  \BibitemOpen
  \bibfield  {author} {\bibinfo {author} {\bibfnamefont {F.~C.}\ \bibnamefont
  {Spano}}\ and\ \bibinfo {author} {\bibfnamefont {S.}~\bibnamefont
  {Mukamel}},\ }\bibfield  {title} {\bibinfo {title} {Superradiance in
  molecular aggregates},\ }\href {https://doi.org/10.1063/1.457174} {\bibfield
  {journal} {\bibinfo  {journal} {J. Chem. Phys.}\ }\textbf {\bibinfo {volume}
  {91}},\ \bibinfo {pages} {683} (\bibinfo {year} {1989})}\BibitemShut
  {NoStop}%
\bibitem [{\citenamefont {Spano}\ \emph {et~al.}(1990)\citenamefont {Spano},
  \citenamefont {Kuklinski},\ and\ \citenamefont
  {Mukamel}}]{spano1990temperature}%
  \BibitemOpen
  \bibfield  {author} {\bibinfo {author} {\bibfnamefont {F.~C.}\ \bibnamefont
  {Spano}}, \bibinfo {author} {\bibfnamefont {J.~R.}\ \bibnamefont
  {Kuklinski}},\ and\ \bibinfo {author} {\bibfnamefont {S.}~\bibnamefont
  {Mukamel}},\ }\bibfield  {title} {\bibinfo {title} {Temperature-dependent
  superradiant decay of excitons in small aggregates},\ }\href
  {https://doi.org/10.1103/PhysRevLett.65.211} {\bibfield  {journal} {\bibinfo
  {journal} {Phys. Rev. Lett.}\ }\textbf {\bibinfo {volume} {65}},\ \bibinfo
  {pages} {211} (\bibinfo {year} {1990})}\BibitemShut {NoStop}%
\bibitem [{\citenamefont {Spano}\ \emph {et~al.}(1991)\citenamefont {Spano},
  \citenamefont {Kuklinski},\ and\ \citenamefont
  {Mukamel}}]{spano1991cooperative}%
  \BibitemOpen
  \bibfield  {author} {\bibinfo {author} {\bibfnamefont {F.~C.}\ \bibnamefont
  {Spano}}, \bibinfo {author} {\bibfnamefont {J.~R.}\ \bibnamefont
  {Kuklinski}},\ and\ \bibinfo {author} {\bibfnamefont {S.}~\bibnamefont
  {Mukamel}},\ }\bibfield  {title} {\bibinfo {title} {Cooperative radiative
  dynamics in molecular aggregates},\ }\href {https://doi.org/10.1063/1.460185}
  {\bibfield  {journal} {\bibinfo  {journal} {J. Chem. Phys.}\ }\textbf
  {\bibinfo {volume} {94}},\ \bibinfo {pages} {7534} (\bibinfo {year}
  {1991})}\BibitemShut {NoStop}%
\bibitem [{\citenamefont {Meier}\ \emph {et~al.}(1997)\citenamefont {Meier},
  \citenamefont {Zhao}, \citenamefont {Chernyak},\ and\ \citenamefont
  {Mukamel}}]{meier1997temperature}%
  \BibitemOpen
  \bibfield  {author} {\bibinfo {author} {\bibfnamefont {T.}~\bibnamefont
  {Meier}}, \bibinfo {author} {\bibfnamefont {Y.}~\bibnamefont {Zhao}},
  \bibinfo {author} {\bibfnamefont {V.}~\bibnamefont {Chernyak}},\ and\
  \bibinfo {author} {\bibfnamefont {S.}~\bibnamefont {Mukamel}},\ }\bibfield
  {title} {\bibinfo {title} {Polarons, localization, and excitonic coherence in
  superradiance of biological antenna complexes},\ }\href
  {https://doi.org/10.1063/1.474746} {\bibfield  {journal} {\bibinfo  {journal}
  {J. Chem. Phys.}\ }\textbf {\bibinfo {volume} {107}},\ \bibinfo {pages}
  {3876} (\bibinfo {year} {1997})}\BibitemShut {NoStop}%
\bibitem [{\citenamefont {Scheibner}\ \emph {et~al.}(2007)\citenamefont
  {Scheibner}, \citenamefont {Schmidt}, \citenamefont {Worschech},
  \citenamefont {Forchel}, \citenamefont {Bacher}, \citenamefont {Passow},\
  and\ \citenamefont {Hommel}}]{Scheibner2007superradiance}%
  \BibitemOpen
  \bibfield  {author} {\bibinfo {author} {\bibfnamefont {M.}~\bibnamefont
  {Scheibner}}, \bibinfo {author} {\bibfnamefont {T.}~\bibnamefont {Schmidt}},
  \bibinfo {author} {\bibfnamefont {L.}~\bibnamefont {Worschech}}, \bibinfo
  {author} {\bibfnamefont {A.}~\bibnamefont {Forchel}}, \bibinfo {author}
  {\bibfnamefont {G.}~\bibnamefont {Bacher}}, \bibinfo {author} {\bibfnamefont
  {T.}~\bibnamefont {Passow}},\ and\ \bibinfo {author} {\bibfnamefont
  {D.}~\bibnamefont {Hommel}},\ }\bibfield  {title} {\bibinfo {title}
  {Superradiance of quantum dots},\ }\href {https://doi.org/10.1038/nphys494}
  {\bibfield  {journal} {\bibinfo  {journal} {Nat. Phys.}\ }\textbf {\bibinfo
  {volume} {3}},\ \bibinfo {pages} {106} (\bibinfo {year} {2007})}\BibitemShut
  {NoStop}%
\bibitem [{\citenamefont {Yukalov}\ and\ \citenamefont
  {Yukalova}(2010)}]{Yukalov2010dynamics}%
  \BibitemOpen
  \bibfield  {author} {\bibinfo {author} {\bibfnamefont {V.~I.}\ \bibnamefont
  {Yukalov}}\ and\ \bibinfo {author} {\bibfnamefont {E.~P.}\ \bibnamefont
  {Yukalova}},\ }\bibfield  {title} {\bibinfo {title} {Dynamics of quantum dot
  superradiance},\ }\href {https://doi.org/10.1103/PhysRevB.81.075308}
  {\bibfield  {journal} {\bibinfo  {journal} {Phys. Rev. B}\ }\textbf {\bibinfo
  {volume} {81}},\ \bibinfo {pages} {075308} (\bibinfo {year}
  {2010})}\BibitemShut {NoStop}%
\bibitem [{\citenamefont {Bradac}\ \emph {et~al.}(2017)\citenamefont {Bradac},
  \citenamefont {Johnsson}, \citenamefont {Breugel}, \citenamefont {Baragiola},
  \citenamefont {Martin}, \citenamefont {Juan}, \citenamefont {Brennen},\ and\
  \citenamefont {Volz}}]{Bradac2017room}%
  \BibitemOpen
  \bibfield  {author} {\bibinfo {author} {\bibfnamefont {C.}~\bibnamefont
  {Bradac}}, \bibinfo {author} {\bibfnamefont {M.~T.}\ \bibnamefont
  {Johnsson}}, \bibinfo {author} {\bibfnamefont {M.~v.}\ \bibnamefont
  {Breugel}}, \bibinfo {author} {\bibfnamefont {B.~Q.}\ \bibnamefont
  {Baragiola}}, \bibinfo {author} {\bibfnamefont {R.}~\bibnamefont {Martin}},
  \bibinfo {author} {\bibfnamefont {M.~L.}\ \bibnamefont {Juan}}, \bibinfo
  {author} {\bibfnamefont {G.~K.}\ \bibnamefont {Brennen}},\ and\ \bibinfo
  {author} {\bibfnamefont {T.}~\bibnamefont {Volz}},\ }\bibfield  {title}
  {\bibinfo {title} {Room-temperature spontaneous superradiance from single
  diamond nanocrystals},\ }\href {https://doi.org/10.1038/s41467-017-01397-4}
  {\bibfield  {journal} {\bibinfo  {journal} {Nat. Commun.}\ }\textbf {\bibinfo
  {volume} {8}},\ \bibinfo {pages} {1205} (\bibinfo {year} {2017})}\BibitemShut
  {NoStop}%
\bibitem [{\citenamefont {Haider}\ \emph {et~al.}(2021)\citenamefont {Haider},
  \citenamefont {Sampathkumar}, \citenamefont {Verhagen}, \citenamefont
  {Nádvorník}, \citenamefont {Sonia}, \citenamefont {Valeš}, \citenamefont
  {Sýkora}, \citenamefont {Kapusta}, \citenamefont {Němec}, \citenamefont
  {Hof}, \citenamefont {Frank}, \citenamefont {Chen}, \citenamefont
  {Vejpravová},\ and\ \citenamefont {Kalbáč}}]{Haider2021superradiant}%
  \BibitemOpen
  \bibfield  {author} {\bibinfo {author} {\bibfnamefont {G.}~\bibnamefont
  {Haider}}, \bibinfo {author} {\bibfnamefont {K.}~\bibnamefont
  {Sampathkumar}}, \bibinfo {author} {\bibfnamefont {T.}~\bibnamefont
  {Verhagen}}, \bibinfo {author} {\bibfnamefont {L.}~\bibnamefont
  {Nádvorník}}, \bibinfo {author} {\bibfnamefont {F.~J.}\ \bibnamefont
  {Sonia}}, \bibinfo {author} {\bibfnamefont {V.}~\bibnamefont {Valeš}},
  \bibinfo {author} {\bibfnamefont {J.}~\bibnamefont {Sýkora}}, \bibinfo
  {author} {\bibfnamefont {P.}~\bibnamefont {Kapusta}}, \bibinfo {author}
  {\bibfnamefont {P.}~\bibnamefont {Němec}}, \bibinfo {author} {\bibfnamefont
  {M.}~\bibnamefont {Hof}}, \bibinfo {author} {\bibfnamefont {O.}~\bibnamefont
  {Frank}}, \bibinfo {author} {\bibfnamefont {Y.-F.}\ \bibnamefont {Chen}},
  \bibinfo {author} {\bibfnamefont {J.}~\bibnamefont {Vejpravová}},\ and\
  \bibinfo {author} {\bibfnamefont {M.}~\bibnamefont {Kalbáč}},\ }\bibfield
  {title} {\bibinfo {title} {Superradiant emission from coherent excitons in
  van der waals heterostructures},\ }\href
  {https://doi.org/https://doi.org/10.1002/adfm.202102196} {\bibfield
  {journal} {\bibinfo  {journal} {Adv. Funct. Mater.}\ }\textbf {\bibinfo
  {volume} {31}},\ \bibinfo {pages} {2102196} (\bibinfo {year}
  {2021})}\BibitemShut {NoStop}%
\bibitem [{\citenamefont {Mello}\ \emph {et~al.}(2022)\citenamefont {Mello},
  \citenamefont {Li}, \citenamefont {Camayd-Muñoz}, \citenamefont {DeVault},
  \citenamefont {Lobet}, \citenamefont {Tang}, \citenamefont {Lonçar},\ and\
  \citenamefont {Mazur}}]{Mello2022extended}%
  \BibitemOpen
  \bibfield  {author} {\bibinfo {author} {\bibfnamefont {O.}~\bibnamefont
  {Mello}}, \bibinfo {author} {\bibfnamefont {Y.}~\bibnamefont {Li}}, \bibinfo
  {author} {\bibfnamefont {S.~A.}\ \bibnamefont {Camayd-Muñoz}}, \bibinfo
  {author} {\bibfnamefont {C.}~\bibnamefont {DeVault}}, \bibinfo {author}
  {\bibfnamefont {M.}~\bibnamefont {Lobet}}, \bibinfo {author} {\bibfnamefont
  {H.}~\bibnamefont {Tang}}, \bibinfo {author} {\bibfnamefont {M.}~\bibnamefont
  {Lonçar}},\ and\ \bibinfo {author} {\bibfnamefont {E.}~\bibnamefont
  {Mazur}},\ }\bibfield  {title} {\bibinfo {title} {Extended many-body
  superradiance in diamond epsilon near-zero metamaterials},\ }\href
  {https://doi.org/10.1063/5.0062869} {\bibfield  {journal} {\bibinfo
  {journal} {Appl. Phys. Lett.}\ }\textbf {\bibinfo {volume} {120}},\ \bibinfo
  {pages} {061105} (\bibinfo {year} {2022})}\BibitemShut {NoStop}%
\bibitem [{\citenamefont {Wang}\ \emph {et~al.}(2007)\citenamefont {Wang},
  \citenamefont {Yelin}, \citenamefont {C\^ot\'e}, \citenamefont {Eyler},
  \citenamefont {Farooqi}, \citenamefont {Gould}, \citenamefont
  {Ko\ifmmode~\check{s}\else \v{s}\fi{}trun}, \citenamefont {Tong},\ and\
  \citenamefont {Vrinceanu}}]{Wang2007superradiance}%
  \BibitemOpen
  \bibfield  {author} {\bibinfo {author} {\bibfnamefont {T.}~\bibnamefont
  {Wang}}, \bibinfo {author} {\bibfnamefont {S.~F.}\ \bibnamefont {Yelin}},
  \bibinfo {author} {\bibfnamefont {R.}~\bibnamefont {C\^ot\'e}}, \bibinfo
  {author} {\bibfnamefont {E.~E.}\ \bibnamefont {Eyler}}, \bibinfo {author}
  {\bibfnamefont {S.~M.}\ \bibnamefont {Farooqi}}, \bibinfo {author}
  {\bibfnamefont {P.~L.}\ \bibnamefont {Gould}}, \bibinfo {author}
  {\bibfnamefont {M.}~\bibnamefont {Ko\ifmmode~\check{s}\else \v{s}\fi{}trun}},
  \bibinfo {author} {\bibfnamefont {D.}~\bibnamefont {Tong}},\ and\ \bibinfo
  {author} {\bibfnamefont {D.}~\bibnamefont {Vrinceanu}},\ }\bibfield  {title}
  {\bibinfo {title} {Superradiance in ultracold rydberg gases},\ }\href
  {https://doi.org/10.1103/PhysRevA.75.033802} {\bibfield  {journal} {\bibinfo
  {journal} {Phys. Rev. A}\ }\textbf {\bibinfo {volume} {75}},\ \bibinfo
  {pages} {033802} (\bibinfo {year} {2007})}\BibitemShut {NoStop}%
\bibitem [{\citenamefont {Ferioli}\ \emph {et~al.}(2021)\citenamefont
  {Ferioli}, \citenamefont {Glicenstein}, \citenamefont {Robicheaux},
  \citenamefont {Sutherland}, \citenamefont {Browaeys},\ and\ \citenamefont
  {Ferrier-Barbut}}]{Ferioli2021laser}%
  \BibitemOpen
  \bibfield  {author} {\bibinfo {author} {\bibfnamefont {G.}~\bibnamefont
  {Ferioli}}, \bibinfo {author} {\bibfnamefont {A.}~\bibnamefont
  {Glicenstein}}, \bibinfo {author} {\bibfnamefont {F.}~\bibnamefont
  {Robicheaux}}, \bibinfo {author} {\bibfnamefont {R.~T.}\ \bibnamefont
  {Sutherland}}, \bibinfo {author} {\bibfnamefont {A.}~\bibnamefont
  {Browaeys}},\ and\ \bibinfo {author} {\bibfnamefont {I.}~\bibnamefont
  {Ferrier-Barbut}},\ }\bibfield  {title} {\bibinfo {title} {Laser-driven
  superradiant ensembles of two-level atoms near dicke regime},\ }\href
  {https://doi.org/10.1103/PhysRevLett.127.243602} {\bibfield  {journal}
  {\bibinfo  {journal} {Phys. Rev. Lett.}\ }\textbf {\bibinfo {volume} {127}},\
  \bibinfo {pages} {243602} (\bibinfo {year} {2021})}\BibitemShut {NoStop}%
\bibitem [{\citenamefont {Liedl}\ \emph {et~al.}(2022)\citenamefont {Liedl},
  \citenamefont {Pucher}, \citenamefont {Tebbenjohanns}, \citenamefont
  {Schneeweiss},\ and\ \citenamefont {Rauschenbeutel}}]{Liedl2022collective}%
  \BibitemOpen
  \bibfield  {author} {\bibinfo {author} {\bibfnamefont {C.}~\bibnamefont
  {Liedl}}, \bibinfo {author} {\bibfnamefont {S.}~\bibnamefont {Pucher}},
  \bibinfo {author} {\bibfnamefont {F.}~\bibnamefont {Tebbenjohanns}}, \bibinfo
  {author} {\bibfnamefont {P.}~\bibnamefont {Schneeweiss}},\ and\ \bibinfo
  {author} {\bibfnamefont {A.}~\bibnamefont {Rauschenbeutel}},\ }\href@noop {}
  {\bibinfo {title} {Collective excitation and decay of waveguide-coupled
  atoms: from timed dicke states to inverted ensembles}} (\bibinfo {year}
  {2022}),\ \Eprint {https://arxiv.org/abs/arXiv:2204.04106} {arXiv:2204.04106}
  \BibitemShut {NoStop}%
\bibitem [{\citenamefont {Trebbia}\ \emph {et~al.}(2022)\citenamefont
  {Trebbia}, \citenamefont {Deplano}, \citenamefont {Tamarat},\ and\
  \citenamefont {Lounis}}]{trebbia2022tailoring}%
  \BibitemOpen
  \bibfield  {author} {\bibinfo {author} {\bibfnamefont {J.-B.}\ \bibnamefont
  {Trebbia}}, \bibinfo {author} {\bibfnamefont {Q.}~\bibnamefont {Deplano}},
  \bibinfo {author} {\bibfnamefont {P.}~\bibnamefont {Tamarat}},\ and\ \bibinfo
  {author} {\bibfnamefont {B.}~\bibnamefont {Lounis}},\ }\bibfield  {title}
  {\bibinfo {title} {Tailoring the superradiant and subradiant nature of two
  coherently coupled quantum emitters},\ }\href
  {https://doi.org/10.1038/s41467-022-30672-2} {\bibfield  {journal} {\bibinfo
  {journal} {Nat. Commun.}\ }\textbf {\bibinfo {volume} {13}},\ \bibinfo
  {pages} {2962} (\bibinfo {year} {2022})}\BibitemShut {NoStop}%
\bibitem [{\citenamefont {Lambert}\ \emph {et~al.}(2016)\citenamefont
  {Lambert}, \citenamefont {Matsuzaki}, \citenamefont {Kakuyanagi},
  \citenamefont {Ishida}, \citenamefont {Saito},\ and\ \citenamefont
  {Nori}}]{Lambert2016superradiance}%
  \BibitemOpen
  \bibfield  {author} {\bibinfo {author} {\bibfnamefont {N.}~\bibnamefont
  {Lambert}}, \bibinfo {author} {\bibfnamefont {Y.}~\bibnamefont {Matsuzaki}},
  \bibinfo {author} {\bibfnamefont {K.}~\bibnamefont {Kakuyanagi}}, \bibinfo
  {author} {\bibfnamefont {N.}~\bibnamefont {Ishida}}, \bibinfo {author}
  {\bibfnamefont {S.}~\bibnamefont {Saito}},\ and\ \bibinfo {author}
  {\bibfnamefont {F.}~\bibnamefont {Nori}},\ }\bibfield  {title} {\bibinfo
  {title} {Superradiance with an ensemble of superconducting flux qubits},\
  }\href {https://doi.org/10.1103/PhysRevB.94.224510} {\bibfield  {journal}
  {\bibinfo  {journal} {Phys. Rev. B}\ }\textbf {\bibinfo {volume} {94}},\
  \bibinfo {pages} {224510} (\bibinfo {year} {2016})}\BibitemShut {NoStop}%
\bibitem [{\citenamefont {Wang}\ \emph {et~al.}(2020)\citenamefont {Wang},
  \citenamefont {Li}, \citenamefont {Feng}, \citenamefont {Song}, \citenamefont
  {Song}, \citenamefont {Liu}, \citenamefont {Guo}, \citenamefont {Zhang},
  \citenamefont {Dong}, \citenamefont {Zheng}, \citenamefont {Wang},\ and\
  \citenamefont {Wang}}]{Wang2020controllable}%
  \BibitemOpen
  \bibfield  {author} {\bibinfo {author} {\bibfnamefont {Z.}~\bibnamefont
  {Wang}}, \bibinfo {author} {\bibfnamefont {H.}~\bibnamefont {Li}}, \bibinfo
  {author} {\bibfnamefont {W.}~\bibnamefont {Feng}}, \bibinfo {author}
  {\bibfnamefont {X.}~\bibnamefont {Song}}, \bibinfo {author} {\bibfnamefont
  {C.}~\bibnamefont {Song}}, \bibinfo {author} {\bibfnamefont {W.}~\bibnamefont
  {Liu}}, \bibinfo {author} {\bibfnamefont {Q.}~\bibnamefont {Guo}}, \bibinfo
  {author} {\bibfnamefont {X.}~\bibnamefont {Zhang}}, \bibinfo {author}
  {\bibfnamefont {H.}~\bibnamefont {Dong}}, \bibinfo {author} {\bibfnamefont
  {D.}~\bibnamefont {Zheng}}, \bibinfo {author} {\bibfnamefont
  {H.}~\bibnamefont {Wang}},\ and\ \bibinfo {author} {\bibfnamefont {D.-W.}\
  \bibnamefont {Wang}},\ }\bibfield  {title} {\bibinfo {title} {Controllable
  switching between superradiant and subradiant states in a 10-qubit
  superconducting circuit},\ }\href
  {https://doi.org/10.1103/PhysRevLett.124.013601} {\bibfield  {journal}
  {\bibinfo  {journal} {Phys. Rev. Lett.}\ }\textbf {\bibinfo {volume} {124}},\
  \bibinfo {pages} {013601} (\bibinfo {year} {2020})}\BibitemShut {NoStop}%
\bibitem [{\citenamefont {Orell}\ \emph {et~al.}(2021)\citenamefont {Orell},
  \citenamefont {Zanner}, \citenamefont {Juan}, \citenamefont {Sharafiev},
  \citenamefont {Albert}, \citenamefont {Oleschko}, \citenamefont {Kirchmair},\
  and\ \citenamefont {Silveri}}]{Orell2021collective}%
  \BibitemOpen
  \bibfield  {author} {\bibinfo {author} {\bibfnamefont {T.}~\bibnamefont
  {Orell}}, \bibinfo {author} {\bibfnamefont {M.}~\bibnamefont {Zanner}},
  \bibinfo {author} {\bibfnamefont {M.~L.}\ \bibnamefont {Juan}}, \bibinfo
  {author} {\bibfnamefont {A.}~\bibnamefont {Sharafiev}}, \bibinfo {author}
  {\bibfnamefont {R.}~\bibnamefont {Albert}}, \bibinfo {author} {\bibfnamefont
  {S.}~\bibnamefont {Oleschko}}, \bibinfo {author} {\bibfnamefont
  {G.}~\bibnamefont {Kirchmair}},\ and\ \bibinfo {author} {\bibfnamefont
  {M.}~\bibnamefont {Silveri}},\ }\href@noop {} {\bibinfo {title} {Collective
  bosonic effects in an array of transmon devices}} (\bibinfo {year} {2021}),\
  \Eprint {https://arxiv.org/abs/arXiv:2112.08134} {arXiv:2112.08134}
  \BibitemShut {NoStop}%
\bibitem [{\citenamefont {Gonz\'alez-Tudela}\ \emph {et~al.}(2015)\citenamefont
  {Gonz\'alez-Tudela}, \citenamefont {Paulisch}, \citenamefont {Chang},
  \citenamefont {Kimble},\ and\ \citenamefont
  {Cirac}}]{gonzalez2015deterministic}%
  \BibitemOpen
  \bibfield  {author} {\bibinfo {author} {\bibfnamefont {A.}~\bibnamefont
  {Gonz\'alez-Tudela}}, \bibinfo {author} {\bibfnamefont {V.}~\bibnamefont
  {Paulisch}}, \bibinfo {author} {\bibfnamefont {D.~E.}\ \bibnamefont {Chang}},
  \bibinfo {author} {\bibfnamefont {H.~J.}\ \bibnamefont {Kimble}},\ and\
  \bibinfo {author} {\bibfnamefont {J.~I.}\ \bibnamefont {Cirac}},\ }\bibfield
  {title} {\bibinfo {title} {Deterministic generation of arbitrary photonic
  states assisted by dissipation},\ }\href
  {https://doi.org/10.1103/PhysRevLett.115.163603} {\bibfield  {journal}
  {\bibinfo  {journal} {Phys. Rev. Lett.}\ }\textbf {\bibinfo {volume} {115}},\
  \bibinfo {pages} {163603} (\bibinfo {year} {2015})}\BibitemShut {NoStop}%
\bibitem [{\citenamefont {Paulisch}\ \emph {et~al.}(2019)\citenamefont
  {Paulisch}, \citenamefont {Perarnau-Llobet}, \citenamefont
  {Gonz\'alez-Tudela},\ and\ \citenamefont {Cirac}}]{paulisch2019quantum}%
  \BibitemOpen
  \bibfield  {author} {\bibinfo {author} {\bibfnamefont {V.}~\bibnamefont
  {Paulisch}}, \bibinfo {author} {\bibfnamefont {M.}~\bibnamefont
  {Perarnau-Llobet}}, \bibinfo {author} {\bibfnamefont {A.}~\bibnamefont
  {Gonz\'alez-Tudela}},\ and\ \bibinfo {author} {\bibfnamefont {J.~I.}\
  \bibnamefont {Cirac}},\ }\bibfield  {title} {\bibinfo {title} {Quantum
  metrology with one-dimensional superradiant photonic states},\ }\href
  {https://doi.org/10.1103/PhysRevA.99.043807} {\bibfield  {journal} {\bibinfo
  {journal} {Phys. Rev. A}\ }\textbf {\bibinfo {volume} {99}},\ \bibinfo
  {pages} {043807} (\bibinfo {year} {2019})}\BibitemShut {NoStop}%
\bibitem [{\citenamefont {Groiseau}\ \emph {et~al.}(2021)\citenamefont
  {Groiseau}, \citenamefont {Elliott}, \citenamefont {Masson},\ and\
  \citenamefont {Parkins}}]{groiseau2021proposal}%
  \BibitemOpen
  \bibfield  {author} {\bibinfo {author} {\bibfnamefont {C.}~\bibnamefont
  {Groiseau}}, \bibinfo {author} {\bibfnamefont {A.~E.~J.}\ \bibnamefont
  {Elliott}}, \bibinfo {author} {\bibfnamefont {S.~J.}\ \bibnamefont
  {Masson}},\ and\ \bibinfo {author} {\bibfnamefont {S.}~\bibnamefont
  {Parkins}},\ }\bibfield  {title} {\bibinfo {title} {Proposal for a
  deterministic single-atom source of quasisuperradiant $n$-photon pulses},\
  }\href {https://doi.org/10.1103/PhysRevLett.127.033602} {\bibfield  {journal}
  {\bibinfo  {journal} {Phys. Rev. Lett.}\ }\textbf {\bibinfo {volume} {127}},\
  \bibinfo {pages} {033602} (\bibinfo {year} {2021})}\BibitemShut {NoStop}%
\bibitem [{\citenamefont {Perarnau-Llobet}\ \emph {et~al.}(2020)\citenamefont
  {Perarnau-Llobet}, \citenamefont {Gonz\'alez-Tudela},\ and\ \citenamefont
  {Cirac}}]{perarnau2020multimode}%
  \BibitemOpen
  \bibfield  {author} {\bibinfo {author} {\bibfnamefont {M.}~\bibnamefont
  {Perarnau-Llobet}}, \bibinfo {author} {\bibfnamefont {A.}~\bibnamefont
  {Gonz\'alez-Tudela}},\ and\ \bibinfo {author} {\bibfnamefont {J.~I.}\
  \bibnamefont {Cirac}},\ }\bibfield  {title} {\bibinfo {title} {Multimode fock
  states with large photon number: effective descriptions and applications in
  quantum metrology},\ }\href {https://doi.org/10.1088/2058-9565/ab6ce5}
  {\bibfield  {journal} {\bibinfo  {journal} {Quantum Sci. Technol.}\ }\textbf
  {\bibinfo {volume} {5}},\ \bibinfo {pages} {025003} (\bibinfo {year}
  {2020})}\BibitemShut {NoStop}%
\bibitem [{\citenamefont {Monshouwer}\ \emph {et~al.}(1997)\citenamefont
  {Monshouwer}, \citenamefont {Abrahamsson}, \citenamefont {van Mourik},\ and\
  \citenamefont {van Grondelle}}]{monshouwer1997superradiance}%
  \BibitemOpen
  \bibfield  {author} {\bibinfo {author} {\bibfnamefont {R.}~\bibnamefont
  {Monshouwer}}, \bibinfo {author} {\bibfnamefont {M.}~\bibnamefont
  {Abrahamsson}}, \bibinfo {author} {\bibfnamefont {F.}~\bibnamefont {van
  Mourik}},\ and\ \bibinfo {author} {\bibfnamefont {R.}~\bibnamefont {van
  Grondelle}},\ }\bibfield  {title} {\bibinfo {title} {Superradiance and
  exciton delocalization in bacterial photosynthetic light-harvesting
  systems},\ }\href {https://doi.org/10.1021/jp963377t} {\bibfield  {journal}
  {\bibinfo  {journal} {The Journal of Physical Chemistry B}\ }\textbf
  {\bibinfo {volume} {101}},\ \bibinfo {pages} {7241} (\bibinfo {year}
  {1997})}\BibitemShut {NoStop}%
\bibitem [{\citenamefont {Scholes}(2002)}]{scholes2002designing}%
  \BibitemOpen
  \bibfield  {author} {\bibinfo {author} {\bibfnamefont {G.~D.}\ \bibnamefont
  {Scholes}},\ }\bibfield  {title} {\bibinfo {title} {Designing
  light-harvesting antenna systems based on superradiant molecular
  aggregates},\ }\href
  {https://doi.org/https://doi.org/10.1016/S0301-0104(01)00533-X} {\bibfield
  {journal} {\bibinfo  {journal} {Chem. Phys.}\ }\textbf {\bibinfo {volume}
  {275}},\ \bibinfo {pages} {373} (\bibinfo {year} {2002})},\ \bibinfo {note}
  {photoprocesses in Multichromophoric Molecular Assemblies}\BibitemShut
  {NoStop}%
\bibitem [{\citenamefont {Celardo}\ \emph {et~al.}(2014)\citenamefont
  {Celardo}, \citenamefont {Poli}, \citenamefont {Lussardi},\ and\
  \citenamefont {Borgonovi}}]{celardo2014cooperative}%
  \BibitemOpen
  \bibfield  {author} {\bibinfo {author} {\bibfnamefont {G.~L.}\ \bibnamefont
  {Celardo}}, \bibinfo {author} {\bibfnamefont {P.}~\bibnamefont {Poli}},
  \bibinfo {author} {\bibfnamefont {L.}~\bibnamefont {Lussardi}},\ and\
  \bibinfo {author} {\bibfnamefont {F.}~\bibnamefont {Borgonovi}},\ }\bibfield
  {title} {\bibinfo {title} {Cooperative robustness to dephasing:
  Single-exciton superradiance in a nanoscale ring to model natural
  light-harvesting systems},\ }\href
  {https://doi.org/10.1103/PhysRevB.90.085142} {\bibfield  {journal} {\bibinfo
  {journal} {Phys. Rev. B}\ }\textbf {\bibinfo {volume} {90}},\ \bibinfo
  {pages} {085142} (\bibinfo {year} {2014})}\BibitemShut {NoStop}%
\bibitem [{\citenamefont {Rainò}\ \emph {et~al.}(2020)\citenamefont {Rainò},
  \citenamefont {Utzat}, \citenamefont {Bawendi},\ and\ \citenamefont
  {Kovalenko}}]{raino2020superradiant}%
  \BibitemOpen
  \bibfield  {author} {\bibinfo {author} {\bibfnamefont {G.}~\bibnamefont
  {Rainò}}, \bibinfo {author} {\bibfnamefont {H.}~\bibnamefont {Utzat}},
  \bibinfo {author} {\bibfnamefont {M.}~\bibnamefont {Bawendi}},\ and\ \bibinfo
  {author} {\bibfnamefont {M.}~\bibnamefont {Kovalenko}},\ }\bibfield  {title}
  {\bibinfo {title} {Superradiant emission from self-assembled light emitters:
  From molecules to quantum dots},\ }\href
  {https://doi.org/10.1557/mrs.2020.250} {\bibfield  {journal} {\bibinfo
  {journal} {MRS Bulletin}\ }\textbf {\bibinfo {volume} {45}},\ \bibinfo
  {pages} {841–848} (\bibinfo {year} {2020})}\BibitemShut {NoStop}%
\bibitem [{\citenamefont {Yang}\ \emph {et~al.}(2021)\citenamefont {Yang},
  \citenamefont {Oh}, \citenamefont {Han}, \citenamefont {Son}, \citenamefont
  {Kim}, \citenamefont {Kim}, \citenamefont {Lee},\ and\ \citenamefont
  {An}}]{yang2021realization}%
  \BibitemOpen
  \bibfield  {author} {\bibinfo {author} {\bibfnamefont {D.}~\bibnamefont
  {Yang}}, \bibinfo {author} {\bibfnamefont {S.-h.}\ \bibnamefont {Oh}},
  \bibinfo {author} {\bibfnamefont {J.}~\bibnamefont {Han}}, \bibinfo {author}
  {\bibfnamefont {G.}~\bibnamefont {Son}}, \bibinfo {author} {\bibfnamefont
  {J.}~\bibnamefont {Kim}}, \bibinfo {author} {\bibfnamefont {J.}~\bibnamefont
  {Kim}}, \bibinfo {author} {\bibfnamefont {M.}~\bibnamefont {Lee}},\ and\
  \bibinfo {author} {\bibfnamefont {K.}~\bibnamefont {An}},\ }\bibfield
  {title} {\bibinfo {title} {Realization of superabsorption by time reversal of
  superradiance},\ }\href {https://doi.org/10.1038/s41566-021-00770-6}
  {\bibfield  {journal} {\bibinfo  {journal} {Nat. Photonics}\ }\textbf
  {\bibinfo {volume} {15}},\ \bibinfo {pages} {272} (\bibinfo {year}
  {2021})}\BibitemShut {NoStop}%
\bibitem [{\citenamefont {Higgins}\ \emph {et~al.}(2014)\citenamefont
  {Higgins}, \citenamefont {Benjamin}, \citenamefont {Stace}, \citenamefont
  {Milburn}, \citenamefont {Lovett},\ and\ \citenamefont
  {Gauger}}]{higgins2014superabsorption}%
  \BibitemOpen
  \bibfield  {author} {\bibinfo {author} {\bibfnamefont {K.~D.~B.}\
  \bibnamefont {Higgins}}, \bibinfo {author} {\bibfnamefont {S.~C.}\
  \bibnamefont {Benjamin}}, \bibinfo {author} {\bibfnamefont {T.~M.}\
  \bibnamefont {Stace}}, \bibinfo {author} {\bibfnamefont {G.~J.}\ \bibnamefont
  {Milburn}}, \bibinfo {author} {\bibfnamefont {B.~W.}\ \bibnamefont
  {Lovett}},\ and\ \bibinfo {author} {\bibfnamefont {E.~M.}\ \bibnamefont
  {Gauger}},\ }\bibfield  {title} {\bibinfo {title} {Superabsorption of light
  via quantum engineering},\ }\href {https://doi.org/10.1038/ncomms5705}
  {\bibfield  {journal} {\bibinfo  {journal} {Nat. Commun.}\ }\textbf {\bibinfo
  {volume} {5}},\ \bibinfo {pages} {4705} (\bibinfo {year} {2014})}\BibitemShut
  {NoStop}%
\bibitem [{\citenamefont {Masson}\ and\ \citenamefont
  {Asenjo-Garcia}(2020)}]{Masson2020atomic}%
  \BibitemOpen
  \bibfield  {author} {\bibinfo {author} {\bibfnamefont {S.~J.}\ \bibnamefont
  {Masson}}\ and\ \bibinfo {author} {\bibfnamefont {A.}~\bibnamefont
  {Asenjo-Garcia}},\ }\bibfield  {title} {\bibinfo {title} {Atomic-waveguide
  quantum electrodynamics},\ }\href
  {https://doi.org/10.1103/PhysRevResearch.2.043213} {\bibfield  {journal}
  {\bibinfo  {journal} {Phys. Rev. Research}\ }\textbf {\bibinfo {volume}
  {2}},\ \bibinfo {pages} {043213} (\bibinfo {year} {2020})}\BibitemShut
  {NoStop}%
\bibitem [{\citenamefont {Ruks}\ and\ \citenamefont
  {Busch}(2022)}]{ruks2022greens}%
  \BibitemOpen
  \bibfield  {author} {\bibinfo {author} {\bibfnamefont {L.}~\bibnamefont
  {Ruks}}\ and\ \bibinfo {author} {\bibfnamefont {T.}~\bibnamefont {Busch}},\
  }\bibfield  {title} {\bibinfo {title} {Green's functions of and emission into
  discrete anisotropic and hyperbolic baths},\ }\href
  {https://doi.org/10.1103/PhysRevResearch.4.023044} {\bibfield  {journal}
  {\bibinfo  {journal} {Phys. Rev. Research}\ }\textbf {\bibinfo {volume}
  {4}},\ \bibinfo {pages} {023044} (\bibinfo {year} {2022})}\BibitemShut
  {NoStop}%
\bibitem [{\citenamefont {Cardenas-Lopez}\ \emph {et~al.}(2022)\citenamefont
  {Cardenas-Lopez}, \citenamefont {Masson}, \citenamefont {Zager},\ and\
  \citenamefont {Asenjo-Garcia}}]{silvia2022many}%
  \BibitemOpen
  \bibfield  {author} {\bibinfo {author} {\bibfnamefont {S.}~\bibnamefont
  {Cardenas-Lopez}}, \bibinfo {author} {\bibfnamefont {S.~J.}\ \bibnamefont
  {Masson}}, \bibinfo {author} {\bibfnamefont {Z.}~\bibnamefont {Zager}},\ and\
  \bibinfo {author} {\bibfnamefont {A.}~\bibnamefont {Asenjo-Garcia}},\
  }\href@noop {} {\bibinfo {title} {Many-body superradiance and dynamical
  symmetry breaking in waveguide qed}} (\bibinfo {year} {2022}),\ \Eprint
  {https://arxiv.org/abs/arXiv:2209.12970} {arXiv:2209.12970} \BibitemShut
  {NoStop}%
\bibitem [{\citenamefont {Carmichael}\ and\ \citenamefont
  {Kim}(2000)}]{carmichael2000quantum}%
  \BibitemOpen
  \bibfield  {author} {\bibinfo {author} {\bibfnamefont {H.}~\bibnamefont
  {Carmichael}}\ and\ \bibinfo {author} {\bibfnamefont {K.}~\bibnamefont
  {Kim}},\ }\bibfield  {title} {\bibinfo {title} {A quantum trajectory
  unraveling of the superradiance master equation1we dedicate this paper to
  marlan scully on the occasion of his 60th birthday.1},\ }\href
  {https://doi.org/https://doi.org/10.1016/S0030-4018(99)00694-X} {\bibfield
  {journal} {\bibinfo  {journal} {Opt. Commun.}\ }\textbf {\bibinfo {volume}
  {179}},\ \bibinfo {pages} {417} (\bibinfo {year} {2000})}\BibitemShut
  {NoStop}%
\bibitem [{\citenamefont {Mattiotti}\ \emph {et~al.}(2020)\citenamefont
  {Mattiotti}, \citenamefont {Kuno}, \citenamefont {Borgonovi}, \citenamefont
  {Jankó},\ and\ \citenamefont {Celardo}}]{mattiotti2020thermal}%
  \BibitemOpen
  \bibfield  {author} {\bibinfo {author} {\bibfnamefont {F.}~\bibnamefont
  {Mattiotti}}, \bibinfo {author} {\bibfnamefont {M.}~\bibnamefont {Kuno}},
  \bibinfo {author} {\bibfnamefont {F.}~\bibnamefont {Borgonovi}}, \bibinfo
  {author} {\bibfnamefont {B.}~\bibnamefont {Jankó}},\ and\ \bibinfo {author}
  {\bibfnamefont {G.~L.}\ \bibnamefont {Celardo}},\ }\bibfield  {title}
  {\bibinfo {title} {Thermal decoherence of superradiance in lead halide
  perovskite nanocrystal superlattices},\ }\href
  {https://doi.org/10.1021/acs.nanolett.0c02784} {\bibfield  {journal}
  {\bibinfo  {journal} {Nano Lett.}\ }\textbf {\bibinfo {volume} {20}},\
  \bibinfo {pages} {7382} (\bibinfo {year} {2020})}\BibitemShut {NoStop}%
\bibitem [{\citenamefont {Bellomo}\ \emph {et~al.}(2017)\citenamefont
  {Bellomo}, \citenamefont {Giorgi}, \citenamefont {Palma},\ and\ \citenamefont
  {Zambrini}}]{bellomo2017quantum}%
  \BibitemOpen
  \bibfield  {author} {\bibinfo {author} {\bibfnamefont {B.}~\bibnamefont
  {Bellomo}}, \bibinfo {author} {\bibfnamefont {G.~L.}\ \bibnamefont {Giorgi}},
  \bibinfo {author} {\bibfnamefont {G.~M.}\ \bibnamefont {Palma}},\ and\
  \bibinfo {author} {\bibfnamefont {R.}~\bibnamefont {Zambrini}},\ }\bibfield
  {title} {\bibinfo {title} {Quantum synchronization as a local signature of
  super- and subradiance},\ }\href {https://doi.org/10.1103/PhysRevA.95.043807}
  {\bibfield  {journal} {\bibinfo  {journal} {Phys. Rev. A}\ }\textbf {\bibinfo
  {volume} {95}},\ \bibinfo {pages} {043807} (\bibinfo {year}
  {2017})}\BibitemShut {NoStop}%
\bibitem [{\citenamefont {El-Nashar}\ \emph {et~al.}(2003)\citenamefont
  {El-Nashar}, \citenamefont {Zhang}, \citenamefont {Cerdeira},\ and\
  \citenamefont {Ibiyinka~A.}}]{hassan2003synchronization}%
  \BibitemOpen
  \bibfield  {author} {\bibinfo {author} {\bibfnamefont {H.~F.}\ \bibnamefont
  {El-Nashar}}, \bibinfo {author} {\bibfnamefont {Y.}~\bibnamefont {Zhang}},
  \bibinfo {author} {\bibfnamefont {H.~A.}\ \bibnamefont {Cerdeira}},\ and\
  \bibinfo {author} {\bibfnamefont {F.}~\bibnamefont {Ibiyinka~A.}},\
  }\bibfield  {title} {\bibinfo {title} {Synchronization in a chain of nearest
  neighbors coupled oscillators with fixed ends},\ }\href
  {https://doi.org/10.1063/1.1611851} {\bibfield  {journal} {\bibinfo
  {journal} {Chaos}\ }\textbf {\bibinfo {volume} {13}},\ \bibinfo {pages}
  {1216} (\bibinfo {year} {2003})}\BibitemShut {NoStop}%
\bibitem [{\citenamefont {Lindblad}(1976)}]{Lindblad1976on}%
  \BibitemOpen
  \bibfield  {author} {\bibinfo {author} {\bibfnamefont {G.}~\bibnamefont
  {Lindblad}},\ }\bibfield  {title} {\bibinfo {title} {On the generators of
  quantum dynamical semigroups},\ }\href {https://doi.org/10.1007/BF01608499}
  {\bibfield  {journal} {\bibinfo  {journal} {Commun. Math. Phys.}\ }\textbf
  {\bibinfo {volume} {48}},\ \bibinfo {pages} {119} (\bibinfo {year}
  {1976})}\BibitemShut {NoStop}%
\bibitem [{\citenamefont {Breuer}\ and\ \citenamefont
  {Petruccione}(2007)}]{Breuer2007theory}%
  \BibitemOpen
  \bibfield  {author} {\bibinfo {author} {\bibfnamefont {H.-P.}\ \bibnamefont
  {Breuer}}\ and\ \bibinfo {author} {\bibfnamefont {F.}~\bibnamefont
  {Petruccione}},\ }\href
  {https://doi.org/10.1093/acprof:oso/9780199213900.001.0001} {\emph {\bibinfo
  {title} {The Theory of Open Quantum Systems}}}\ (\bibinfo  {publisher}
  {Oxford University Press},\ \bibinfo {address} {Oxford},\ \bibinfo {year}
  {2007})\ p.\ \bibinfo {pages} {656}\BibitemShut {NoStop}%
\bibitem [{\citenamefont {Lidar}(2019)}]{Lidar2019lecture}%
  \BibitemOpen
  \bibfield  {author} {\bibinfo {author} {\bibfnamefont {D.~A.}\ \bibnamefont
  {Lidar}},\ }\href@noop {} {\bibinfo {title} {Lecture notes on the theory of
  open quantum systems}} (\bibinfo {year} {2019}),\ \Eprint
  {https://arxiv.org/abs/arXiv:1902.00967} {arXiv:1902.00967} \BibitemShut
  {NoStop}%
\bibitem [{sup()}]{supp}%
  \BibitemOpen
  \href@noop {} {}\bibinfo {note} {See Supplementary Materials for additional
  calculation details, which includes Ref. [68-77]}\BibitemShut {NoStop}%
\bibitem [{\citenamefont {Golub}\ and\ \citenamefont {van
  Loan}(2013)}]{golub2013matrix}%
  \BibitemOpen
  \bibfield  {author} {\bibinfo {author} {\bibfnamefont {G.~H.}\ \bibnamefont
  {Golub}}\ and\ \bibinfo {author} {\bibfnamefont {C.~F.}\ \bibnamefont {van
  Loan}},\ }\href {http://www.cs.cornell.edu/cv/GVL4/golubandvanloan.htm}
  {\emph {\bibinfo {title} {Matrix Computations}}},\ \bibinfo {edition} {4th}\
  ed.\ (\bibinfo  {publisher} {JHU Press},\ \bibinfo {year} {2013})\BibitemShut
  {NoStop}%
\bibitem [{\citenamefont {Periwal}\ \emph {et~al.}(2021)\citenamefont
  {Periwal}, \citenamefont {Cooper}, \citenamefont {Kunkel}, \citenamefont
  {Wienand}, \citenamefont {Davis},\ and\ \citenamefont
  {Schleier-Smith}}]{periwal2021programmable}%
  \BibitemOpen
  \bibfield  {author} {\bibinfo {author} {\bibfnamefont {A.}~\bibnamefont
  {Periwal}}, \bibinfo {author} {\bibfnamefont {E.~S.}\ \bibnamefont {Cooper}},
  \bibinfo {author} {\bibfnamefont {P.}~\bibnamefont {Kunkel}}, \bibinfo
  {author} {\bibfnamefont {J.~F.}\ \bibnamefont {Wienand}}, \bibinfo {author}
  {\bibfnamefont {E.~J.}\ \bibnamefont {Davis}},\ and\ \bibinfo {author}
  {\bibfnamefont {M.}~\bibnamefont {Schleier-Smith}},\ }\bibfield  {title}
  {\bibinfo {title} {Programmable interactions and emergent geometry in an
  array{\^a} of atom clouds},\ }\href
  {https://doi.org/10.1038/s41586-021-04156-0} {\bibfield  {journal} {\bibinfo
  {journal} {Nature}\ }\textbf {\bibinfo {volume} {600}},\ \bibinfo {pages}
  {630} (\bibinfo {year} {2021})}\BibitemShut {NoStop}%
\bibitem [{\citenamefont {Koll\'{a}r}\ \emph {et~al.}(2019)\citenamefont
  {Koll\'{a}r}, \citenamefont {Fitzpatrick},\ and\ \citenamefont
  {Houck}}]{Kollar2019hyperbolic}%
  \BibitemOpen
  \bibfield  {author} {\bibinfo {author} {\bibfnamefont {A.~J.}\ \bibnamefont
  {Koll\'{a}r}}, \bibinfo {author} {\bibfnamefont {M.}~\bibnamefont
  {Fitzpatrick}},\ and\ \bibinfo {author} {\bibfnamefont {A.~A.}\ \bibnamefont
  {Houck}},\ }\bibfield  {title} {\bibinfo {title} {Hyperbolic lattices in
  circuit quantum electrodynamics},\ }\href
  {https://doi.org/10.1038/s41586-019-1348-3} {\bibfield  {journal} {\bibinfo
  {journal} {Nature}\ }\textbf {\bibinfo {volume} {571}},\ \bibinfo {pages}
  {45} (\bibinfo {year} {2019})}\BibitemShut {NoStop}%
\bibitem [{\citenamefont {Douglas}\ \emph {et~al.}(2015)\citenamefont
  {Douglas}, \citenamefont {Habibian}, \citenamefont {Hung}, \citenamefont
  {Gorshkov}, \citenamefont {Kimble},\ and\ \citenamefont
  {Chang}}]{Douglas2015quantum}%
  \BibitemOpen
  \bibfield  {author} {\bibinfo {author} {\bibfnamefont {J.~S.}\ \bibnamefont
  {Douglas}}, \bibinfo {author} {\bibfnamefont {H.}~\bibnamefont {Habibian}},
  \bibinfo {author} {\bibfnamefont {C.-L.}\ \bibnamefont {Hung}}, \bibinfo
  {author} {\bibfnamefont {A.~V.}\ \bibnamefont {Gorshkov}}, \bibinfo {author}
  {\bibfnamefont {H.~J.}\ \bibnamefont {Kimble}},\ and\ \bibinfo {author}
  {\bibfnamefont {D.~E.}\ \bibnamefont {Chang}},\ }\bibfield  {title} {\bibinfo
  {title} {Quantum many-body models with cold atoms coupled to photonic
  crystals},\ }\href {https://doi.org/10.1038/nphoton.2015.57} {\bibfield
  {journal} {\bibinfo  {journal} {Nat. Photonics}\ }\textbf {\bibinfo {volume}
  {9}},\ \bibinfo {pages} {326} (\bibinfo {year} {2015})}\BibitemShut {NoStop}%
\bibitem [{\citenamefont {Munro}\ \emph {et~al.}(2018)\citenamefont {Munro},
  \citenamefont {Asenjo-Garcia}, \citenamefont {Lin}, \citenamefont {Kwek},
  \citenamefont {Regal},\ and\ \citenamefont {Chang}}]{Munro2018population}%
  \BibitemOpen
  \bibfield  {author} {\bibinfo {author} {\bibfnamefont {E.}~\bibnamefont
  {Munro}}, \bibinfo {author} {\bibfnamefont {A.}~\bibnamefont
  {Asenjo-Garcia}}, \bibinfo {author} {\bibfnamefont {Y.}~\bibnamefont {Lin}},
  \bibinfo {author} {\bibfnamefont {L.~C.}\ \bibnamefont {Kwek}}, \bibinfo
  {author} {\bibfnamefont {C.~A.}\ \bibnamefont {Regal}},\ and\ \bibinfo
  {author} {\bibfnamefont {D.~E.}\ \bibnamefont {Chang}},\ }\bibfield  {title}
  {\bibinfo {title} {Population mixing due to dipole-dipole interactions in a
  one-dimensional array of multilevel atoms},\ }\href
  {https://doi.org/10.1103/PhysRevA.98.033815} {\bibfield  {journal} {\bibinfo
  {journal} {Phys. Rev. A}\ }\textbf {\bibinfo {volume} {98}},\ \bibinfo
  {pages} {033815} (\bibinfo {year} {2018})}\BibitemShut {NoStop}%
\bibitem [{\citenamefont {Solano}\ \emph {et~al.}(2017)\citenamefont {Solano},
  \citenamefont {Barberis-Blostein}, \citenamefont {Fatemi}, \citenamefont
  {Orozco},\ and\ \citenamefont {Rolston}}]{Solano2017super}%
  \BibitemOpen
  \bibfield  {author} {\bibinfo {author} {\bibfnamefont {P.}~\bibnamefont
  {Solano}}, \bibinfo {author} {\bibfnamefont {P.}~\bibnamefont
  {Barberis-Blostein}}, \bibinfo {author} {\bibfnamefont {F.~K.}\ \bibnamefont
  {Fatemi}}, \bibinfo {author} {\bibfnamefont {L.~A.}\ \bibnamefont {Orozco}},\
  and\ \bibinfo {author} {\bibfnamefont {S.~L.}\ \bibnamefont {Rolston}},\
  }\bibfield  {title} {\bibinfo {title} {Super-radiance reveals infinite-range
  dipole interactions through a nanofiber},\ }\href
  {https://doi.org/10.1038/s41467-017-01994-3} {\bibfield  {journal} {\bibinfo
  {journal} {Nat. Commun.}\ }\textbf {\bibinfo {volume} {8}},\ \bibinfo {pages}
  {1857} (\bibinfo {year} {2017})}\BibitemShut {NoStop}%
\bibitem [{\citenamefont {Lodahl}\ \emph {et~al.}(2017)\citenamefont {Lodahl},
  \citenamefont {Mahmoodian}, \citenamefont {Stobbe}, \citenamefont
  {Rauschenbeutel}, \citenamefont {Schneeweiss}, \citenamefont {Volz},
  \citenamefont {Pichler},\ and\ \citenamefont {Zoller}}]{Lodahl2017chiral}%
  \BibitemOpen
  \bibfield  {author} {\bibinfo {author} {\bibfnamefont {P.}~\bibnamefont
  {Lodahl}}, \bibinfo {author} {\bibfnamefont {S.}~\bibnamefont {Mahmoodian}},
  \bibinfo {author} {\bibfnamefont {S.}~\bibnamefont {Stobbe}}, \bibinfo
  {author} {\bibfnamefont {A.}~\bibnamefont {Rauschenbeutel}}, \bibinfo
  {author} {\bibfnamefont {P.}~\bibnamefont {Schneeweiss}}, \bibinfo {author}
  {\bibfnamefont {J.}~\bibnamefont {Volz}}, \bibinfo {author} {\bibfnamefont
  {H.}~\bibnamefont {Pichler}},\ and\ \bibinfo {author} {\bibfnamefont
  {P.}~\bibnamefont {Zoller}},\ }\bibfield  {title} {\bibinfo {title} {Chiral
  quantum optics},\ }\href {https://doi.org/10.1038/nature21037} {\bibfield
  {journal} {\bibinfo  {journal} {Nature}\ }\textbf {\bibinfo {volume} {541}},\
  \bibinfo {pages} {473} (\bibinfo {year} {2017})}\BibitemShut {NoStop}%
\bibitem [{\citenamefont {Pichler}\ \emph {et~al.}(2015)\citenamefont
  {Pichler}, \citenamefont {Ramos}, \citenamefont {Daley},\ and\ \citenamefont
  {Zoller}}]{Pichler2015quantum}%
  \BibitemOpen
  \bibfield  {author} {\bibinfo {author} {\bibfnamefont {H.}~\bibnamefont
  {Pichler}}, \bibinfo {author} {\bibfnamefont {T.}~\bibnamefont {Ramos}},
  \bibinfo {author} {\bibfnamefont {A.~J.}\ \bibnamefont {Daley}},\ and\
  \bibinfo {author} {\bibfnamefont {P.}~\bibnamefont {Zoller}},\ }\bibfield
  {title} {\bibinfo {title} {Quantum optics of chiral spin networks},\ }\href
  {https://doi.org/10.1103/PhysRevA.91.042116} {\bibfield  {journal} {\bibinfo
  {journal} {Phys. Rev. A}\ }\textbf {\bibinfo {volume} {91}},\ \bibinfo
  {pages} {042116} (\bibinfo {year} {2015})}\BibitemShut {NoStop}%
\bibitem [{\citenamefont {Mok}\ \emph {et~al.}(2020{\natexlab{a}})\citenamefont
  {Mok}, \citenamefont {Aghamalyan}, \citenamefont {You}, \citenamefont {Haug},
  \citenamefont {Zhang}, \citenamefont {Png},\ and\ \citenamefont
  {Kwek}}]{mok2020long}%
  \BibitemOpen
  \bibfield  {author} {\bibinfo {author} {\bibfnamefont {W.-K.}\ \bibnamefont
  {Mok}}, \bibinfo {author} {\bibfnamefont {D.}~\bibnamefont {Aghamalyan}},
  \bibinfo {author} {\bibfnamefont {J.-B.}\ \bibnamefont {You}}, \bibinfo
  {author} {\bibfnamefont {T.}~\bibnamefont {Haug}}, \bibinfo {author}
  {\bibfnamefont {W.}~\bibnamefont {Zhang}}, \bibinfo {author} {\bibfnamefont
  {C.~E.}\ \bibnamefont {Png}},\ and\ \bibinfo {author} {\bibfnamefont {L.-C.}\
  \bibnamefont {Kwek}},\ }\bibfield  {title} {\bibinfo {title} {Long-distance
  dissipation-assisted transport of entangled states via a chiral waveguide},\
  }\href {https://doi.org/10.1103/PhysRevResearch.2.013369} {\bibfield
  {journal} {\bibinfo  {journal} {Phys. Rev. Research}\ }\textbf {\bibinfo
  {volume} {2}},\ \bibinfo {pages} {013369} (\bibinfo {year}
  {2020}{\natexlab{a}})}\BibitemShut {NoStop}%
\bibitem [{\citenamefont {Mok}\ \emph {et~al.}(2020{\natexlab{b}})\citenamefont
  {Mok}, \citenamefont {You}, \citenamefont {Kwek},\ and\ \citenamefont
  {Aghamalyan}}]{mok2020microresonators}%
  \BibitemOpen
  \bibfield  {author} {\bibinfo {author} {\bibfnamefont {W.-K.}\ \bibnamefont
  {Mok}}, \bibinfo {author} {\bibfnamefont {J.-B.}\ \bibnamefont {You}},
  \bibinfo {author} {\bibfnamefont {L.-C.}\ \bibnamefont {Kwek}},\ and\
  \bibinfo {author} {\bibfnamefont {D.}~\bibnamefont {Aghamalyan}},\ }\bibfield
   {title} {\bibinfo {title} {Microresonators enhancing long-distance dynamical
  entanglement generation in chiral quantum networks},\ }\href
  {https://doi.org/10.1103/PhysRevA.101.053861} {\bibfield  {journal} {\bibinfo
   {journal} {Phys. Rev. A}\ }\textbf {\bibinfo {volume} {101}},\ \bibinfo
  {pages} {053861} (\bibinfo {year} {2020}{\natexlab{b}})}\BibitemShut
  {NoStop}%
\bibitem [{\citenamefont {Mahmoodian}\ \emph {et~al.}(2020)\citenamefont
  {Mahmoodian}, \citenamefont {Calaj\'o}, \citenamefont {Chang}, \citenamefont
  {Hammerer},\ and\ \citenamefont {S\o{}rensen}}]{mahmoodian2020dynamics}%
  \BibitemOpen
  \bibfield  {author} {\bibinfo {author} {\bibfnamefont {S.}~\bibnamefont
  {Mahmoodian}}, \bibinfo {author} {\bibfnamefont {G.}~\bibnamefont
  {Calaj\'o}}, \bibinfo {author} {\bibfnamefont {D.~E.}\ \bibnamefont {Chang}},
  \bibinfo {author} {\bibfnamefont {K.}~\bibnamefont {Hammerer}},\ and\
  \bibinfo {author} {\bibfnamefont {A.~S.}\ \bibnamefont {S\o{}rensen}},\
  }\bibfield  {title} {\bibinfo {title} {Dynamics of many-body photon bound
  states in chiral waveguide qed},\ }\href
  {https://doi.org/10.1103/PhysRevX.10.031011} {\bibfield  {journal} {\bibinfo
  {journal} {Phys. Rev. X}\ }\textbf {\bibinfo {volume} {10}},\ \bibinfo
  {pages} {031011} (\bibinfo {year} {2020})}\BibitemShut {NoStop}%
\bibitem [{\citenamefont {Robicheaux}\ and\ \citenamefont
  {Suresh}(2021)}]{robicheaux2021beyond}%
  \BibitemOpen
  \bibfield  {author} {\bibinfo {author} {\bibfnamefont {F.}~\bibnamefont
  {Robicheaux}}\ and\ \bibinfo {author} {\bibfnamefont {D.~A.}\ \bibnamefont
  {Suresh}},\ }\bibfield  {title} {\bibinfo {title} {Beyond lowest order
  mean-field theory for light interacting with atom arrays},\ }\href
  {https://doi.org/10.1103/PhysRevA.104.023702} {\bibfield  {journal} {\bibinfo
   {journal} {Phys. Rev. A}\ }\textbf {\bibinfo {volume} {104}},\ \bibinfo
  {pages} {023702} (\bibinfo {year} {2021})}\BibitemShut {NoStop}%
\bibitem [{\citenamefont {Rubies-Bigorda}\ \emph {et~al.}(2022)\citenamefont
  {Rubies-Bigorda}, \citenamefont {Ostermann},\ and\ \citenamefont
  {Yelin}}]{oriol2022characterizing}%
  \BibitemOpen
  \bibfield  {author} {\bibinfo {author} {\bibfnamefont {O.}~\bibnamefont
  {Rubies-Bigorda}}, \bibinfo {author} {\bibfnamefont {S.}~\bibnamefont
  {Ostermann}},\ and\ \bibinfo {author} {\bibfnamefont {S.~F.}\ \bibnamefont
  {Yelin}},\ }\href@noop {} {\bibinfo {title} {Characterizing superradiant
  dynamics in atomic arrays via a cumulant expansion approach}} (\bibinfo
  {year} {2022}),\ \Eprint {https://arxiv.org/abs/arXiv:2211.11895}
  {arXiv:2211.11895} \BibitemShut {NoStop}%
\end{thebibliography}%


\begin{thebibliography}{18}%
\makeatletter
\providecommand \@ifxundefined [1]{%
 \@ifx{#1\undefined}
}%
\providecommand \@ifnum [1]{%
 \ifnum #1\expandafter \@firstoftwo
 \else \expandafter \@secondoftwo
 \fi
}%
\providecommand \@ifx [1]{%
 \ifx #1\expandafter \@firstoftwo
 \else \expandafter \@secondoftwo
 \fi
}%
\providecommand \natexlab [1]{#1}%
\providecommand \enquote  [1]{``#1''}%
\providecommand \bibnamefont  [1]{#1}%
\providecommand \bibfnamefont [1]{#1}%
\providecommand \citenamefont [1]{#1}%
\providecommand \href@noop [0]{\@secondoftwo}%
\providecommand \href [0]{\begingroup \@sanitize@url \@href}%
\providecommand \@href[1]{\@@startlink{#1}\@@href}%
\providecommand \@@href[1]{\endgroup#1\@@endlink}%
\providecommand \@sanitize@url [0]{\catcode `\\12\catcode `\$12\catcode
  `\&12\catcode `\#12\catcode `\^12\catcode `\_12\catcode `\%12\relax}%
\providecommand \@@startlink[1]{}%
\providecommand \@@endlink[0]{}%
\providecommand \url  [0]{\begingroup\@sanitize@url \@url }%
\providecommand \@url [1]{\endgroup\@href {#1}{\urlprefix }}%
\providecommand \urlprefix  [0]{URL }%
\providecommand \Eprint [0]{\href }%
\providecommand \doibase [0]{http://dx.doi.org/}%
\providecommand \selectlanguage [0]{\@gobble}%
\providecommand \bibinfo  [0]{\@secondoftwo}%
\providecommand \bibfield  [0]{\@secondoftwo}%
\providecommand \translation [1]{[#1]}%
\providecommand \BibitemOpen [0]{}%
\providecommand \bibitemStop [0]{}%
\providecommand \bibitemNoStop [0]{.\EOS\space}%
\providecommand \EOS [0]{\spacefactor3000\relax}%
\providecommand \BibitemShut  [1]{\csname bibitem#1\endcsname}%
\let\auto@bib@innerbib\@empty
\bibitem [{\citenamefont {Masson}\ and\ \citenamefont
  {Asenjo-Garcia}(2022)}]{Masson2022universality}%
  \BibitemOpen
  \bibfield  {author} {\bibinfo {author} {\bibfnamefont {S.~J.}\ \bibnamefont
  {Masson}}\ and\ \bibinfo {author} {\bibfnamefont {A.}~\bibnamefont
  {Asenjo-Garcia}},\ }\href {\doibase 10.1038/s41467-022-29805-4} {\bibfield
  {journal} {\bibinfo  {journal} {Nat. Commun.}\ }\textbf {\bibinfo {volume}
  {13}},\ \bibinfo {pages} {2285} (\bibinfo {year} {2022})}\BibitemShut
  {NoStop}%
\bibitem [{\citenamefont {Robicheaux}(2021)}]{Robicheaux2021theoretical}%
  \BibitemOpen
  \bibfield  {author} {\bibinfo {author} {\bibfnamefont {F.}~\bibnamefont
  {Robicheaux}},\ }\href {\doibase 10.1103/PhysRevA.104.063706} {\bibfield
  {journal} {\bibinfo  {journal} {Phys. Rev. A}\ }\textbf {\bibinfo {volume}
  {104}},\ \bibinfo {pages} {063706} (\bibinfo {year} {2021})}\BibitemShut
  {NoStop}%
\bibitem [{\citenamefont {Sierra}\ \emph {et~al.}(2022)\citenamefont {Sierra},
  \citenamefont {Masson},\ and\ \citenamefont
  {Asenjo-Garcia}}]{Sierra2021dicke}%
  \BibitemOpen
  \bibfield  {author} {\bibinfo {author} {\bibfnamefont {E.}~\bibnamefont
  {Sierra}}, \bibinfo {author} {\bibfnamefont {S.~J.}\ \bibnamefont {Masson}},
  \ and\ \bibinfo {author} {\bibfnamefont {A.}~\bibnamefont {Asenjo-Garcia}},\
  }\href {\doibase 10.1103/PhysRevResearch.4.023207} {\bibfield  {journal}
  {\bibinfo  {journal} {Phys. Rev. Research}\ }\textbf {\bibinfo {volume}
  {4}},\ \bibinfo {pages} {023207} (\bibinfo {year} {2022})}\BibitemShut
  {NoStop}%
\bibitem [{\citenamefont {Douglas}\ \emph {et~al.}(2015)\citenamefont
  {Douglas}, \citenamefont {Habibian}, \citenamefont {Hung}, \citenamefont
  {Gorshkov}, \citenamefont {Kimble},\ and\ \citenamefont
  {Chang}}]{Douglas2015quantum}%
  \BibitemOpen
  \bibfield  {author} {\bibinfo {author} {\bibfnamefont {J.~S.}\ \bibnamefont
  {Douglas}}, \bibinfo {author} {\bibfnamefont {H.}~\bibnamefont {Habibian}},
  \bibinfo {author} {\bibfnamefont {C.-L.}\ \bibnamefont {Hung}}, \bibinfo
  {author} {\bibfnamefont {A.~V.}\ \bibnamefont {Gorshkov}}, \bibinfo {author}
  {\bibfnamefont {H.~J.}\ \bibnamefont {Kimble}}, \ and\ \bibinfo {author}
  {\bibfnamefont {D.~E.}\ \bibnamefont {Chang}},\ }\href {\doibase
  10.1038/nphoton.2015.57} {\bibfield  {journal} {\bibinfo  {journal} {Nat.
  Photonics}\ }\textbf {\bibinfo {volume} {9}},\ \bibinfo {pages} {326}
  (\bibinfo {year} {2015})}\BibitemShut {NoStop}%
\bibitem [{\citenamefont {Masson}\ \emph {et~al.}(2020)\citenamefont {Masson},
  \citenamefont {Ferrier-Barbut}, \citenamefont {Orozco}, \citenamefont
  {Browaeys},\ and\ \citenamefont {Asenjo-Garcia}}]{Masson2020many}%
  \BibitemOpen
  \bibfield  {author} {\bibinfo {author} {\bibfnamefont {S.~J.}\ \bibnamefont
  {Masson}}, \bibinfo {author} {\bibfnamefont {I.}~\bibnamefont
  {Ferrier-Barbut}}, \bibinfo {author} {\bibfnamefont {L.~A.}\ \bibnamefont
  {Orozco}}, \bibinfo {author} {\bibfnamefont {A.}~\bibnamefont {Browaeys}}, \
  and\ \bibinfo {author} {\bibfnamefont {A.}~\bibnamefont {Asenjo-Garcia}},\
  }\href {\doibase 10.1103/PhysRevLett.125.263601} {\bibfield  {journal}
  {\bibinfo  {journal} {Phys. Rev. Lett.}\ }\textbf {\bibinfo {volume} {125}},\
  \bibinfo {pages} {263601} (\bibinfo {year} {2020})}\BibitemShut {NoStop}%
\bibitem [{\citenamefont {Munro}\ \emph {et~al.}(2018)\citenamefont {Munro},
  \citenamefont {Asenjo-Garcia}, \citenamefont {Lin}, \citenamefont {Kwek},
  \citenamefont {Regal},\ and\ \citenamefont {Chang}}]{Munro2018population}%
  \BibitemOpen
  \bibfield  {author} {\bibinfo {author} {\bibfnamefont {E.}~\bibnamefont
  {Munro}}, \bibinfo {author} {\bibfnamefont {A.}~\bibnamefont
  {Asenjo-Garcia}}, \bibinfo {author} {\bibfnamefont {Y.}~\bibnamefont {Lin}},
  \bibinfo {author} {\bibfnamefont {L.~C.}\ \bibnamefont {Kwek}}, \bibinfo
  {author} {\bibfnamefont {C.~A.}\ \bibnamefont {Regal}}, \ and\ \bibinfo
  {author} {\bibfnamefont {D.~E.}\ \bibnamefont {Chang}},\ }\href {\doibase
  10.1103/PhysRevA.98.033815} {\bibfield  {journal} {\bibinfo  {journal} {Phys.
  Rev. A}\ }\textbf {\bibinfo {volume} {98}},\ \bibinfo {pages} {033815}
  (\bibinfo {year} {2018})}\BibitemShut {NoStop}%
\bibitem [{\citenamefont {Solano}\ \emph {et~al.}(2017)\citenamefont {Solano},
  \citenamefont {Barberis-Blostein}, \citenamefont {Fatemi}, \citenamefont
  {Orozco},\ and\ \citenamefont {Rolston}}]{Solano2017super}%
  \BibitemOpen
  \bibfield  {author} {\bibinfo {author} {\bibfnamefont {P.}~\bibnamefont
  {Solano}}, \bibinfo {author} {\bibfnamefont {P.}~\bibnamefont
  {Barberis-Blostein}}, \bibinfo {author} {\bibfnamefont {F.~K.}\ \bibnamefont
  {Fatemi}}, \bibinfo {author} {\bibfnamefont {L.~A.}\ \bibnamefont {Orozco}},
  \ and\ \bibinfo {author} {\bibfnamefont {S.~L.}\ \bibnamefont {Rolston}},\
  }\href {\doibase 10.1038/s41467-017-01994-3} {\bibfield  {journal} {\bibinfo
  {journal} {Nat. Commun.}\ }\textbf {\bibinfo {volume} {8}},\ \bibinfo {pages}
  {1857} (\bibinfo {year} {2017})}\BibitemShut {NoStop}%
\bibitem [{\citenamefont {Lodahl}\ \emph {et~al.}(2017)\citenamefont {Lodahl},
  \citenamefont {Mahmoodian}, \citenamefont {Stobbe}, \citenamefont
  {Rauschenbeutel}, \citenamefont {Schneeweiss}, \citenamefont {Volz},
  \citenamefont {Pichler},\ and\ \citenamefont {Zoller}}]{Lodahl2017chiral}%
  \BibitemOpen
  \bibfield  {author} {\bibinfo {author} {\bibfnamefont {P.}~\bibnamefont
  {Lodahl}}, \bibinfo {author} {\bibfnamefont {S.}~\bibnamefont {Mahmoodian}},
  \bibinfo {author} {\bibfnamefont {S.}~\bibnamefont {Stobbe}}, \bibinfo
  {author} {\bibfnamefont {A.}~\bibnamefont {Rauschenbeutel}}, \bibinfo
  {author} {\bibfnamefont {P.}~\bibnamefont {Schneeweiss}}, \bibinfo {author}
  {\bibfnamefont {J.}~\bibnamefont {Volz}}, \bibinfo {author} {\bibfnamefont
  {H.}~\bibnamefont {Pichler}}, \ and\ \bibinfo {author} {\bibfnamefont
  {P.}~\bibnamefont {Zoller}},\ }\href {\doibase 10.1038/nature21037}
  {\bibfield  {journal} {\bibinfo  {journal} {Nature}\ }\textbf {\bibinfo
  {volume} {541}},\ \bibinfo {pages} {473} (\bibinfo {year}
  {2017})}\BibitemShut {NoStop}%
\bibitem [{\citenamefont {Pichler}\ \emph {et~al.}(2015)\citenamefont
  {Pichler}, \citenamefont {Ramos}, \citenamefont {Daley},\ and\ \citenamefont
  {Zoller}}]{Pichler2015quantum}%
  \BibitemOpen
  \bibfield  {author} {\bibinfo {author} {\bibfnamefont {H.}~\bibnamefont
  {Pichler}}, \bibinfo {author} {\bibfnamefont {T.}~\bibnamefont {Ramos}},
  \bibinfo {author} {\bibfnamefont {A.~J.}\ \bibnamefont {Daley}}, \ and\
  \bibinfo {author} {\bibfnamefont {P.}~\bibnamefont {Zoller}},\ }\href
  {\doibase 10.1103/PhysRevA.91.042116} {\bibfield  {journal} {\bibinfo
  {journal} {Phys. Rev. A}\ }\textbf {\bibinfo {volume} {91}},\ \bibinfo
  {pages} {042116} (\bibinfo {year} {2015})}\BibitemShut {NoStop}%
\bibitem [{\citenamefont {Dicke}(1954)}]{Dicke1954coherence}%
  \BibitemOpen
  \bibfield  {author} {\bibinfo {author} {\bibfnamefont {R.~H.}\ \bibnamefont
  {Dicke}},\ }\href {\doibase 10.1103/PhysRev.93.99} {\bibfield  {journal}
  {\bibinfo  {journal} {Phys. Rev.}\ }\textbf {\bibinfo {volume} {93}},\
  \bibinfo {pages} {99} (\bibinfo {year} {1954})}\BibitemShut {NoStop}%
\bibitem [{\citenamefont {Gross}\ and\ \citenamefont
  {Haroche}(1982)}]{Gross1982superradiance}%
  \BibitemOpen
  \bibfield  {author} {\bibinfo {author} {\bibfnamefont {M.}~\bibnamefont
  {Gross}}\ and\ \bibinfo {author} {\bibfnamefont {S.}~\bibnamefont
  {Haroche}},\ }\href {\doibase https://doi.org/10.1016/0370-1573(82)90102-8}
  {\bibfield  {journal} {\bibinfo  {journal} {Phys. Rep.}\ }\textbf {\bibinfo
  {volume} {93}},\ \bibinfo {pages} {301} (\bibinfo {year} {1982})}\BibitemShut
  {NoStop}%
\bibitem [{\citenamefont {Mok}\ \emph {et~al.}(2020{\natexlab{a}})\citenamefont
  {Mok}, \citenamefont {Aghamalyan}, \citenamefont {You}, \citenamefont {Haug},
  \citenamefont {Zhang}, \citenamefont {Png},\ and\ \citenamefont
  {Kwek}}]{mok2020long}%
  \BibitemOpen
  \bibfield  {author} {\bibinfo {author} {\bibfnamefont {W.-K.}\ \bibnamefont
  {Mok}}, \bibinfo {author} {\bibfnamefont {D.}~\bibnamefont {Aghamalyan}},
  \bibinfo {author} {\bibfnamefont {J.-B.}\ \bibnamefont {You}}, \bibinfo
  {author} {\bibfnamefont {T.}~\bibnamefont {Haug}}, \bibinfo {author}
  {\bibfnamefont {W.}~\bibnamefont {Zhang}}, \bibinfo {author} {\bibfnamefont
  {C.~E.}\ \bibnamefont {Png}}, \ and\ \bibinfo {author} {\bibfnamefont
  {L.-C.}\ \bibnamefont {Kwek}},\ }\href {\doibase
  10.1103/PhysRevResearch.2.013369} {\bibfield  {journal} {\bibinfo  {journal}
  {Phys. Rev. Research}\ }\textbf {\bibinfo {volume} {2}},\ \bibinfo {pages}
  {013369} (\bibinfo {year} {2020}{\natexlab{a}})}\BibitemShut {NoStop}%
\bibitem [{\citenamefont {Mok}\ \emph {et~al.}(2020{\natexlab{b}})\citenamefont
  {Mok}, \citenamefont {You}, \citenamefont {Kwek},\ and\ \citenamefont
  {Aghamalyan}}]{mok2020microresonators}%
  \BibitemOpen
  \bibfield  {author} {\bibinfo {author} {\bibfnamefont {W.-K.}\ \bibnamefont
  {Mok}}, \bibinfo {author} {\bibfnamefont {J.-B.}\ \bibnamefont {You}},
  \bibinfo {author} {\bibfnamefont {L.-C.}\ \bibnamefont {Kwek}}, \ and\
  \bibinfo {author} {\bibfnamefont {D.}~\bibnamefont {Aghamalyan}},\ }\href
  {\doibase 10.1103/PhysRevA.101.053861} {\bibfield  {journal} {\bibinfo
  {journal} {Phys. Rev. A}\ }\textbf {\bibinfo {volume} {101}},\ \bibinfo
  {pages} {053861} (\bibinfo {year} {2020}{\natexlab{b}})}\BibitemShut
  {NoStop}%
\bibitem [{\citenamefont {Mahmoodian}\ \emph {et~al.}(2020)\citenamefont
  {Mahmoodian}, \citenamefont {Calaj\'o}, \citenamefont {Chang}, \citenamefont
  {Hammerer},\ and\ \citenamefont {S\o{}rensen}}]{mahmoodian2020dynamics}%
  \BibitemOpen
  \bibfield  {author} {\bibinfo {author} {\bibfnamefont {S.}~\bibnamefont
  {Mahmoodian}}, \bibinfo {author} {\bibfnamefont {G.}~\bibnamefont
  {Calaj\'o}}, \bibinfo {author} {\bibfnamefont {D.~E.}\ \bibnamefont {Chang}},
  \bibinfo {author} {\bibfnamefont {K.}~\bibnamefont {Hammerer}}, \ and\
  \bibinfo {author} {\bibfnamefont {A.~S.}\ \bibnamefont {S\o{}rensen}},\
  }\href {\doibase 10.1103/PhysRevX.10.031011} {\bibfield  {journal} {\bibinfo
  {journal} {Phys. Rev. X}\ }\textbf {\bibinfo {volume} {10}},\ \bibinfo
  {pages} {031011} (\bibinfo {year} {2020})}\BibitemShut {NoStop}%
\bibitem [{\citenamefont {Malz}\ \emph {et~al.}(2022)\citenamefont {Malz},
  \citenamefont {Trivedi},\ and\ \citenamefont {Cirac}}]{malz2022large}%
  \BibitemOpen
  \bibfield  {author} {\bibinfo {author} {\bibfnamefont {D.}~\bibnamefont
  {Malz}}, \bibinfo {author} {\bibfnamefont {R.}~\bibnamefont {Trivedi}}, \
  and\ \bibinfo {author} {\bibfnamefont {J.~I.}\ \bibnamefont {Cirac}},\ }\href
  {\doibase 10.1103/PhysRevA.106.013716} {\bibfield  {journal} {\bibinfo
  {journal} {Phys. Rev. A}\ }\textbf {\bibinfo {volume} {106}},\ \bibinfo
  {pages} {013716} (\bibinfo {year} {2022})}\BibitemShut {NoStop}%
\bibitem [{\citenamefont {Cardenas-Lopez}\ \emph {et~al.}(2022)\citenamefont
  {Cardenas-Lopez}, \citenamefont {Masson}, \citenamefont {Zager},\ and\
  \citenamefont {Asenjo-Garcia}}]{silvia2022many}%
  \BibitemOpen
  \bibfield  {author} {\bibinfo {author} {\bibfnamefont {S.}~\bibnamefont
  {Cardenas-Lopez}}, \bibinfo {author} {\bibfnamefont {S.~J.}\ \bibnamefont
  {Masson}}, \bibinfo {author} {\bibfnamefont {Z.}~\bibnamefont {Zager}}, \
  and\ \bibinfo {author} {\bibfnamefont {A.}~\bibnamefont {Asenjo-Garcia}},\
  }\href@noop {} {\enquote {\bibinfo {title} {Many-body superradiance and
  dynamical symmetry breaking in waveguide qed},}\ } (\bibinfo {year} {2022}),\
  \Eprint {http://arxiv.org/abs/arXiv:2209.12970} {arXiv:2209.12970}
  \BibitemShut {NoStop}%
\bibitem [{\citenamefont {Robicheaux}\ and\ \citenamefont
  {Suresh}(2021)}]{robicheaux2021beyond}%
  \BibitemOpen
  \bibfield  {author} {\bibinfo {author} {\bibfnamefont {F.}~\bibnamefont
  {Robicheaux}}\ and\ \bibinfo {author} {\bibfnamefont {D.~A.}\ \bibnamefont
  {Suresh}},\ }\href {\doibase 10.1103/PhysRevA.104.023702} {\bibfield
  {journal} {\bibinfo  {journal} {Phys. Rev. A}\ }\textbf {\bibinfo {volume}
  {104}},\ \bibinfo {pages} {023702} (\bibinfo {year} {2021})}\BibitemShut
  {NoStop}%
\bibitem [{\citenamefont {Rubies-Bigorda}\ \emph {et~al.}(2022)\citenamefont
  {Rubies-Bigorda}, \citenamefont {Ostermann},\ and\ \citenamefont
  {Yelin}}]{oriol2022characterizing}%
  \BibitemOpen
  \bibfield  {author} {\bibinfo {author} {\bibfnamefont {O.}~\bibnamefont
  {Rubies-Bigorda}}, \bibinfo {author} {\bibfnamefont {S.}~\bibnamefont
  {Ostermann}}, \ and\ \bibinfo {author} {\bibfnamefont {S.~F.}\ \bibnamefont
  {Yelin}},\ }\href@noop {} {\enquote {\bibinfo {title} {Characterizing
  superradiant dynamics in atomic arrays via a cumulant expansion approach},}\
  } (\bibinfo {year} {2022}),\ \Eprint {http://arxiv.org/abs/arXiv:2211.11895}
  {arXiv:2211.11895} \BibitemShut {NoStop}%
\end{thebibliography}%
\end{document}


\title{Supplementary Material: Dicke superradiance requires interactions beyond nearest-neighbors} 

\author{Wai-Keong Mok}
\affiliation{Centre for Quantum Technologies, National University of Singapore, 3 Science Drive 2, Singapore 117543}
\affiliation{California Institute of Technology, Pasadena, CA 91125, USA}
\author{Ana Asenjo-Garcia}
\affiliation{Department of Physics, Columbia University, New York, New York 10027, USA}
\author{Tze Chien Sum}
\affiliation{Division of Physics and Applied Physics, School of Physical and Mathematical Sciences, Nanyang Technological University,
Singapore 637371}
\author{Leong-Chuan Kwek}
\affiliation{Centre for Quantum Technologies, National University of Singapore, 3 Science Drive 2, Singapore 117543}
\affiliation{MajuLab, CNRS-UNS-NUS-NTU International Joint Research Unit, Singapore UMI 3654, Singapore}
\affiliation{National Institute of Education, Nanyang Technological University, Singapore 637616, Singapore}
\affiliation{Quantum Science and Engineering Centre (QSec), Nanyang Technological University, Singapore}

\maketitle
\tableofcontents
\section{Superradiance conditions}
\label{sec:superradiance}
\subsection{Second-order correlation function $g^{(2)}(0)$}
\label{sec:g2}
The collective jump operators can be constructed as
\begin{equation}
    \hat{c}_\nu = \sum_{j=1}^{N} v_{j \nu} \sigma_j^-
\label{eq:jumpops}
\end{equation}
where $\{(v_{1 \nu} \ldots v_{N \nu})^T\}$ are the normalized eigenvectors of the decoherence matrix $\mathbf{\Gamma}$, which are mutually orthogonal. In this new basis, the master equation can be rewritten in the standard Lindblad form as
\begin{equation}
\dot{\rho}= -i\left[ \sum_{i,j = 1}^{N} H_{ij} \sigma_i^+ \sigma_j^- , \rho \right] + \sum_{\nu=1}^N \Gamma_\nu \mathcal{D}[\hat{c}_\nu]\rho    
\label{eq:ME_diagonalized}
\end{equation}
where we have defined the Lindblad superoperator as $\mathcal{D}[\hat{O}]\rho = \hat{O}\rho\hat{O}^\dag - \{\hat{O}^\dag \hat{O},\rho\}/2$.
In Ref.~\cite{Masson2022universality}, Masson and Asenjo-Garcia proposed to use the zero-delay second-order correlation function $g^{(2)}(0)$, defined here as
\begin{equation}
    g^{(2)}(0) = \frac{\sum_{\mu,\nu=1}^N \Gamma_\mu \Gamma_\nu \braket{\hat{c}_\mu^\dag \hat{c}_\nu^\dag \hat{c}_\nu \hat{c}_\mu}}{\left( \sum_{\mu=1}^N \Gamma_\mu \braket{\hat{c}_\mu^\dag \hat{c}_\mu} \right)^2},
\label{eq:g20}
\end{equation}
as the minimal superradiance condition, where $g^{(2)}(0) > 1$ for the fully-excited initial state is the signature for superradiant burst. Intuitively, in a superradiant burst, the emission of the first photon enhances the emission of the second photon, which leads to photon bunching, i.e., $g^{(2)}(0) > 1$. The same effect can also be interpreted as the synchronization of the emitters at short times, which manifests as nonzero correlations $\braket{\sigma_j^+ \sigma_k^-}$ between them.

For pedagogical reasons, we derive an
explicit formula that relates $\dot{R}(0)$ to $g^{(2)}(0)$. Using the master equation \eqref{eq:ME_diagonalized}, we can calculate $\dot{R}(0)$ as
\begin{equation}
\begin{split}
    \dot{R}(0) &= \sum_{\mu,\nu=1}^N \Gamma_\mu \Gamma_\nu \left( \braket{\hat{c}_\mu^\dag \hat{c}_\nu^\dag \hat{c}_\nu \hat{c_\mu}} - \braket{\hat{c}_\mu^\dag \hat{c}_\mu \hat{c}_\nu^\dag \hat{c}_\nu} \right) + i \sum_{\mu=1}^N \Gamma_\mu \braket{[H, \hat{c}_\mu^\dag \hat{c}_\mu]},
\end{split}
\end{equation}
where $H$ is the Hamiltonian in Eq.~\eqref{eq:ME_diagonalized}. Using the definition in Eq.~\eqref{eq:g20}, the first sum is simply $R^2(0) g^{(2)}(0)$. The second and third sums do not give simple results for arbitrary initial states. However, for the fully-excited initial state $\ket{e}^{\otimes N}$, it can be shown that the second sum factorizes exactly to
\begin{equation}
\sum_{\mu,\nu=1}^N \Gamma_\mu \Gamma_\nu \braket{\hat{c}_\mu^\dag \hat{c}_\mu \hat{c}_\nu^\dag \hat{c}_\nu} = \left(\sum_{\mu=1} \Gamma_\mu \braket{\hat{c}_\mu^\dag \hat{c}_\mu}\right)^2 = R^2 (0)
\end{equation}
and the third sum vanishes for any arbitrary $H_{ij}$. We can justify both of these claims by direct calculations:
\begin{equation}
\begin{split}
    \braket{\hat{c}_\mu^\dag \hat{c}_\mu \hat{c}_\nu^\dag \hat{c}_\nu} &= \sum_{m,n,p,q=1}^{N} v_{m\mu}^* v_{n\mu} v_{p\nu}^* v_{q\nu} \braket{e| \sigma_m^+ \sigma_n^- \sigma_p^+ \sigma_q^- |e} = \sum_{m,n,p,q=1}^{N} v_{m\mu}^* v_{n\mu} v_{p\nu}^* v_{q\nu} \delta_{mn} \delta_{pq} \braket{e| \sigma_m^+ \sigma_n^- \sigma_p^+ \sigma_q^- |e} \\&= \sum_{m,p=1}^{N} v_{m\mu}^* v_{m\mu} v_{p\nu}^* v_{p\nu} \braket{e| \sigma_m^+ \sigma_m^- \sigma_p^+ \sigma_p^- |e} = \braket{\hat{c}_\mu^\dag \hat{c}_\mu} \braket{\hat{c}_\nu^\dag \hat{c}_\nu}.
\end{split}
\end{equation}
Note that this is not true in general for arbitrary product states. 

Next, note that the Hamiltonian contributes the terms like $i\braket{[H,\hat{c}_\mu^\dag \hat{c}_\mu]}$ in $\dot{R}(0)$. For a generic Hamiltonian, $H = \sum_{i,j} H_{ij} \sigma_i^+ \sigma_j^-$, expanding the commutator yields
\begin{equation}
\begin{split}
    \braket{[H,\hat{c}_\mu^\dag \hat{c}_\mu]} &= \sum_{i,j,k,l=1}^{N} H_{ij} v_{k\mu}^* v_{l\mu} (\braket{e| \sigma_i^+ \sigma_j^- \sigma_k^+ \sigma_l^- |e} - \braket{e| \sigma_k^+ \sigma_l^- \sigma_i^+ \sigma_j^- |e}) =  \sum_{i,j,k,l=1}^{N} H_{ij} v_{k\mu}^* v_{l\mu} (\delta_{ij}\delta_{kl} - \delta_{kl}\delta_{ij}) = 0, 
\end{split}
\end{equation}
which shows that $H$ does not affect the presence or absence of a superradiant burst in the system. This provides a simple explanation for the observation in Fig. 3(a) of Ref.~\cite{Masson2022universality} which shows no significant effect of the Hamiltonian on superradiance. Thus, we obtain the relation
\begin{equation}
    \dot{R}(0) = R^2(0) \left[g^{(2)}(0) - 1 \right],
\label{eq:rdot0}
\end{equation}
which reveals the close connection between superradiance and photon bunching, thereby justifying the results in Ref.~\cite{Masson2022universality}. A more detailed analysis can be found in a recent work by F. Robicheaux~\cite{Robicheaux2021theoretical}, which includes additional considerations beyond the scope of this work such as directional emission as well as arbitrary product initial states.

\subsection{Efficient calculation of $g^{(2)}(0)$}
Evaluating Eq.~\eqref{eq:g20} using the initial state $\ket{e}^{\otimes N}$ yields the a simple expression~\cite{Masson2022universality, Robicheaux2021theoretical}
\begin{equation}
    g^{(2)}(0) = 1 - \frac{2}{N} + \frac{ \sum_{\nu=1}^N \Gamma_\nu^2}{\left( \sum_{\nu=1}^N \Gamma_\nu\right)^2} = 1 - \frac{2}{N} + \frac{\text{Tr} \, (\mathbf{\Gamma}^2)}{(\text{Tr} \, (\mathbf{\Gamma}))^2},
\label{eq:g2simple_supp}
\end{equation}
written purely in terms of the eigenvalues of the decoherence matrix $\mathbf{\Gamma}$. Diagonalization can be avoided by taking traces of $\mathbf{\Gamma}$ and $\mathbf{\Gamma}^2$. Further reduction in complexity can be achieved by harnessing the Hermiticity of $\Gamma$ which yields $\text{Tr} \, (\mathbf{\Gamma}^2) = \lVert \mathbf{\Gamma} \rVert_F^2$~\cite{Sierra2021dicke}. For ordered arrays, the $g^{(2)}(0)$ function can be efficiently computed in $\mathcal{O}(N)$ steps due to the periodicity of the array~\cite{Robicheaux2021theoretical,Sierra2021dicke}.

Although the diagonalization of $\mathbf{\Gamma}$ is not necessary to compute $g^{(2)}(0)$, there are several reasons why it is still useful. First, the eigenvectors allow us to construct the jump operators $\hat{c}_\nu$ in Eq.~\eqref{eq:jumpops}, which provide insights about the dominant decay channels in the system. Moreover, since the eigenvalues $\Gamma_\nu$ must be non-negative, calculating the eigenvalues allow us to determine the physically valid regime for the generic $\mathbf{\Gamma}$ that we use throughout the paper.

\section{Explicit calculations for nearest-neighbor emitter arrays}
\label{app:NN_d_array}
\subsection{1D chain}
\label{sec:nogo_1D}
To begin, let us consider a chain of $N$ dissipatively-coupled emitters with open boundary conditions, initialized in the fully-excited state. We neglect the Hamiltonian interactions between the emitters since they do not affect the existence of superradiant burst as explained in Sec.~\ref{sec:g2}. Furthermore, we restrict to the case of NN interactions, leading to the master equation with $H_{ij} = 0$ and $\gamma_{ij} = \delta_{ij} + \gamma (\delta_{i,j+1}+\delta_{i,j-1})$ for arbitrary $0 \leq \gamma \leq 1$. Note that without loss of generality, we have assumed $\gamma_{ij}$ to be real and positive, since $g^{(2)}(0)$ is invariant under the transformation $\gamma \to \gamma e^{i\phi}$ for any $\phi \in \mathbb{R}$. Physically, if $\gamma = e^{-\kappa d}$ where $\kappa > 0$ is the attenuation coefficient and $d$ is the constant separation between each emitter, then this model describes the exponentially-decaying interactions which can arise from evanescent coupling in photonic crystals \cite{Douglas2015quantum}.

The decoherence matrix $\mathbf{\Gamma}$ in this case is a tridiagonal Toeplitz matrix with all diagonal elements $1$ and all super-/sub-diagonal elements $\gamma$, which can be exactly diagonalized to yield the eigenvalues
\begin{equation}
    \Gamma_\nu = 1 + 2 \gamma \cos\left(\frac{\nu \pi}{N+1} \right), \quad \nu = 1,\ldots,N.
\label{eq:eigval_1dNNchain}
\end{equation}
The $g^{(2)}(0)$ function thus reads
\begin{equation}
    g^{(2)}(0) = 1 - \frac{1}{N} + \frac{2(N-1)}{N^2} \gamma^2.
\label{eq:g2_1DNNchain}
\end{equation}
Let us denote $\gamma_s$ as the critical $\gamma$ for a superradiant transition, where $g^{(2)}(0) = 1$.
Applying the condition $g^{(2)}(0) > 1$ directly gives the superradiant condition $\gamma > \gamma_s$, where
\begin{equation}
    \gamma_s = \frac{N}{\sqrt{2N(N-1)}}.
\end{equation}
However, enforcing that $\Gamma_\nu \geq 0$ for all $\nu = 1, \ldots N$ gives the physically valid regime as $\gamma \leq \gamma_p$, where
\begin{equation}
    \gamma_p = \frac{1}{2} \sec\left( \frac{\pi}{N+1} \right).
\end{equation}
It is easy to show that for a positive integer $N$, $\gamma_p < \gamma_s$ for all $N > 2$, with equality at exactly $N = 2$. This implies that no superradiance is possible within the physically valid regime, except in the trivial case of $N=2$ where the system reduces to the original Dicke model for two emitters with the collective jump operator $\hat{c}_1 = \sigma_1^- + \sigma_2^-$. 

This example of a 1D NN model demonstrates a potential pitfall of naively applying the $g^{(2)}(0) > 1$ condition for an arbitrary decoherence matrix $\mathbf{\Gamma}$, where the predicted superradiant regime falls outside of the physically valid regime and is thus not possible. This is not a concern for previous works~\cite{Masson2022universality,Robicheaux2021theoretical,Sierra2021dicke} where the electromagnetic interactions are already physically valid by construction. While obtaining $g^{(2)}(0)$ can be done in $\mathcal{O}(N)$ steps for ordered arrays, verifying that $\mathbf{\Gamma}$ is positive-semidefinite typically requires $\mathcal{O}(N^3)$ steps, such as by performing a Cholesky decomposition or by finding the smallest eigenvalue of $\mathbf{\Gamma}$.

\subsection{Generalization to non-identical nearest-neighbor coupling}
In the main text, we have assumed that all the nearest-neighbor coupling are the same, $\gamma_{ij} = \gamma$ for $|i-j|=1$. Now, we can lift the assumption and consider a more general case where the nearest-neighbor coupling in a 1D chain are not the same, so
\begin{equation}
    \gamma_{ij} = \begin{cases} 1, \quad &i=j \\ \gamma_{\text{min}(i,j)}, \quad &|i-j|=1 \\ 0, \quad &\text{else}. \end{cases}
\end{equation}
Let us also define $A = \mathbf{\Gamma} - I_{N}$ which has zero diagonal elements. The superradiance condition $g^{(2)}(0) > 1$ becomes
\begin{equation}
    ||A||_F > \sqrt{N}
\end{equation}
where $||A||_F$ is the Frobenius norm. Recall that for a physically valid model, the eigenvalues of $\mathbf{\Gamma}$ must be non-negative. Moreover, since $A$ is a symmetric tridiagonal matrix with zero on the main diagonal, the spectrum of $A$ is symmetrical with respect to zero. Hence, the spectral radius $\rho(A) = -\lambda_{\text{min}}(A) = \lambda_{\text{max}}(A) = \sigma_{\text{max}}(A) = ||A||_2$, where $\lambda_{\text{max}}(A)$ and $\sigma_{\text{max}}(A)$ are the maximum eigenvalue and singular value of $A$ respectively. For $N$ even, we have $\text{rank}(A) = N$ while for $N$ odd we have $\text{rank}(A) = N-1$ (due to 1 zero eigenvalue). Using the inequality $||A||_F \leq \sqrt{\text{rank}(A)} ||A||_2$, we obtain
\begin{equation}
    \lambda_{\text{min}}(\mathbf{\Gamma}) \geq 0 \implies ||A||_2 = \sigma_{\text{max}}(A) \leq 1 \implies ||A||_F \leq \begin{cases} \sqrt{N}, \quad &N \, \text{even} \\ \sqrt{N-1}, \quad &N \, \text{odd} \end{cases}
\end{equation}
which is a necessary condition for physical validity. Comparing with the superradiance condition above, we once show that the superradiance regime is unphysical.

\subsection{D-dimensional hypercube array with nearest-neighbor interactions}
We can generalize the above calculations by considering an arbitrary array dimensionality $D$. The number of pairwise interactions is just $D n^{D-1} (n-1)$ where $N = n^D$, which means that the number of nonzero off-diagonal elements $\gamma$ is simply twice of that: $2D n^{D-1} (n-1)$. Thus,
\begin{equation}
\text{Tr} \, (\mathbf{\Gamma}^2) = \sum_{i,j = 1}^N |\gamma_{ij}|^2 = N + 2D n^{D-1} (n-1) \gamma^2 = N + 2D(N - N^{1-1/D}) \gamma^2,
\end{equation}
which leads to
\begin{equation}
    g^{(2)}(0) = 1 - \frac{2}{N} + \frac{1}{N^2} (N + 2D n^{D-1} (n-1) \gamma^2) = 1 - \frac{1}{N} + \frac{2D(N - N^{1-1/D}) \gamma^2}{N^2} = 1 + \frac{2D\gamma^2 - 1}{N} - \frac{2D \gamma^2}{N^{1+1/D}}.\end{equation}
In the infinite-dimensional case $D \to \infty$ for a fixed $N$, $g^{(2)}(0) \to 1 - 1/N < 1$ for $N > 1$ and the emitters behave independently. The $D$-dimensional hypercube configuration of emitters with nearest-neighbor coupling can be mapped onto a generalized grid graph $G_{n^D}$ with adjacency matrix $\mathbf{A}$. The decoherence matrix is simply $\mathbf{\Gamma} = \mathbf{I}_N + \gamma \mathbf{A}$. $G_{n^D}$ is the Cartesian product of $D$ path graphs of length-$n$ $P_n$. The eigenvalues of the adjacency matrix of $P_n$ are $2 \cos{\frac{j\pi}{n+1}}$, $j = 1, \ldots, n$. 

Hence, the $n^D$ eigenvalues of $\mathbf{A}$ are given by the sum
\begin{equation}
    a_{\{j_\alpha \}} = 2 \sum_{\alpha=1}^{D} \cos \left( \frac{j_\alpha \pi}{n+1} \right), \quad j_\alpha = 1, \ldots, n.
\end{equation}
The minimum value occurs when all $j_\alpha = n$, such that
\begin{equation}
    \min_{\{j_\alpha\}} a_{\{j_\alpha \}} = 2D \cos \left( \frac{n \pi}{n+1} \right) = -2D \cos \left( \frac{\pi}{n+1} \right).
\end{equation}
From this, we obtain the minimum eigenvalue of $\mathbf{\Gamma}$ as
\begin{equation}
    \Gamma_{\text{min}} = 1 - 2D\gamma  \cos \left( \frac{\pi}{n+1} \right) = 1 - 2D\gamma  \cos \left( \frac{\pi}{N^{1/D}+1} \right).
\end{equation}
For the smallest non-trivial array, we have $n = 2$ with $\Gamma_{\text{min}} = 1 - D \gamma$. For an infinite-array, we have $\Gamma_{\text{min}} = 1 - 2D \gamma$. In both cases, the coefficient of $\gamma$ is simply the coordination number of each emitter in the array (ignoring boundaries when $N \to \infty$).

\subsection{Generalization to unequal edge lengths}
\label{app:hyperrectangle}
The generalization to the `hyperrectangle' configuration where $n$ can be different for each dimension is straightforward. The grid graph for $N = \prod_j n_j$ emitters is now $G_N = P_{n_1} \Osq \cdots \Osq P_{n_D}$, leading to
\begin{equation}
    \Gamma_{\text{min}} = 1 - 2 \gamma \sum_{j=1}^{D} \cos \left( \frac{\pi}{n_j+1} \right)
\end{equation}
and hence
\begin{equation}
    \gamma_p = \left[ 2 \sum_{j=1}^D \cos \left( \frac{\pi}{n_j+1} \right) \right]^{-1}.
\end{equation}
In this generalized array, there are $2N (D - \sum_j n_j^{-1})$ nearest-neighbor interactions which results in
\begin{equation}
    g^{(2)}(0) = 1 + \frac{2D\gamma^2-1}{N} - \frac{2\gamma^2}{N} \sum_{j=1}^{D} \frac{1}{n_j}
\end{equation}
and the superradiant transition at
\begin{equation}
    \gamma_s = \left[ 2 \left( D - \sum_{j=1}^D \frac{1}{j} \right) \right]^{-1/2} = \left[ 2  \sum_{j=1}^D \left(1 - \frac{1}{n_j} \right) \right]^{-1/2}.
\end{equation}
Since $\cos(\pi/(n_j+1)) > 1-1/n_j$ for $n_j > 2$ (with equality at $n_j = 2$), we conclude that $\gamma_p \leq \gamma_s^2$ and therefore no superradiant burst can be observed.

\subsection{Generalization to product initial states}
So far, we have considered the fully-excited initial state $\ket{e}^{\otimes N}$. We can show that our main result of `no superradiance for NN interactions' is valid even for product initial states $\bigotimes_n (\cos{\theta/2} \ket{g_n} + e^{i\phi} \sin{\theta/2} \ket{e_n})$. 

Let us also consider the extreme case of the all-to-all Hamiltonian $H = J \sum_{m,n} \sigma_m^+ \sigma_n^-$. For the dissipative coupling, we choose the phase convention $\gamma_{mn} \geq 0$, $\gamma_{mn} = \gamma_{nm}$. The time-derivative of the emission rate at time $t=0$ is given by~\cite{Robicheaux2021theoretical}:
\begin{equation}
    \dot{R}(0) = -N \sin^2 \frac{\theta}{2} - \frac{1}{2} \sin^2 \theta \sideset{}{'}\sum_{m,n} \gamma_{mn} + 2 \sin^2 \frac{\theta}{2} \left(\sin^2 \frac{\theta}{2} - \frac{1}{2} \right) \sideset{}{'}\sum_{m,n} \gamma_{mn}^2 + \frac{1}{4}\sin^2 \theta \left(\sin^2 \frac{\theta}{2} - \frac{1}{2} \right) \sideset{}{'}\sum_{l,m,n} \gamma_{mn} (\gamma_{ml} + \gamma_{nl}),
\end{equation}
where the notations $\sideset{}{'}\sum_{m,n}$ and $\sideset{}{'}\sum_{l,m,n}$ indicate that the dummy indices must all be different. Notice that $\dot{R}(0)$ is independent of $J$, which means that the Hamiltonian does not contribute to the superradiance burst (at initial time). This is also valid even if we use a more general spin-spin Hamiltonian. As a check, we can set $\theta = \pi$ to recover the result for the fully-excited initial state: $\dot{R}(0) = -N + \sideset{}{'}\sum_{m,n} \gamma_{mn}^2$. 

Now, we consider the nearest-neighbor dissipative interactions $\gamma_{\braket{m,n}} = \gamma \in [0,1/(2D)]$. This simplifies the expression to give
\begin{equation}
    \frac{\dot{R}(0)}{N} = \frac{1}{8} [-4+4D\gamma(-3+4\gamma)+(4+D\gamma(8-(15+2D)\gamma))\cos \theta + D\gamma(4\cos 2\theta + (2D-1)\gamma \cos 3\theta)].
\end{equation}
Choosing $\gamma = 1/(2D)$, we have
\begin{equation}
    \frac{\dot{R}(0)}{N} = \frac{\sin^2 \frac{\theta}{2}}{8D}[9-26D-2(6D-1)\cos \theta - (2D-1)\cos 2\theta] \leq 0, 
\end{equation}
independent of $\phi$ and $J$. Since $D \geq 1$, we have $\dot{R}(0) < 0$ except for $\dot{R}(0) = 0$ at $\theta = 0$ (corresponding to all emitters in the ground state).

\section{Derivations of $g^{(2)}(0)$ for various models}
\label{app:g2}

Here, we derive explicitly the various $g^{(2)}(0)$ expressions used throughout the paper. From the main text, the only calculation needed is $\text{Tr} \, (\mathbf{\Gamma}^2)$ which is equal to the Frobenium norm $\lVert \mathbf{\Gamma} \rVert_F^2 = \sum_{i,j} |\gamma_{ij}|^2$ using the Hermiticity of $\mathbf{\Gamma}$. Without loss of generality, we consider $\gamma_{ij}$ to be real and positive, and also rescale $\mathbf{\Gamma}$ appropriately such that $\text{Tr} \, (\mathbf{\Gamma}) = N$. Since we assume that the atoms decay identically, this amounts to setting the diagonal elements to $\gamma_{ii} = 1$. 

\subsection{1D chain with exponentially-decaying interactions}
\label{sec:1D_chain}
For open boundary conditions, we label the emitters sequentially along the chain by the index $i$, starting from one end of the chain. Without loss of generality, we align the 1D chain to the $x$-axis, giving the spatial coordinate $x_i = i d$ for the $i^{\text{th}}$ emitter. The matrix elements of $\mathbf{\Gamma}$ are then given by $\gamma_{ij} = e^{-\kappa d|i-j|} \equiv \gamma^{|i-j|}$. In the short-range limit, where we neglect terms like $\gamma^2$ or smaller, the results agree with the nearest-neighbor model up to $\mathcal{O}(\gamma)$. 

Without diagonalizing $\mathbf{\Gamma}$ analytically, we can check that it has a Cholesky decomposition for all $0 \leq \gamma \leq 1$ which asserts that $\mathbf{\Gamma}$ is positive semidefinite and thus physically valid for any $d \geq 0$. To calculate the superradiant transition, we first work out the $g^{(2)}(0)$ function. Starting from $\gamma_{ij} = e^{-\kappa d|i-j|} \equiv \gamma^{|i-j|}$, we have $2(N-m)$ off-diagonal matrix elements with the value $\gamma^{m}$, $m = 1,\ldots, N-1$, and $N$ diagonal elements with the value $1$. Thus,
\begin{equation}
\text{Tr} \, (\mathbf{\Gamma}^2) = \sum_{i,j = 1}^N |\gamma_{ij}|^2 = N + 2 \sum_{m=1}^{N-1} (N-m) \gamma^{2m} = N - \frac{2\gamma^2(1-\gamma^{2N}-N(1-\gamma^2))}{(1-\gamma^2)^2}
\end{equation}
and
\begin{equation}
    g^{(2)}(0) = 1 - \frac{2}{N} + \frac{1}{N^2} \left(N - \frac{2\gamma^2(1-\gamma^{2N}-N(1-\gamma^2))}{(1-\gamma^2)^2} \right) = 1 - \frac{1}{N} - \frac{2\gamma^2(1-\gamma^{2N}-N(1-\gamma^2))}{N^2(1-\gamma^2)^2}. 
\end{equation}

For a fixed $N$, we can obtain the superradiant transition point $\gamma_s$ numerically by solving $g^{(2)}(0) = 1$. We can proceed further analytically by considering the regime where $N \gg 2/\kappa d$ such that $\gamma^{2N} = e^{-2N\kappa d} \ll 1$. This gives the asymptotic expansion
\begin{equation}
    g^{(2)}(0) = 1 + \frac{3\gamma^2-1}{N(1-\gamma^2)} + \mathcal{O}\left(\frac{1}{N^2}\right),
\label{eq:g2_1Dchain_approx}
\end{equation}
which leads to $\gamma_s \sim 1/\sqrt{3}$. In terms of $\kappa d$, this means that superradiance occurs when
\begin{equation}
    \kappa d < \frac{1}{2} \ln 3 \approx 0.549.
\end{equation}
Interestingly, the superradiant transition point is approximately independent of $N$ for sufficient large $N$, which is fully consistent with the numerical simulations in Refs.~\cite{Masson2022universality,Robicheaux2021theoretical,Masson2020many,Sierra2021dicke} which used a more realistic interaction model. Thus, our result provides a good qualitative explanation for the previous observations on superradiant burst in 1D arrays despite its simplicity.

At first glance, by comparing the large-$N$ expressions for $g^{(2)}(0)$ in Eq.~\eqref{eq:g2_1Dchain_approx} with Eq.~\eqref{eq:g2_1DNNchain}, there appears to be an inconsistency. However, this can be easily resolved by recalling that the two models only agree up to $\mathcal{O}(\gamma)$, while the discrepancy only arises in the $\mathcal{O}(\gamma^2)$ terms. It is worth emphasizing that the nearest-neighbor interaction model in Sec.~\ref{sec:nogo_1D} and the NN interaction model here are fundamentally different: With nearest neighbor interactions, physical constraints require $\gamma < \gamma_p \sim 1/2$. On the other hand, adding long-range interactions ensure that the model is physically valid for all $0 \leq \gamma \leq 1$, which allows for superradiance.

\subsection{1D ring with exponentially-decaying interactions}
Next, we consider the case of periodic boundary conditions where the emitters are arranged in a ring configuration with a constant separation $d$. For simplicity, we consider the case where $N$ is odd. The decoherence matrix $\mathbf{\Gamma}$ for the 1D ring turns out to be a circulant matrix with the first column given by $(1,\gamma, \ldots, \gamma^{(N-1)/2}, \gamma^{(N-1)/2}, \ldots, \gamma)^T$, or $\gamma_{i1} = \max\{ \gamma^{i-1}, \gamma^{N-i+1}\}$. The subsequent columns are simply cyclic permutations of the first column. Thus,
\begin{equation}
\text{Tr} \, (\mathbf{\Gamma}^2) = \sum_{i,j = 1}^N |\gamma_{ij}|^2 = N (1 + \gamma^2 + \ldots + \gamma^{(N-1)} + \gamma^{(N-1)} + \ldots + \gamma^2 ) = N + 2N \sum_{m=1}^{(N-1)/2} \gamma^{2m} = N + 2 N \frac{\gamma^2 (1-\gamma^{N-1})}{1-\gamma^2}
\end{equation}
and we obtain
\begin{equation}
\begin{split}
    g^{(2)}(0) &= 1 - \frac{1}{N} + \frac{2u^2(1-\gamma^{N-1})}{N(1-\gamma^2)} = 1 + \frac{3\gamma^2 - 1}{N(1-\gamma^2)} + \mathcal{O}\left(\frac{1}{N^2}\right)
\end{split}
\label{eq:g2_1Dring}
\end{equation}
for $N \gg 1/\kappa d$. Note that for sufficiently large $N$, the superradiant transition $\gamma_s = 1/\sqrt{3}$ is the same for both the chain and ring configurations, which physically means that the boundary contributions become negligible. This again provides a simple explanation for the observations in Ref.~\cite{Masson2022universality}, where the significant discrepancy between the 1D chain and ring only occurs for small $N < 10$.

Alternatively, we can exploit the nice properties of circulant matrices to diagonalize $\mathbf{\Gamma}$ exactly, which yields the eigenvalues
\begin{equation}
\begin{split}
    \Gamma_j &= (1-\gamma)\frac{1+\gamma-2\gamma^{(N+1)/2} \cos(j\pi) \cos(j\pi/N)}{1+\gamma^2 - 2 \gamma \cos(2j\pi/N)} \approx \frac{1-\gamma^2}{1+\gamma^2 - 2 \gamma \cos(2j\pi/N)}, \quad j=0,\ldots,N-1
\end{split}
\label{eq:eigvals_1Dring}
\end{equation}
for large $N$. The sum $\sum_j \Gamma_j^2$ can be approximated by integral
\begin{equation}
    \sum_{j=0}^{N-1} \Gamma_j^2 \approx \frac{N}{2\pi} \int_{0}^{2\pi} d\theta \, \left( \frac{1-\gamma^2}{1+\gamma^2 - 2\gamma\cos\theta} \right)^2 = \frac{N(1+\gamma^2)}{(1-\gamma^2)},
\end{equation}
which gives the same limiting value for $g^{(2)}(0)$ as Eqs.~\eqref{eq:g2_1Dchain_approx} and ~\eqref{eq:g2_1Dring}. From the exact eigenvalues in Eq.~\eqref{eq:eigvals_1Dring} we can argue that $\mathbf{\Gamma}$ is positive semidefinite for all $0 \leq \gamma \leq 1$: It can be easily verified that $1 + \gamma \geq 2\gamma^{(N+1)/2}$ for $N > 1$, with equality achieved at exactly $\gamma = 1$. This implies that the numerator of $\Gamma_j$ is positive. From the law of cosines, the denominator represents geometrically the length-squared of the triangular side opposite the angle $2j\pi/N$, subtended by the other two sides of lengths $1$ and $\gamma$, which is also positive. Hence, the superradiant regime $\gamma > \gamma_s$ is physically valid. 

\subsection{D-dimensional hypercube array with exponentially-decaying interactions}
Assuming a large number of emitters, each emitter approximately contributes
\begin{equation}
    \frac{1}{N}\text{Tr} \, \mathbf{\Gamma}^2 \approx 1 + 2D (\gamma^2 + \gamma^4 + \gamma^6 + \ldots) + 2^D \sum_{n_1,n_2,\ldots,n_D=1} \gamma^{2\sqrt{n_1^2 + \ldots + n_D^2}},
\end{equation}
where the second term comes from the emitters displaced by exactly integer multiples of the Cartesian unit vectors, and the third term comes from the rest of the emitters which can be divided into $2^D$ equal contributions by symmetry. The final sum can be approximated by converting it to an integral in hyperspherical coordinates and exploiting the rotational symmetry to get
\begin{equation}
    \sum_{n_1,n_2,\ldots,n_D=1} \gamma^{2\sqrt{n_1^2 + \ldots + n_D^2}} \sim C \int_{0}^{\infty}dr \, r^{D-1} \gamma^{2r} = \frac{C}{(-\ln \gamma)^D}
\end{equation}
for some constant $C$. Note that this is asymptotically independent of $N$.

\subsection{$1/r$-interactions in 1D, 2D and 3D arrays}
Next, we consider dissipative long-range interactions that decay as $1/r$, with $\gamma_{ij} \propto \frac{1}{\kappa r_{ij}}$. Although an accurate modelling of dipole interactions require additional short- and medium-range contributions (like $1/r^3$ and $1/r^2$ terms), we will show that including only the $1/r$ interactions results in qualitative agreement with previous numerical simulations for 1D, 2D and 3D arrays in the large-$N$ regime~\cite{Robicheaux2021theoretical,Sierra2021dicke}.
\subsubsection{1D chain}
Here, $\gamma_{ij} = \delta_{ij} + (1-\delta_{ij}) \frac{\gamma}{|i-j|}$, and thus the second order correlation function is readily found to be
\begin{equation}
    g^{(2)}(0) = 1 - \frac{2}{N} + \frac{1}{N}\left(1+2\gamma^2 \sum_{n=1}^\infty \frac{1}{n^2}\right) \approx 1 + \frac{1}{N} \left( \frac{\pi^2}{3} \gamma^2 - 1 \right),
\end{equation}
which gives the superradiant transition point $\gamma_s = \sqrt{3}/\pi$ which is once again independent of $N$.
\subsubsection{2D square array}
Labelling the emitters' Cartesian coordinates on the square grid as $\vec{r}_i = (x_i,y_i)$, where $x_i, y_i$ are integers, we have $\gamma_{ij} = \delta_{ij} + (1-\delta_{ij}) \frac{\gamma}{|\vec{r}_i-\vec{r}_j|}$. Therefore,
\begin{equation}
    \text{Tr}({\mathbf{\Gamma}^2}) \approx N\gamma^2 \left(1 + \frac{2\pi^2}{3} + 4 \sum_{x,y=1}^{n} \frac{1}{x^2 + y^2}\right) \approx N\gamma^2 ( A + B \ln{N} ),
\end{equation}
where $N = n^2$, and $A$ and $B$ are real constants which can be determined numerically. From this, we can deduce that superradiance condition takes the form
\begin{equation}
    \gamma > \frac{2}{\sqrt{A+B \ln{N}}} \sim \frac{1}{\sqrt{\ln {N}}}.
\end{equation}
\subsubsection{3D cubic array}
The calculations are similar to the 2D case, but with $\vec{x}_i = (x_i,y_i,z_i)$. We have 
\begin{equation}
\text{Tr}({\mathbf{\Gamma}^2}) \approx N\gamma^2 \left(1 + \pi^2 + 8 \sum_{x,y,z=1}^{n} \frac{1}{x^2 + y^2 + z^2}\right) \approx N\gamma^2 (A + B N^{1/3}),
\end{equation}
where $N = n^3$. This gives the superradiance condition of the form
\begin{equation}
    \gamma > \frac{2}{\sqrt{A+B N^{1/3}}} \sim \frac{1}{N^{1/6}}.
\end{equation}
Note that these couplings are consistent with those found in Ref.~\cite{Sierra2021dicke}.

\subsection{1D chain with infinite-range interactions}
On the other extreme, we consider the case of infinite-range interactions between emitters on a 1D array, i.e. $|\gamma_{ij}| > 0$ in the limit $|x_i - x_j| \to \infty$. This can be realized for example in waveguide quantum electrodynamics by coupling atoms to a 1D photonic crystal or a nanophotonic waveguide, where the photons in the waveguide mediate the infinite-range interactions between the spins~\cite{Munro2018population,Solano2017super}. We also consider the more general case of a chiral waveguide, where the photon emission rate is direction-dependent~\cite{Lodahl2017chiral}. Denoting the left(right) emission rate by $\gamma_L(\gamma_R)$ and neglecting the coherent interactions, we can model the dissipative emitter dynamics (after tracing out the environment) with the master equation~\cite{Pichler2015quantum}
\begin{equation}
    \dot{\rho} = -i[H,\rho] + \gamma_L \mathcal{D}[\hat{c}_L]\rho + \gamma_R \mathcal{D}[\hat{c}_R]\rho,
\end{equation}
where $\hat{c}_{L/R} = \sum_j e^{\pm i k x_j} \sigma_j^-$ is the collective spin operator. $H$ contains the individual emitter Hamiltonian as well as coherent interactions between the emitters. We will not specify the form of $H$ here since it is not relevant for the purposes of calculating $g^{(2)}(0)$. The decoherence matrix elements are
\begin{equation}
    \gamma_{jl} = \cos((kd(j-l)) - i \chi \sin(kd(j-l))
\label{eq:chiral_matrix}
\end{equation}
where $\chi \equiv (\gamma_R - \gamma_L)/(\gamma_R + \gamma_L)$ is the chirality parameter. Normalizing the decoherence matrix by dividing $\gamma_L + \gamma_R$ throughout, we obtain the matrix elements
\begin{equation}
    \gamma_{jl} = \frac{\gamma_L}{\gamma_L + \gamma_R} e^{ik (x_j - x_l)} + \frac{\gamma_R}{\gamma_L + \gamma_R} e^{-ik(x_j - x_l)} = \cos(k(x_j - x_l)) - i \chi \sin(k(x_j - x_l)).
\end{equation}
We now assume a constant emitter separation of $d$, and calculate in a similar fashion as the case of exponentially-decaying interactions
\begin{equation}
\text{Tr} \, (\mathbf{\Gamma}^2) = \sum_{i,j = 1}^N |\gamma_{ij}|^2 = \frac{1}{4} \csc^2(kd) \left[ (1-\chi^2) (1-\cos(2Nkd)) + N^2 (1+\chi^2) (1-\cos(2kd)) \right]
\end{equation}
and
\begin{equation}
    g^{(2)}(0) = \frac{1}{2}\left(3+\chi^2 - \frac{4}{N}\right) + \frac{1-\chi^2}{2N^2} \csc^2(kd) \sin^2(Nkd).
\end{equation}
Note that since
\begin{equation}
    \partial_{|\chi|} g^{(2)}(0) = \frac{|\chi| \csc^2(kd)}{2N^2} \left[ N^2(1-\cos(2kd)) + \cos(2Nkd) - 1\right] \geq 0
\end{equation}
we conclude that chirality always enhances the superradiance effect, except at points when $g^{(2)}(0)$ is already at maximal value of $2(1-1/N)$.

For a consistency check, the limit $kd \to 0$ yields the well-known result $g^{(2)}(0) = 2(1-1/N)$ for the original Dicke model~\cite{Dicke1954coherence,Gross1982superradiance}, where $g^{(2)}(0) > 1$ for all $N > 2$ can be interpreted as a spontaneous superradiance~\cite{Gross1982superradiance}. Physically, when the emitters become sufficiently close together (compared to the wavelength) such that they become indistinguishable, and the only decay channel is that associated with the collective jump operator $\hat{c}_1 = \sum_j \sigma_j^-$ at a rate $N$ times of the individual decay rate.

\begin{figure}
\centering
\subfloat{%
  \includegraphics[width=0.6\linewidth]{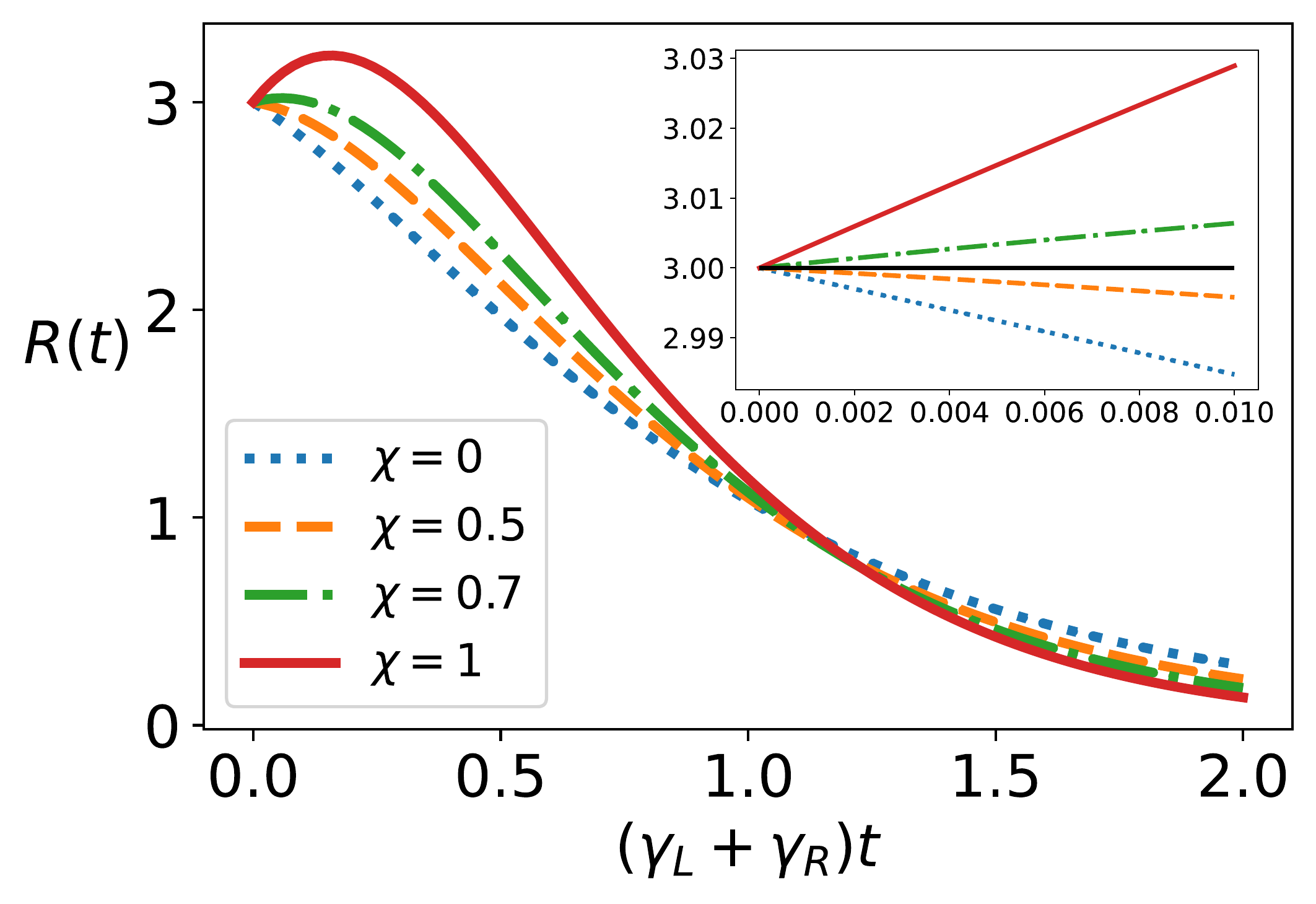}%
  \label{}%
}
	\caption{Total emission rate $R(t)$ for $N=3$ emitters coupled to a 1D chiral waveguide with various chirality parameters $\chi$. Analytical calculations predict $\dot{R}(0) > 0$ (superradiant burst) for $\chi > 1/\sqrt{3} \approx 0.577$. Inset shows the short-time behavior of $R(t)$. The black horizontal line at $R = 3 $ is added for reference. The numerical results demonstrate that a minimum chirality is required for superradiance.}
	\label{fig:chiral_N=3}
\end{figure}

Noting that $\partial_{|\chi|} g^{(2)}(0) \geq 0$, we find that a larger chirality $|\chi|$ always results in a stronger superradiance effect. The limiting case $|\chi| = 1$ corresponds to a perfectly unidirectional emission, where the dynamics become independent of the emitter separation~\cite{mok2020long,mok2020microresonators,mahmoodian2020dynamics}. When $Nkd = m\pi$ where $m \in \mathbb{Z}$ and $m/N \notin \mathbb{Z}$, $g^{(2)}(0)$ is minimal with the value $(3+\chi^2 - 4/N)/2$. For $N > 3$, superradiance always occurs regardless of chirality. However, for $N = 3$, we require a minimum chirality of $|\chi| = 1/\sqrt{3}$ such that $g^{(2)}(0) > 1$, as confirmed via numerical simulations shown in Fig.~\ref{fig:chiral_N=3}. Of course, with only $N = 3$ emitters, the superradiance effect just above the threshold chirality will be quite weak.

\subsection{All-to-all interactions}
The all-to-all dissipatively-coupled system of emitters can be described a complete graph $G$ with $N$ vertices. The decoherence matrix is $\mathbf{\Gamma} = \mathbf{I}_N + \gamma \mathbf{A}$, where $\gamma$ is the interaction strength and $\mathbf{A}$ is the corresponding adjacency matrix. This leads to
\begin{equation}
    \text{Tr} \, (\mathbf{\Gamma}^2) = N + 2 \gamma^2 E(G) = N + \gamma^2 N(N-1)
\end{equation}
and
\begin{equation}
    g^{(2)}(0) = 1 + \frac{\gamma^2(N-1)-1}{N}.
\end{equation}
The case of $\gamma=1$ is exactly the original Dicke model which always results in superradiance. For arbitrary $\gamma$, the transition point is $\gamma > \gamma_s$, where $\gamma_s = 1/\sqrt{N-1}$. Such a model is always physically valid, i.e. $\gamma_p = 1$. To see this, we note that $\mathbf{\Gamma}$ corresponds to the master equation
\begin{equation}
    \dot{\rho} = \gamma\mathcal{D}\bigg[\sum_j \sigma_j^-\bigg]\rho + (1-\gamma) \sum_{j} \mathcal{D}[\sigma_j^-]\rho,
\label{eq:dickewithlocaldecay}
\end{equation}
which is in the standard Lindblad form. Physically, this corresponds to the Dicke model with local dissipation serving as an imperfection. The condition $\gamma > \gamma_s$ therefore sets an upper bound to the amount of local dissipation such that superradiance is preserved. Intuitively, superradiance becomes more robust to local dissipation for large $N$. This also agrees with the exact solution of Eq.~\eqref{eq:dickewithlocaldecay} obtained recently in the large-N limit~\cite{malz2022large}.

\section{Third-order correlation function, $g^{(3)}(0)$}
\label{sec:g3}
We can characterize the superradiant burst beyond the initial rate $\dot{R}(0)$ by calculating the third-order correlation function $g^{(3)}(0)$~\cite{Masson2022universality},
\begin{equation}
    g^{(3)}(0) = 1 - \frac{6}{N} + \frac{12}{N^3} + 3\left(1- \frac{4}{N}\right) \frac{\text{Tr} \, \mathbf{\Gamma}^2}{(\text{Tr} \, \mathbf{\Gamma})^2} + 2 \frac{\text{Tr} \, \mathbf{\Gamma}^3}{(\text{Tr} \, \mathbf{\Gamma})^3}.
\label{eq:g30}
\end{equation}
From Ref.~\cite{Robicheaux2021theoretical}, the second derivative of $R(t)$ evaluated at the initial time was found to be 
\begin{equation}
    \ddot{R}(0) = \frac{8}{N^2} (\text{Tr} \, \mathbf{\Gamma})^3 - \frac{8}{N} (\text{Tr} \, \mathbf{\Gamma}) (\text{Tr} \, \mathbf{\Gamma}^2) + (\text{Tr} \, \mathbf{\Gamma}^3)
\end{equation}
which can be expressed in terms of the correlation functions as
\begin{equation}
\begin{split}
    \ddot{R}(0) &= (\text{Tr} \, \mathbf{\Gamma})^3 \bigg[1 + \frac{2}{N} - \frac{2}{N^2} - \left( \frac{3}{2} + \frac{2}{N} \right) g^{(2)}(0) + \frac{1}{2} g^{(3)}(0) \bigg].
\end{split}
\end{equation}
This can then be used to estimate the peak time and intensity of the superradiant burst.

In Ref.~\cite{Masson2022universality}, it was observed that for the fully excited emitters, the third-order correlation function $g^{(3)}(0)$ cannot be greater than $1$ without superradiance, i.e., $g^{(2)}(0) > 1$. The physical meaning is that the three-photon emission cannot be enhanced unless the two-photon emission is also enhanced. However, as we will see, this is not true in general and depends greatly on the dissipative model used.

\begin{figure}
\centering
\subfloat{%
  \includegraphics[width=0.6\linewidth]{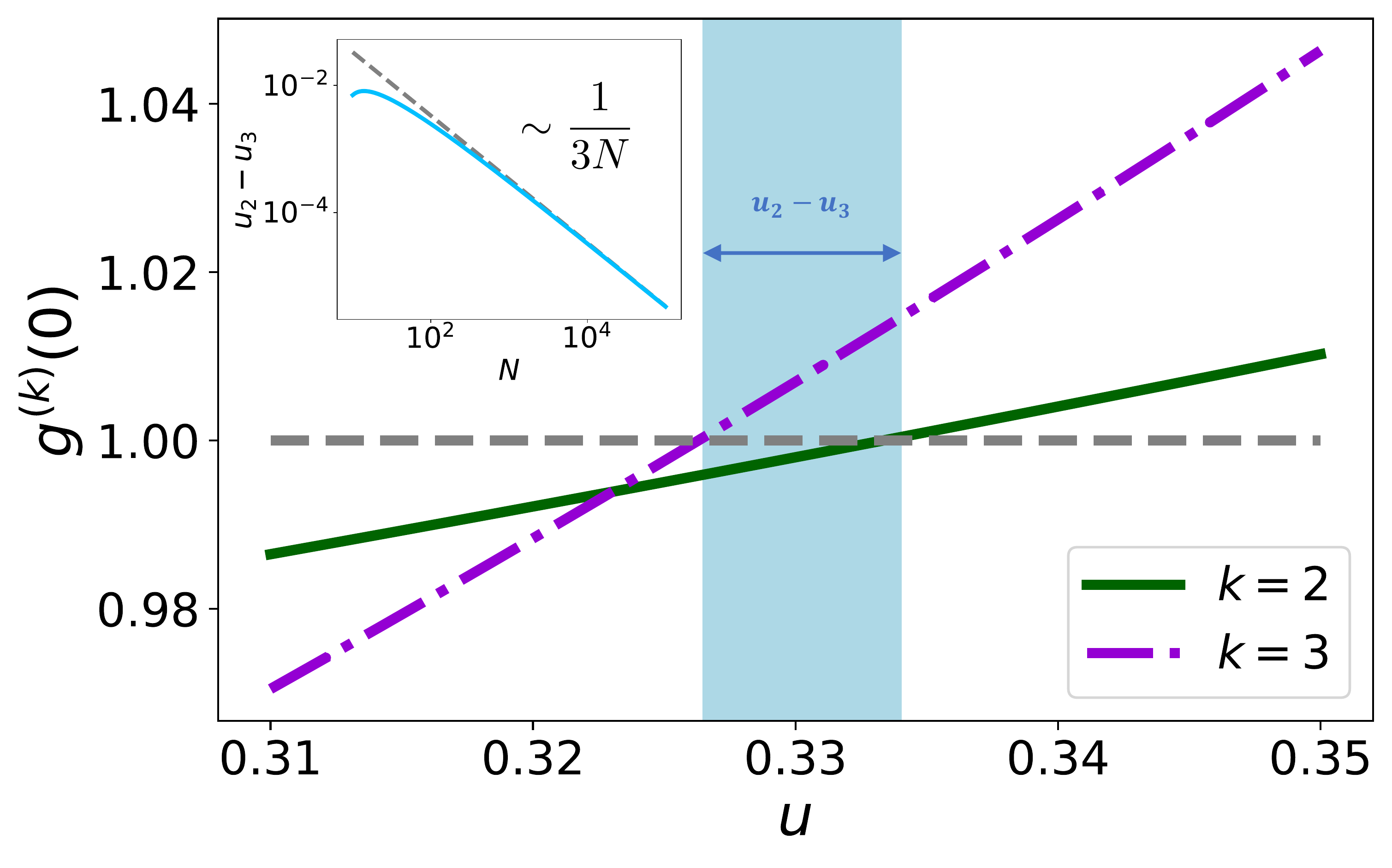}%
  \label{}%
}
	\caption{Second- and third-order correlation functions for the all-to-all interaction model with $N=10$ emitters. By increasing the collective coupling $u$, $g^{(3)}(0) > 1$ is achieved (at $u = u_3$) before $g^{(2)}(0) > 1$ (at $u = \gamma_2$). The inset shows the gap $\gamma_2 - u_3$ against $N$, with the asymptotic behavior $1/3N$ as illustrated by the gray dashed line.}
	\label{fig:dicke_N=10}
\end{figure}

Without loss of generality, let $\text{Tr} \, \mathbf{\Gamma} = N$. Calculating $g^{(3)}(0)$ subject to the constraint $g^{2}(0) = 1$ yields the simple condition 
\begin{equation}
    g^{(3)}(0) > 1 \implies \text{Tr} \, \mathbf{\Gamma}^3 > 6N.
\label{eq:g3>1}
\end{equation}
By the Cauchy-Schwarz inequality, we must have $\text{Tr} \, \mathbf{\Gamma}^3 \leq (\text{Tr} \, \mathbf{\Gamma})(\text{Tr} \, \mathbf{\Gamma}^2) = 2N^2$, which allows for the condition in Eq.~\eqref{eq:g3>1} to hold. Let us now test this for some of the previously introduced models:

\begin{itemize}
    \item 1D ring with exponentially-decaying interactions: At the superradiant transition $\gamma = \gamma_s = 1/\sqrt{3}$ and using the decay rates in Eq.~\eqref{eq:eigvals_1Dring}, we have $\text{Tr} \, \mathbf{\Gamma}^3 \approx 11N/2 < 6N$. Thus, $g^{(3)}(0) < 1$, which agrees with our intuition.
    \item 1D ring with NN and NNN interactions: At the superradiant transition $\gamma_2 = \sqrt{(1-2\gamma_1^2)/2}$ and $\gamma_1 \in [0,1/\sqrt{2}]$, we have $\text{Tr} \, \mathbf{\Gamma}^3 = N(4 + 3 \sqrt{2} \gamma_1^2 \sqrt{1-2\gamma_1^2}) < 6N$. Thus, $g^{(3)}(0) < 1$, which agrees with our intuition.
    \item All-to-all interactions: At the superradiant transition $\gamma = \gamma_s = 1/\sqrt{N-1}$, we have
    \begin{equation}
        \text{Tr} \, \mathbf{\Gamma}^3 = \frac{N(N-2 + 4 \sqrt{N-1})}{\sqrt{N-1}}
    \end{equation}
    which can exceed $6N$ for $N > 2(2+\sqrt{2}) \approx 6.8$. Thus, for a sufficiently large number of emitters, we can enhance three-photon emission ($g^{(3)}(0) > 1$) while suppressing two-photon emission $(g^{(2)}(0) < 1)$ (see Fig.~\ref{fig:dicke_N=10}). The range of $\gamma$ for which this occurs decreases as $1/3N$ for large $N$, such that $g^{(2)}(0)$ and $g^{(3)}(0)$ both cross $1$ at the same value of $\gamma$ as $N \to \infty$.
\end{itemize}
The all-to-all interaction model, which is a Dicke model with local dissipation, provides a simple counterexample to the intuitive notion that in the absence of a superradiant burst, three-photon emission is suppressed. The pertinent question, however, is whether the simultaneous two-photon suppression and three-photon enhancement results in a `delayed superradiance', that is, $R(t)$ first decreases for a short time before rising above its initial value. Surprisingly, such a phenomenon is indeed possible, but we found that the resulting superradiant burst is extremely weak. 
\subsection{Delayed superradiance in Dicke model with local dissipation}
\label{app:delayed}
For the Dicke model with local dissipation,
\begin{equation}
    \dot{R}(0) = N^2 \gamma^2 - N(1+\gamma^2)
\end{equation}
and
\begin{equation}
    \ddot{R}(0) = N^3 \gamma^3 + N^2 \gamma^2 (3\gamma + 5) + N(1+\gamma^2(2\gamma+5)).
\end{equation}
\begin{figure}
\centering
\subfloat{%
  \includegraphics[width=0.5\linewidth]{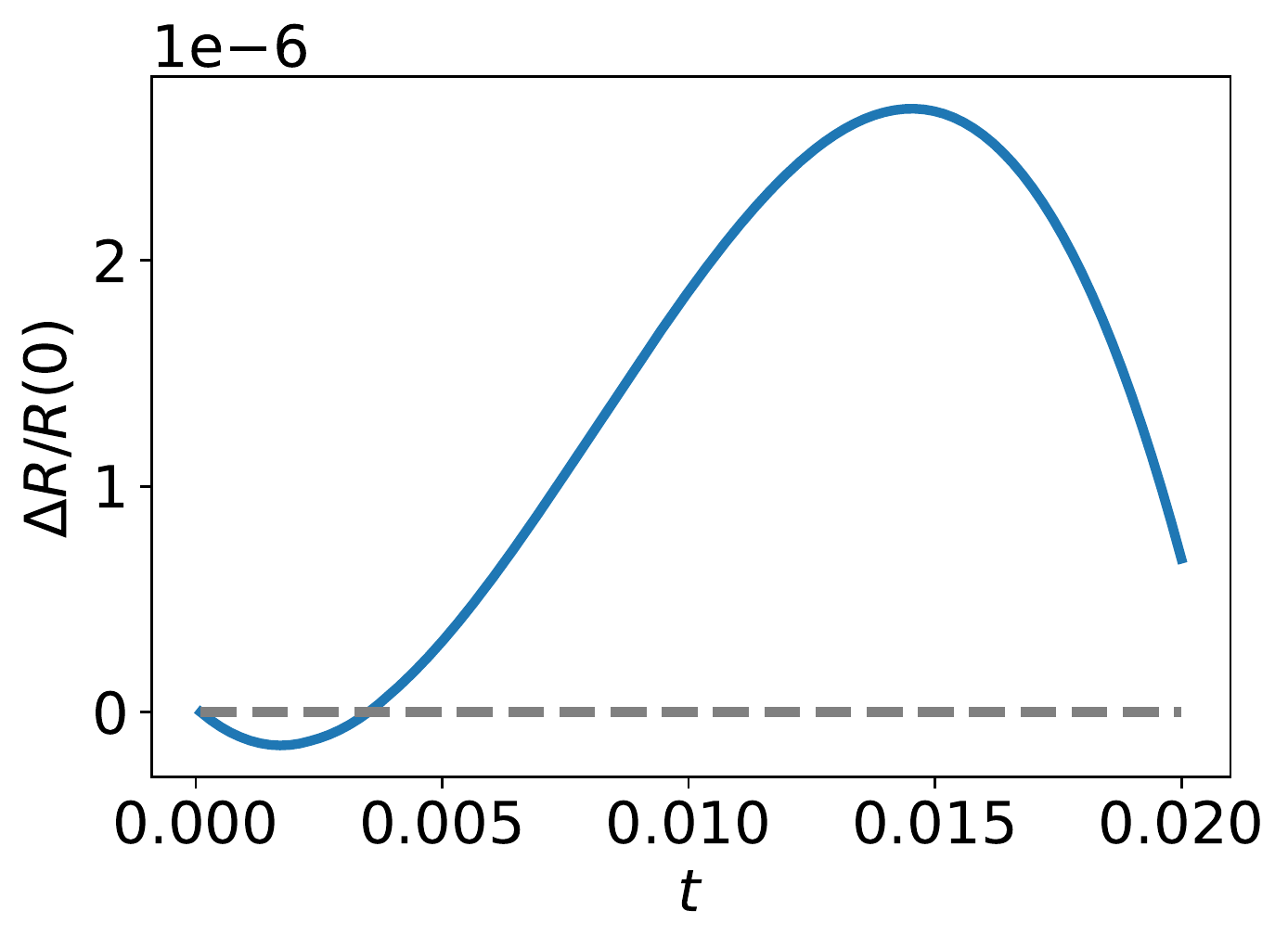}%
  \label{}%
}
	\caption{Fractional change in total emission rate with respect to the initial rate $R(0)$, for $N = 20$ emitters and coupling parameter $u = (1-\epsilon)/\sqrt{N-1}$, $\epsilon = 10^{-4}$. The peak emission rate has a very small fractional increase of approximately $10^{-6}$, and the superradiant burst is thus negligible.}
	\label{fig:dicke_delayed}
\end{figure}
For a suitable choice of $\gamma$, one can satisfy $\dot{R}(0)<0$ and $\ddot{R}(0) > 0$ simultaneously. When $\dot{R}(0) = 0$, $\ddot{R}(0) > 0$ for $N > 2(5+2\sqrt{5}) \approx 18.9$. However, having $\dot{R}(0)<0$ and $\ddot{R}(0) > 0$ alone does not guarantee a delayed superradiance, since the total emission rate $R(t)$ may not exceed the initial value $R(0)$. From numerical simulations, we observe delayed superradiance for a narrow parameter regime, but the resulting superradiant burst is very weak, with the peak emission rate only very slightly above the initial rate (fractional increase on the order of $10^{-6}$). We also remark that this effect is not present in the nearest and next-nearest interaction models, and therefore does not affect the validity of our claims in the main text.

Generically, in the presence of local dissipation (where the physical origin of the local term is different from that of the interactions) one has to be mindful about how to properly account for the distinct decay channels. This can be done by defining a ``directional'' second order correlation function, as done in Ref.~\cite{Robicheaux2021theoretical,silvia2022many}.

\subsection{Physical interpretation of $g^{(2)}(0)$ and $g^{(3)}(0)$}
We can obtain an alternative physical interpretation of the correlation functions $g^{(2)}(0)$ and $g^{(3)}(0)$ by connecting them with quantum jumps. Let us define the normalized $K$-jump state by applying $K$ collective jump operators $\hat{c}_\nu$ on the fully-excited initial state $\ket{e}$
\begin{equation}
    \ket{\Phi_K (\vec{\nu})} = \frac{1}{\mathcal{N}} \prod_{j=1}^{K} \hat{c}_{\nu j} \ket{e},
\end{equation}
where $\vec{\nu} = (\nu_1,\ldots, \nu_K)$ and $\mathcal{N}$ is the normalization factor. For an arbitrary 1-jump state $\ket{\Phi_1(\nu)}$, using the expansion $\hat{c}_\nu = \sum_j v_{\nu j} \sigma_j^-$, we have the emission rate
\begin{equation} 
R_\nu = \sum_{i,j} \gamma_{ij} \braket{\Phi_1(\nu)|   \sigma_i^+ \sigma_j^- |\Phi_1(\nu)} = N-1 + \sum_{\substack{i,j \\ i \neq j}} \gamma_{ij} v_{\nu j}^* v_{\nu i}.
\end{equation}
Taking average of all the 1-jump states weighted by their respective decay rates give
\begin{equation}   
\bar{R}_1 = \frac{1}{N} \sum_{\nu} \Gamma_\nu R_\nu = N - 1 + \frac{1}{N} (\text{Tr}(\mathbf{\Gamma}^2) - N ) = N g^{(2)}(0),
\end{equation}
where we have used $\gamma_{jk} = \gamma_{kj} = \sum_\nu \Gamma_\nu v_{\nu j}^* v_{\nu k}$ from the spectral decomposition of $\mathbf{\Gamma}$. Thus, we can interpret $g^{(2)}(0)$ as proportional to the average emission rate of the 1-jump states, and $g^{(2)}(0) < 1$ implies that these states do not give rise to superradiance. Now, let us consider the normalized 2-jump states
\begin{equation}
\ket{\Phi_2 (\nu,\mu)} = \frac{\hat{c}_\mu \hat{c}_\nu \ket{e}}{\sqrt{1+\delta_{\mu \nu} - 2 \sum_j |v_{\nu j}|^2 |v_{\mu j}|^2}}.
\end{equation}
Taking the weighted average of the rates of all 2-jump states, we have
\begin{equation}
    \bar{R}_2 = \frac{1}{N^2} \sum_{\nu,\mu} \Gamma_\nu \Gamma_\mu \frac{\sum_\chi \Gamma_\chi \braket{e|\hat{c}_\nu^\dag \hat{c}_\mu^\dag \hat{c}_\chi^\dag \hat{c}_\chi \hat{c}_\mu \hat{c}_\nu|e}}{1+\delta_{\mu \nu} - 2 \sum_j |v_{\nu j}|^2 |v_{\mu j}|^2}.
\end{equation}
We now make the assumption that $\sum_j |v_{\nu j}|^2 |v_{\mu j}|^2$ vanishes as $N \to \infty$. This is true for lattices with periodic boundary conditions in which $v_{\nu j}$ are the plane wave coefficients. As an example, we have $v_{\nu j} = \exp{(2\pi i j \nu / N)}/\sqrt{N}$ for the 1D ring which simply comes from the eigenvectors of a circulant matrix. Hence, as $N \to \infty$,
\begin{equation}
    \bar{R}_2 \leq \frac{1}{N^2} \sum_{\nu, \mu,\chi} \Gamma_\nu \Gamma_\mu \Gamma_\chi \braket{e|\hat{c}_\nu^\dag \hat{c}_\mu^\dag \hat{c}_\chi^\dag \hat{c}_\chi \hat{c}_\mu \hat{c}_\nu|e} = N g^{(3)}(0).
\end{equation}
Thus, we can interpret $N g^{(3)}(0)$ as the upper bound for the average emission rate of the 2-jump states, and $g^{(3)}(0) < 1$ implies that these states do not give rise to superradiance.

\section{Upper bounds on the emission rate}
The superradiance condition $g^{(2)}(0) > 1$ is strictly speaking only valid for classifying superradiance at short times up to $\mathcal{O}(\gamma_0 t)$. `Delayed superradiance' is thus technically possible, hence it is not true in general that $R(t)>R(0)$ for some $t \implies g^{(2)}(0) > 1$. Nonetheless, if we can upper bound $R(t)$ and compare it with $R(0) = N$, we can obtain a regime in which the fully-excited initial state maximizes $R$ which guarantees no superradiant burst at arbitrary times and also for arbitrary system Hamiltonian.
\begin{proposition}
The emission rate $R$ can be written as the average of some spin Hamiltonian $H_{\Gamma}$, where the interaction strengths between the $i^{\text{th}}$ and $j^{\text{th}}$ spins are given by the decoherence matrix elements $\gamma_{ij}$. 
\end{proposition}
This is straightforward to show:
\begin{equation}
\begin{split}
    R &= \sum_{\nu=1}^{N} \Gamma_\nu \sum_{j,k=1}^{N} v_{\nu j}^* v_{\nu k} \braket{\sigma_j^+ \sigma_k^-} = \sum_{j,k=1}^{N} \braket{\sigma_j^+ \sigma_k^-} \sum_{\nu=1}^{N} \Gamma_\nu v_{\nu j}^* v_{\nu_k}= \sum_{j,k=1}^{N} \braket{\sigma_j^+ \sigma_k^-} \left( \sum_{\nu=1}^{N} \Gamma_{\nu} \vec{v}_\nu (\vec{v}_\nu^*)^T \right)_{k,j} \\ &= \sum_{j,k=1}^{N} 
 \gamma_{kj} \braket{\sigma_j^+ \sigma_k^-} \equiv \braket{H_\Gamma},
\end{split}
\end{equation}
where we have used the spectral decomposition of $\Gamma$ in the last step. Hence, the problem of calculating $R$ is equivalent to finding the average energy of a state under the Hamiltonian $H_\Gamma$. We remark that $H_{\Gamma}$ introduced here is only used for calculation purposes, and should not be interpreted as a physical Hamiltonian of our system. With this, we now prove that for our $\text{NN}$ interaction model in arbitrary dimensions, $R$ is maximized by the fully excited state $\ket{e}^{\otimes N}$.
\subsection{D-dimensional nearest-neighbor interactions}
\begin{theorem}
Let $\mathbf{\Gamma}$ describe a nearest-neighbor interaction model, with $\gamma_{ij} = \delta_{ij} + \gamma \delta_{\braket{ij}}$, where $\gamma \in [0,1]$. $\delta_{\braket{ij}} = 1$ if the emitters indexed by $i$ and $j$ are nearest-neighbor on the $D-$dimensional regular lattice, and $0$ otherwise. For $\gamma \leq 1/(2D)$, the emission rate $R$ is maximized by the fully-excited state $\ket{e}^{\otimes N}$ with $R=N$.
\end{theorem}
\textit{Proof.} Let us construct the auxiliary Hamiltonian as
\begin{equation}
    H_{\Gamma} = \sum_{j=1}^{N} \sigma_j^+ \sigma_j^- + \gamma \sum_{\braket{j,k}} (\sigma_j^+ \sigma_k^- + \sigma_k^+ \sigma_j^-),
\end{equation}
which is the XY Hamiltonian with transverse magnetic field. For $D = 1$, this can be solved exactly by mapping to free fermions using the Jordan-Wigner transformation. However, such an approach is not necessary for the proof since we only require the highest-energy (dominant) eigenstate of $H_{\Gamma}$. One can easily check that $\ket{e}^{\otimes N}$ is an eigenstate of $H_{\Gamma}$ with eigenenergy $N$. It remains to be shown that this is in fact the dominant eigenstate for $\gamma \leq 1/(2D)$.

To this end, let us consider the matrix representation of $H_{\Gamma}$ in the standard tensor-product basis $\{ \ket{e}, \ket{g} \}^{\otimes N}$. Note that $H_{\Gamma}$ conserves the total excitation $\sum_j \sigma_j^+ \sigma_j^-$, and hence can be block-diagonalized into sub-blocks with different excitation numbers. In $D$-dimensions, each emitter can only at most interact with $2D$ nearest-neighbors. For the basis state with $N - m$ excitations ($0 \leq m \leq N$), the number of interactions is at most $2Dm$, corresponding to the basis states where the $m$ ground-state emitters are not nearest-neighbors with one another. Hence, the row sum of the matrix $H_{\Gamma}$ for that particular basis state is at most $(N - m) + 2Dm \gamma$, with the first term coming from the diagonal element.

Using the Gershgorin circle theorem, we know that all the eigenvalues of $H_{\Gamma}$ must lie within the largest Gershgorin disc. More specifically, we can bound the largest eigenvalue as
\begin{equation}
    \lambda_{\text{max}} \leq \max_{m,\gamma} (N-m + 2Dm\gamma) \leq \max_{m} \left(N-m + 2Dm \times \frac{1}{2D} \right) = N.
\end{equation}
Hence, $\ket{e}^{\otimes N}$ maximizes $\braket{H_\Gamma}$ and thus maximizes $R$. $\Box$

The condition $\gamma \leq 1/(2D)$ is within the physically valid regime. In fact, this is exactly the physically valid regime for either the $N \to \infty$ limit or if the lattice obeys periodic boundary conditions. For open boundary conditions, the exact physically valid regime contains a correction factor of $1/\cos(\pi/(N^{1/D}+1)) > 1$ which makes it slightly larger than $1/(2D)$. This factor tends to $1$ as $\sim \pi^2/(2(N^{1/D})^2)$, hence the small discrepancy would not matter if the number of emitters per spatial dimension $N^{1/D}$ is sufficiently large.

Relating back to our original problem of superradiance, we can write down an obvious consequence of Theorem 1.
\begin{corollary}
For a nearest-neighbor interaction model with $\gamma \leq 1/(2D)$, assuming a fully-excited initial state $\ket{\psi(0)} = \ket{e}^{\otimes N}$, superradiance is impossible for any system Hamiltonian. In other words, $R(t) \leq R(0) \, \forall t>0$.
\end{corollary}

For the nearest-neighbor model, the upper bound on $R(t)$ is tight which allows us to rigorously prove the impossibility of superradiance at all times. However, this is not true in general for other interaction models $\forall \gamma \in [0,\gamma_p]$. Nonetheless, we can still use the bound to find a regime where superradiance is definitely not possible. For simplicity, we assume periodic boundary conditions for the rest of the section.

\subsection{1D exponentially-decaying interactions}
Consider the 1D exponential model with $N$ emitters. For simplicity, we take $N$ to be odd and assume periodic boundary conditions. For the basis state with $N-m$ excitations, the maximum number of couplings between the excited and ground states in the 1D ring is $2 \times m^\prime \times (N-1)/2 = m^\prime (N-1)$ where $m^\prime = \min\{m,N-m\}$ ranges from $0$ to $(N-1)/2$. In general, all the couplings will have a different contribution $\gamma^{|i-j|}$ to the $H_\Gamma$ matrix. Thus, for a fixed $\gamma$,
\begin{equation}
    \lambda_{\text{max}} \leq \max_{m^\prime} \left[ N-m^\prime + 2m^\prime \left(\gamma + \gamma^2 + \ldots + \gamma^{(N-1)/2}\right) \right]
\label{eq:1D_exp_bound}
\end{equation}
For $\gamma = 1$, $\lambda_{\text{max}} \leq N + (N-2)(N-1)/2 \sim N^2/2$ recovers the quadratic scaling of the Dicke model. For $\gamma < 1$, we can bound $\lambda_{\text{max}}$ as
\begin{equation}
    \lambda_{\text{max}} \leq \max_{m^\prime} \left(N - m^\prime + 2m^\prime \frac{\gamma}{1-\gamma}\right) = \begin{cases} N, \quad &\gamma \leq 1/3 \\ N + \frac{N-1}{2} \left( \frac{3\gamma-1}{1-\gamma}\right), \quad &1/3 < \gamma < 1 \end{cases}
\end{equation}
from which we see that $\ket{e}^{\otimes N}$ maximizes $R$ for $\gamma \leq 1/3 < 1/\sqrt{3} = \gamma_s$. For $\gamma < 1$, the bound for the peak emission rate has a linear scaling with $N$.

\subsection{1D power-law interactions}
This is similar to the calculations for the exponential model, but with the replacement $\gamma^{|i-j|} \to \gamma/|i-j|$, $i \neq j$. Performing the same analysis, we have
\begin{equation}
\begin{split}
    \lambda_{\text{max}} &\leq \max_{m^\prime} \left[ N-m^\prime + 2m^\prime \left(\gamma + \frac{\gamma}{2} + \frac{\gamma}{3} + \ldots + \frac{\gamma}{(N-1)/2} \right) \right] \\ &\leq \max_{m^\prime} \left[ N - m^\prime + 2m^\prime \gamma (\log N + \zeta )  \right] \\&= N + \frac{N-1}{2} \left(2 \gamma (\log N + \zeta) - 1 \right) = O(N \log N)
\label{eq:1D_power_bound}
\end{split}
\end{equation}
where $\zeta \approx 0.577$ is the Euler-Mascheroni constant. 

\section{Effect of Hamiltonian on emission rate}
For the fully-excited initial state, it can be shown that $R(t)$ is independent of the Hamiltonian up to quadratic order in time~\cite{Robicheaux2021theoretical}. The situation is more complicated for other initial states, but this is beyond the scope of our work as we have noted throughout the manuscript. This is intuitive since the coherent spin-spin interactions are negligible at early times when the emitters are fully excited.

An exact argument that the effect of Hamiltonian is small (at least insufficient to induce a superradiant burst) beyond a time of $\mathcal{O}\left(t^2\right)$ requires one to solve the master equation analytically for arbitrary times, which is of course a difficult task. Nonetheless, we can provide a reasonable justification of this claim by using a second-order mean-field approach, also known as a second-order cumulant expansion~\cite{robicheaux2021beyond,oriol2022characterizing}. Consider the Hamiltonian $H = \sum_{i,j} J_{ij} \sigma_i^+ \sigma_j^-$. Using mean-field to factorize $\braket{\sigma_i^+ \sigma_j^- e_n} \approx \braket{\sigma_i^+ \sigma_j^-} \braket{e_n}$ where $e_n = \sigma_n^+ \sigma_n^-$, we can write the dynamics for the emission rate as (assuming $\gamma_{ij} = \gamma_{ji}$)
\begin{align}
    \dot{R} &\approx -R + \sum_{\substack{m,n \\ m \neq n}} \gamma_{mn} \left[ -\braket{\sigma_m^+ \sigma_n^-} + 2 \gamma_{mn} \braket{e_m e_n} - \gamma_{mn} \braket{e_m}  \right] \nonumber\\&+ \sum_{\substack{m,n \\ m \neq n}} \sum_{l \neq m,n} \gamma_{mn} (2\braket{e_n} - 1) \left[- 2\text{Re}(J_{nl}) \text{Im}\braket{\sigma_m^+ \sigma_l^-}  
 - 2\text{Im}(J_{nl}) \text{Re}\braket{\sigma_m^+ \sigma_l^-} + \gamma_{nl}\text{Re}\braket{\sigma_m^+ \sigma_l^-} \right].
\label{eq:MF_ratedynamics}
\end{align}
Using this, we can now make the following claim:
\begin{theorem}
Let $H = \sum_{i,j} J_{ij} \sigma_i^+ \sigma_j^-$ be a real Hamiltonian with $J_{ij} = J_{ji}$. In the second-order mean-field theory, if $H$ and $\mathbf{\Gamma}$ are translation-invariant with periodic boundary conditions, the emission rate $R(t)$ is independent of $H$ for any translation-invariant initial state and $t \geq 0$.
\end{theorem}
\textit{Proof.} Notice from Eq.~\eqref{eq:MF_ratedynamics} that the real part of $J_{nl}$ only couples to the imaginary part of the spin-spin correlation function $\braket{\sigma_m^+ \sigma_l^-}$. For any translation-invariant state, we must have $\braket{\sigma_m^+ \sigma_l^-} = \braket{\sigma_l^+ \sigma_m}$ since the correlations only depend on the geometrical separation between the spins $l$ and $m$. Hence, the correlations must be real, and the effect of $H$ on $R$ vanishes. $\Box$

Another simple consequence of a real Hamiltonian is that it does not affect the rate of the $K-$jump states
\begin{equation}
\ket{\psi_{K}} \propto \prod_{\nu=1}^K \hat{c}_\nu \ket{e}^{\otimes N} 
\end{equation}
up to a normalization factor. $\hat{c}_1 \ldots \hat{c}_K$ are arbitrary jump operators obtained by diagonalizing $\mathbf{\Gamma}$. This result is easy to prove. From the Heisenberg equation,
\begin{equation}
    \dot{R} = i \braket{\psi_K| [H,H_{\Gamma}] | \psi_K} = - 2 \text{Im} \braket{\psi_K | H H_{\Gamma} | \psi_K}.
\end{equation}
Since $\mathbf{\Gamma}$ is real and symmetric, the expansion coefficients $v_{\nu j}$ in $\hat{c}_\nu$ can be chosen to be real. Hence $\ket{\psi_K}$ is real, and any real Hamiltonian $H$ will thus contribute $\dot{R} = 0$.

We now comment on the assumptions used in Theorem 2. Firstly, the assumption that $\mathbf{\Gamma}$ is translation-invariant already applies to the models used in the manuscript, where we assume that all the emitters experience the same dissipative interactions. The periodic boundary condition assumption is also not  in the large-N limit, which is usually the case of interest. The assumption on the interaction Hamiltonian, including the reality assumption, is required to ensure that the state remains translationally-invariant which is a key part of the proof. The most extreme example of such a Hamiltonian is the coherent all-to-all interactions with a common arbitrary interaction strength $J$. Theorem 2 implies that even such a strong Hamiltonian would not significantly affect the emission rate, which we have verified numerically by simulating the master equation exactly for the 1D exponential model with $N=13$ emitters. The translational symmetry used in the proof is also reasonable since one expects superradiance to be stronger for symmetric states (with the strongest superradiance achieved for the permutation-symmetric Dicke state).  

The strength of this argument thus boils down to the validity of the second-order mean field theory. It is well-known that the second-order mean field approach breaks down when calculating the strength of the two-body correlations particularly at long times $t\gg\gamma_0^{-1}$, where $\gamma_0$ is the spontaneous emission rate of each emitter. Previous numerical simulations have also shown that the mean-field approach (at least second-order and above) is rather accurate in terms of the emission rate (see Fig. 3 of~\cite{robicheaux2021beyond} and Fig. 2(c) of~\cite{oriol2022characterizing}). We point out two reasons why we believe that the inadequacies of mean-field theory do not greatly affect the validity of Theorem 2:
\begin{enumerate}
\item At long times $t\gg\gamma_0^{-1}$, the total excitation of the emitters is close to zero. Note that the total excitation must decrease monotonically since the Hamiltonian conserves total excitation, and we do not introduce any pumping in the model. Hence, this makes the emission rate small. Another physical interpretation is that the subradiant states are populated at large times which is also why the mean field assumption breaks down.
\item Our argument is based on the translation symmetry of the state which is of course still true in the exact dynamics. Also, our second-order approach already accounts for short-range correlations which are even more so appropriate for the short-range interaction models we use in the manuscript (NNN, exponential), especially in the large-$N$ limit.	
\end{enumerate}

\subsection{No superradiant burst for nearest-neighbor interactions, translation-invariant initial state}
We now argue that in the second-order mean-field theory, superradiant burst for NN interactions cannot occur at any time $t \geq 0$, using a translation-invariant initial state. From Theorem 2, it suffices to only consider dissipative processes. For the $D$-dimensional NN interaction model, the decay rates are
\begin{equation}
    \Gamma_{\bf{k}} = 1 + \gamma \sum_{i=1}^{D} \cos k_i,
\end{equation}
where ${\bf{k}} = (k_1, \ldots, k_D)^T$, and $k_i \in [-\pi,\pi]$. The corresponding jump operators are
\begin{equation}
    c_{\bf{k}} = \frac{1}{\sqrt{N}} \sum_{\bf{x}} e^{-i\bf{k}\cdot \bf{x}} \sigma_{\bf{x}}^-
\end{equation}
where ${\bf{x}} \in \mathbb{Z}^D$. We consider the thermodynamic limit $N \to \infty$. For a translation-invariant state $\ket{\psi}$,
\begin{equation}
\begin{split}
    \dot{R} &= -\sum_{\mathbf{k},\mathbf{k}^\prime} \Gamma_{\mathbf{k}} \Gamma_{\mathbf{k}^\prime} \braket{c_{\mathbf{k}}^\dag [c_\mathbf{k},c_{\mathbf{k}^\prime}^\dag] c_{\mathbf{k}^\prime}} \\&= \frac{1}{N^2} \sum_{\mathbf{k},\mathbf{k}^\prime} \sum_{\mathbf{x}_1,\mathbf{x}_2,\mathbf{x}_3} (1+\gamma \sum_i \cos k_i)(1+\gamma \sum_i \cos k_i^\prime) \exp(i\mathbf{k} \cdot (\mathbf{x}_1 - \mathbf{x}_2))\exp(i\mathbf{k}^\prime \cdot (\mathbf{x}_2 - \mathbf{x}_3)) \braket{\sigma_{\mathbf{x}_1}^\dag \sigma_{\mathbf{x}_2}^z \sigma_{\mathbf{x}_3}^-} \\
    &\approx \frac{1}{(2\pi)^{2D}} \sum_{\mathbf{x}_1,\mathbf{x}_2,\mathbf{x}_3} \braket{\sigma_{\mathbf{x}_1}^\dag \sigma_{\mathbf{x}_2}^z \sigma_{\mathbf{x}_3}^-} \int d\mathbf{k}  (1+\gamma \sum_i \cos k_i) \exp(i\mathbf{k} \cdot (\mathbf{x}_1 - \mathbf{x}_2)) \int d\mathbf{k}^\prime (1+\gamma \sum_i \cos k_i^\prime)\exp(i\mathbf{k}^\prime \cdot (\mathbf{x}_2 - \mathbf{x}_3)) \\
    &= \sum_{\mathbf{x}_1,\mathbf{x}_2,\mathbf{x}_3} \braket{\sigma_{\mathbf{x}_1}^\dag \sigma_{\mathbf{x}_2}^z \sigma_{\mathbf{x}_3}^-} \left(\delta_{\mathbf{x}_1 \mathbf{x}_2} + \frac{\gamma}{2} \delta_{\langle \mathbf{x}_1 \mathbf{x}_2 \rangle}\right)\left(\delta_{\mathbf{x}_2 \mathbf{x}_3} + \frac{\gamma}{2} \delta_{\langle \mathbf{x}_2 \mathbf{x}_3 \rangle}\right)
\end{split}
\end{equation}
where the Kronecker delta $\delta_{\langle \mathbf{x}_1 \mathbf{x}_2 \rangle}$ is 1 if $\mathbf{x}_1$ and $\mathbf{x}_2$ are NN sites, and vanishes otherwise. Since the state remains translation-invariant under time evolution, we can denote the population of each site as $p$, NN and NNN correlations of the form $\braket{\sigma_{\mathbf{x}}^+ \sigma_{\mathbf{x}^\prime}^-}$ as $c_1$ and $c_2$ respectively. Since we are only interested in the possibility of superradiance, and $c_1, c_2$ are real (by translation invariance), we only need to consider the case of $c_1 > 0, c_2 > 0$. Furthermore, since the interactions are NN, it is also reasonable to assume $c_1 > c_2$. Hence, we can simplify the above expression to give
\begin{equation}
\begin{split}
    \dot{R} &= -Np - 2 N D \gamma c + \frac{1}{4} \gamma^2 \sum_{\mathbf{x},\mathbf{x}^\prime} \braket{\sigma_{\mathbf{x}}^+ \sigma_{\mathbf{x}}^- \sigma_{\mathbf{x}^\prime}^z} \delta_{\langle \mathbf{x}_1 \mathbf{x}_2 \rangle} + \frac{1}{4} \gamma^2 \sum_{\substack{\mathbf{x},\mathbf{x}^\prime,\mathbf{x}^{\prime\prime} \\ \mathbf{x} \neq \mathbf{x}^{\prime\prime}}} \braket{\sigma_{\mathbf{x}}^+ \sigma_{\mathbf{x}^\prime}^z \sigma_{\mathbf{x}^{\prime\prime}}^-} \delta_{\langle \mathbf{x}\mathbf{x}^\prime \rangle}\delta_{\langle \mathbf{x}^\prime \mathbf{x}^{\prime \prime} \rangle} \\ &\leq -Np - 2ND\gamma c_1 + \frac{1}{2}ND\gamma^2 p + \frac{1}{4}N(2D)(2D-1)\gamma^2 c_2
\end{split}
\end{equation}
Maximizing the dissipative interactions by setting $\gamma = 1/(2D)$, we have
\begin{equation}
    \dot{R} \leq -N\left(1-\frac{1}{8D}\right) p - N \left(\frac{3}{4} + \frac{1}{2D}\right) c_1 \leq 0
\end{equation}
which shows that SR is not possible for any translation-invariant state $\ket{\psi}$. This also holds for the evolved (mixed) state at arbitrary time since it can be written as a convex mixture of translation-invariant pure states. Hence, we conclude that $\dot{R}(t) \leq 0$ $\forall t$.

Note that apart from omitting Hamiltonian contributions (justified via second-order mean-field theory), the remainder of the argument does not use any mean-field assumption. Hence, this should hold in the exact case if the dissipative processes are dominant.

\newpage
\bibliography{bib}